\documentclass[prx,longbibliography,twocolumn,notitlepage,showpacs,amsmath,amstex,amssymb,citeautoscript,superscriptaddress]{revtex4-1}
\usepackage{etoolbox}
\usepackage{comment}
\usepackage{lmodern}
\usepackage{graphicx}
\usepackage{mathtools}
\usepackage{amsmath}
\usepackage{amssymb}
\usepackage[export]{adjustbox}
\usepackage{dcolumn}
\usepackage{bm}
\usepackage{hyperref}
\hypersetup{linktocpage,colorlinks,citecolor={blue},pdfdisplaydoctitle=true,pdfpagemode=UseOutlines,bookmarksnumbered=true}
\usepackage{mathrsfs,dsfont}
\usepackage{xcolor}
\usepackage[normalem]{ulem}

\usepackage{bbold} 
\usepackage{gensymb}
\usepackage{csquotes}

\usepackage{braket}

\newcommand{\tr}{\mathrm{tr}}

\newcommand{\DD}{\mathcal{D}}
\newcommand{\VV}{\mathcal{V}}

\definecolor{cbl}{rgb}{0,0,1}
 
\definecolor{crd}{rgb}{1,0,0}
 
\definecolor{Blue}{rgb}{0.0, 0.0, 0.5}

\newcommand{\be}{\begin{equation}}
\newcommand{\ee}{\end{equation}}
\newcommand{\bea}{\begin{eqnarray}}
\newcommand{\eea}{\end{eqnarray}}

\newcommand{\up}{\uparrow}
\newcommand{\down}{\downarrow}

\usepackage{cancel}

\begin{document}

\newcommand{\titleinfo}{Discrete time-crystalline response stabilized by domain-wall confinement}

\title{\titleinfo}

\author{Mario Collura}
\email{mcollura@sissa.it}
\affiliation{SISSA and INFN, 
via Bonomea 265, I-34136 Trieste, Italy}

\author{Andrea De Luca}
\email{andrea.de-luca@cyu.fr}
\affiliation{Laboratoire de Physique Th\'eorique et Mod\'elisation (CNRS UMR 8089),
Universit\'e de Cergy-Pontoise, F-95302 Cergy-Pontoise, France}  

\author{Davide Rossini}
\email{davide.rossini@unipi.it}
\affiliation{Dipartimento di Fisica dell’Università di Pisa and INFN, 
Largo Pontecorvo 3, I-56127 Pisa, Italy}

\author{Alessio Lerose}
\email{alessio.lerose@unige.ch}
\affiliation{Department of Theoretical Physics,
University of Geneva, Quai Ernest-Ansermet 30,
1205 Geneva, Switzerland}

\begin{abstract}
  Discrete time crystals represent a paradigmatic nonequilibrium phase of periodically driven matter.
  Protecting its emergent spatiotemporal order necessitates a  mechanism that hinders the spreading of defects, 
  such as localization of domain walls in disordered quantum spin chains.
  In this work, we establish the effectiveness of a different mechanism arising in clean spin chains:
  the confinement of domain walls into ``mesonic'' bound states.
  We consider translationally invariant quantum Ising chains periodically kicked at arbitrary frequency, 
  and discuss two possible routes to domain-wall confinement: longitudinal fields and interactions beyond nearest neighbors. 
  We study the impact of confinement on the  order parameter evolution by constructing domain-wall-conserving 
  effective Hamiltonians and analyzing the resulting  dynamics of domain walls.
  On the one hand, we show that for arbitrary driving frequency the symmetry-breaking-induced confining potential
  gets effectively averaged out by the drive, leading to deconfined dynamics. 
  On the other hand, we rigorously prove that increasing the range $R$ of spin-spin interactions $J_{i,j}$
  beyond nearest neighbors enhances the order-parameter lifetime \textit{exponentially} in $R$.
  Our theory predictions are corroborated by a combination of exact and matrix-product-state simulations for finite
  and infinite chains, respectively.
  The long-lived stability of spatiotemporal order identified in this work does not rely on Floquet prethermalization
  nor on eigenstate order, but rather on the nonperturbative origin of vacuum-decay processes.
  We point out the experimental relevance of this new mechanism for stabilizing a long-lived time-crystalline response
  in Rydberg-dressed spin chains.
\end{abstract}

\date{\today}
\maketitle 
\section{Introduction}

Because of the subtle role played by the temporal dimension, spontaneous breaking of time-translational symmetry has long escaped conclusive theoretical formulations~\cite{WilczekPRL12,khemani2019brief,Sacha17review}.
A meaningful characterization of an extended many-body system as a \textit{time crystal} requires robust stationary macroscopic oscillations, without a net exchange of energy with external devices~\cite{khemani2019brief}. 
A leap forward has been taken with the realization that certain nonequilibrium setups~\cite{SachaPRA15_MeanfieldTC,KhemaniPRL16,ElsePRL16,ElsePRX17_PrethermalDTC,VonKeyserlingkPRB16_AbsoluteStability,RussomannoPRB17_LMGDTC} allow to circumvent the obstacles posed by thermal equilibrium~\cite{BrunoPRL13,WatanabeOshikawaPRL15}.
Discrete time crystals (DTCs) formed by interacting periodically driven quantum spin systems currently represent the theoretical paradigm of large-scale spatiotemporal ordering~\cite{khemani2019brief,ElseReview20}. Signatures of their stable and robust subharmonic response to the drive have been experimentally observed with several state-of-the-art experimental platforms~\cite{choi2017observation,zhang2017observation,kyprianidis2021observation,DTCexperimentGoogle,DTCexperimentNVCenters,DTCexperimentFrey,DTCexperimentXu}.

A major challenge to realizing a DTC is the fact that the external drive tends to repeatedly inject excitation energy into the system, and the resulting heating generally deteriorates large-scale spatiotemporal ordering.
Protecting  order against melting necessitates a mechanism to keep the impact of dynamically generated excitations under control and thus prevent indefinite entropy growth.
To date, many-body localization (MBL)~\cite{AndersonLocalization,BAA,HuseNandkishore,AbaninRMP} represents the single robust mechanism to stabilize 
a persistent subharmonic DTC response: The strong quenched disorder of a MBL system freezes the motion of local excitations, thereby stabilizing long-range order throughout the many-body spectrum of the system~\cite{HusePRB13_LocalizationProtectedQuantumOrder,ChandranPRB14_MBLProtectedSPT}.
This infinite-time stability can be characterized as eigenstate order.
On the other hand, the quest for disorder-free DTCs calls for alternative mechanisms to evade thermalization. 
The crucial observation that energy absorption from the drive is asymptotically suppressed for large driving frequencies~\cite{MoriPRL16,AbaninPRB17_EffectiveHamiltonians} allowed Else et al.~\cite{ElsePRX17_PrethermalDTC} to formulate the notion of a prethermal DTC as a long-lived (as opposed to permanently stable) dynamical phase exhibiting broken time-translational symmetry.
This phase relies on the existence of an effective Hamiltonian governing the transient dynamics in a suitable rotating frame, possessing an emergent Abelian symmetry determined by the driving protocol, and supporting a spontaneous breaking of this symmetry at low enough effective temperature.
This condition guarantees that a relevant set of initial states will display  quasistationary
macroscopic oscillations of the order parameter in the original frame.

Necessary ingredients for prethermal DTC behavior are a high-frequency drive and an interaction structure that supports a thermal phase transition; in one dimension, this requires fat-tailed long-range interactions~\cite{ElsePRX17_PrethermalDTC}.
A natural question is whether one can realize a robust long-lived DTC response exploiting different mechanisms of thermalization breakdown, such as quantum many-body scarring~\cite{turner2018weak,serbyn2021quantum,bluvstein2021controlling}, Hilbert-space fragmentation~\cite{SalaPRX20,KhemaniHermeleNandkishorePRB20}, Stark~\cite{RefaelStarkMBL,SchultzStarkMBL,starkdtc} and other kinds of disorder-free quasi-MBL~\cite{DeRoeck1,DeRoeckHuveneersScenario,CarleoGlassy,AbaninProsen:SlowDynamics,Cirac:QuasiMBLWithoutDisorder,Papic:MBLWithoutDisorder,Schiulaz1,SmithDisorderFreeLocalization,BrenesConfinementLGT,RussomannoPRR20,SlowFermiHubbard}. 
Indeed, Refs.~\cite{bluvstein2021controlling,starkdtc} already discussed signatures of DTC behavior stabilized via related mechanisms.
A possibility that recently attracted considerable interest is provided by confinement of excitations, an archetypal phenomenon of particle physics~\cite{wilson74} which also exists in low-dimensional condensed matter models~\cite{McCoyWuConfinement,ShankarConfinement,DelfinoDecayIFF,RutkevichMesonSpectrumLattice,RutkevichConfinementXXZ}. 
By now, there is mounting evidence and convincing understanding of robust nonthermal behavior in quantum spin chains with confined excitations~\cite{Kormos:2017aa,RobinsonNonthermalStatesShort,MazzaTransport,LeroseSuraceQuasilocalization,Verdel19_ResonantSB,PaiPretkoFractonsLGT,Tagliacozzo2019,GorshkovConfinement,LeroseDWLR,TanConfinementExperiment}, intimately related to characteristic phenomena of (lattice) gauge theories~\cite{RutkevichFalseVacuum,LeroseSuraceQuasilocalization}.

In this paper, we explore the efficacy of confinement of excitations to stabilize the DTC spatiotemporal order. 
Motivated by the main limitations of the theory of prethermal DTC,
we perform our analysis at arbitrary driving frequencies and do not require slowly decaying interactions.
Our main result is that an extremely long-lived DTC response can be stabilized by domain-wall confinement,
without relying on a Floquet-prethermal long-range ordered Gibbs ensemble.
This occurrence is made possible by the intrinsic slowness of the prethermal dynamics: 
The meltdown of transient spatiotemporal ordering involves nonlocal processes such as the dynamical generation of unbound domain walls; the stronger domain-wall confinement, the longer such processes take.

\begin{figure*}
  \includegraphics[width=0.89\textwidth]{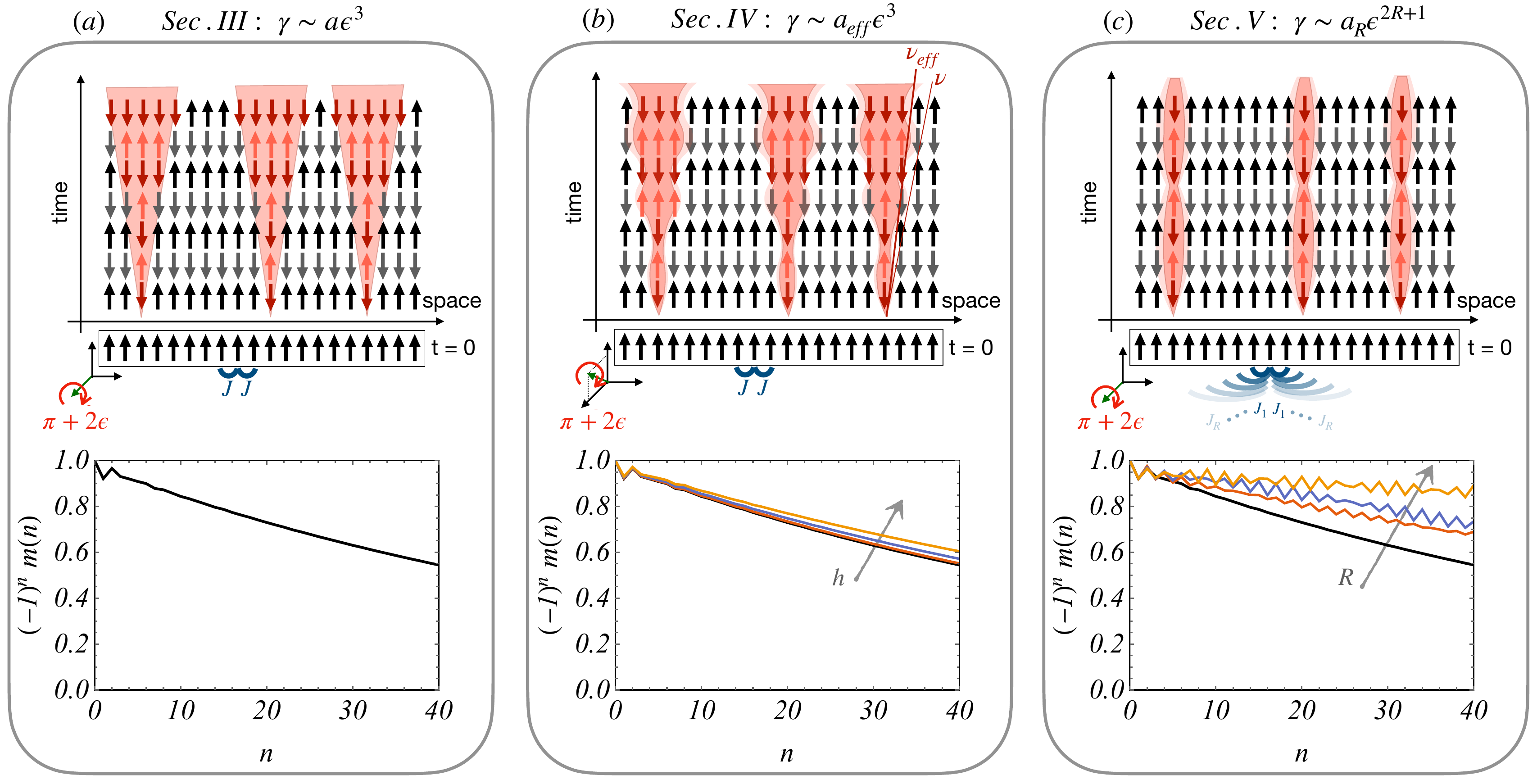}
  \caption{Illustrative summary of the results of this paper for the decay of the stroboscopic order parameter
    $m(n)=\langle Z_j(n) \rangle \sim (-1)^n e^{-\gamma n}$ at integer times $t_n=n$ in kicked Ising-type spin chains.
    {\it (a)} Top: In the standard  transverse-field kicked Ising chain, the decay is driven by the fractionalization
    of isolated spin flips, generated by small imperfections of magnitude $\epsilon$ in the kick, into pairs of unbound
    traveling domain walls. The exact decay rate in Eq.~\eqref{eq_tauexact} is interpreted as $\gamma\sim \rho v$,
    with $\rho \sim \epsilon^2$ the density of spin flips and $v\sim\epsilon$ the spreading velocity of domain walls. 
    Bottom: A representative example [$J = 0.685$, $\epsilon=0.2$ in Eq.~\eqref{eq_toggling}].
    {\it (b)} A tilt of the kick axis (green arrow in bottom left sketches) gives rise to domain-wall confinement,
    hindering the spreading of reversed domains. 
    However, the periodic flips average out the confining potential, resulting in deconfinement.
    Bottom: Mild slowdown of the decay for increasing values of the longitudinal kick component
    [$h=0.2,0.4,0.6$ in Eq.~\eqref{eq_imperfectkick}].
    {\it (c)} Couplings of range $R>1$  beyond nearest neighbors give rise to a form of domain-wall confinement
    completely insensitive to periodic flips. We rigorously establish that, generically, 
    the decay is only triggered by the fractionalization of rare large reversed bubbles into pairs of unbound domain walls.
    This results in a qualitative suppression of the decay rate  as $\gamma\sim\epsilon^{2R+1}$.
    Bottom: Strong enhancement of the order-parameter lifetime upon increasing the couplings' range to a distance
    $R=2$, $3$, $4$ [$J_2=0.144$, $J_3=0.058$, $J_4=0.03$ in Eq.~\eqref{eq_FloquetLR}].
    Note that very weak additional couplings $J_{2,3,4} \ll \epsilon$ suffice to stabilize the response.}
  \label{fig:intro}
\end{figure*}
\begin{figure*}
  \centering
  \includegraphics[width=0.85\textwidth]{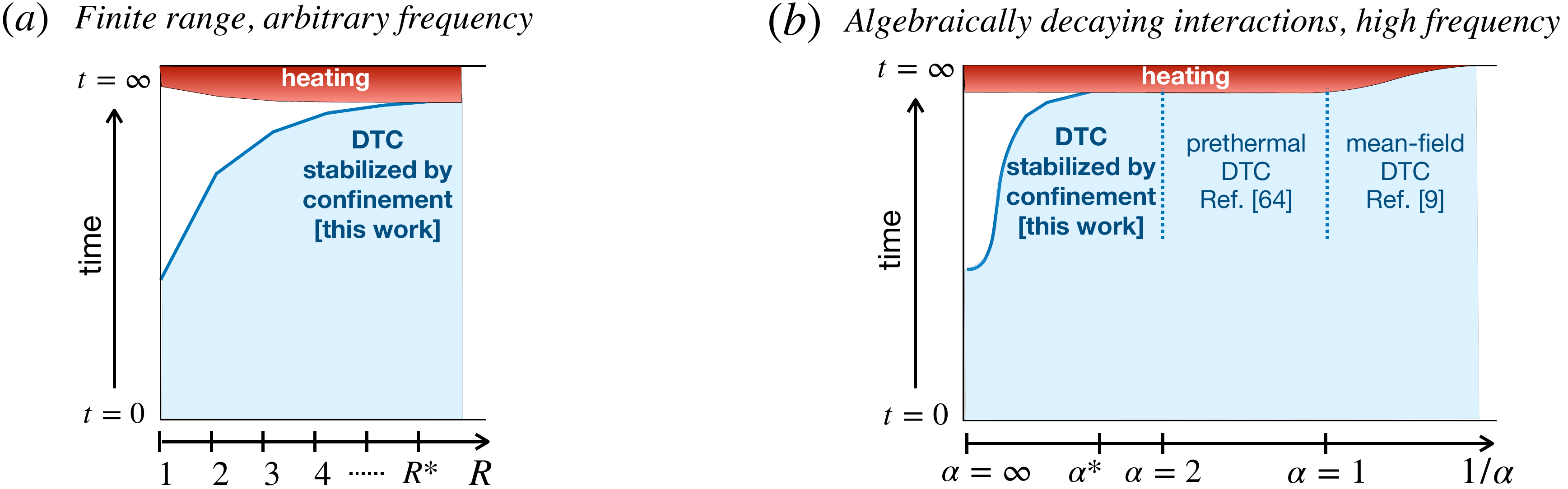}
  \caption{Sketch of the timescales involved in this work and beyond.
    {\it (a)}:  Finite interaction range $R$ and arbitrary driving frequency [Eq.~\eqref{eq_FloquetLR}].
    Here we fix $\epsilon\ll 1$. The blue region denotes the extent of the DTC lifetime window upon increasing $R$;
    the blue upper boundary is the timescale $1/\gamma \sim \epsilon^{-(2R+1)}$ proved in this work [Eq.~\eqref{eq_decayLR}].
    The heating region's lower boundary represents the lower bound $T_{\mathrm{preth}} \ge \exp(C/\epsilon^{1/(2R+1+\delta)})$
    on the heating timescale [Eq.~\eqref{eq_timescaleLRR}]. The crossing point $R^*=R^*(\epsilon)$ is pushed to $\infty$
    (i.e. to the top-right corner) as $\epsilon\to0$. 
    {\it (b)}:  Algebraically decaying interaction and high driving frequency [Eq.~\eqref{eq_FloquetLRHF}].
    Here we take $\epsilon \ll J$ and fix a small $\tau$. The blue region denotes the extent of the DTC lifetime window
    upon decreasing $\alpha$. 
    {\it Left:}  For $\alpha>2$, the blue upper boundary is the timescale $1/\gamma\sim\epsilon^{ -A \epsilon^{-1/(\alpha-2)}}$
    argued in this work [Eq.~\eqref{eq_gammaLRinfty}]. The heating region's lower boundary represents the lower bound
    $T_{\mathrm{preth}} \ge \exp(C/\tau)$ on the heating timescale, which is insensitive to the precise value of $\alpha>1$:
    see, e.g., Ref.~\cite{KUWAHARA201696}. The crossing point $\alpha^*=\alpha^*(\epsilon,\tau)$ is slowly pushed
    to $\infty$ (i.e. to the top-left corner) as $\epsilon\to0$.
    {\it Center:}  Prethermal DTC phase for $1<\alpha\le2$, as established in Ref.~\cite{MachadoPRX20}.
    {\it Right}: Mean-field DTC, established in Ref.~\cite{RussomannoPRB17_LMGDTC} for infinite-range interactions $\alpha=0$.
    For $0<\alpha\le1$, dynamics  preserve the mean-field character for a timescale that diverges with system size, 
    see Refs.~\cite{mori2018prethermalization,LerosePappalardiPRR20}.}
  \label{fig_timeline}
\end{figure*}

\section{Overview of results}

We provide here an overview of the contents and results of this paper, which is meant as guidance to the reader.
Our findings are summarized and illustrated in Figs.~\ref{fig:intro},~\ref{fig_timeline}.

We start (Sec.~\ref{sec:FIM}) by exactly solving for the order-parameter dynamics of  the integrable periodically-kicked transverse-field Ising chain. 
We establish a possibly long but perturbative decay rate $\gamma\sim\epsilon^3$ of DTC response, where $\epsilon$ represents the deviation from a kicking protocol implementing perfect spin flips.
Furthermore, we identify the physical mechanism leading to order-parameter meltdown as the spreading of a small density $\rho\sim\epsilon^2$ of dynamically generated reversed spins, as the domain walls delimiting them freely move at velocity $v\sim\epsilon$ [Fig.~\ref{fig:intro}{\it (a)}].
In passing, we note that this result clarifies previous contradictory findings on finite-size scaling of the DTC signal in this model~\cite{Angelakis_Ising, Choudhury_Ising}. 

Building on this physical intuition, we introduce domain-wall confinement via symmetry-breaking longitudinal fields (Sec.~\ref{sec_mixedfield}), as a mechanism to prevent the spreading of reversed domains.
In absence of perfect flips, confinement individually stabilizes both the positively and negatively magnetized states.
However, similarly to what happens in MBL and high-frequency driven spin chains~\cite{VonKeyserlingkPRB16_AbsoluteStability,ElsePRX17_PrethermalDTC}, we show that symmetry-breaking terms are averaged out by the drive, generating deconfined effective dynamics despite domain walls being instantaneously confined at all times.
As a result, the order-parameter decay rate is only weakly affected, retaining its perturbative nature $\gamma\sim\epsilon^3$  [Fig.~\ref{fig:intro}{\it (b)}].

We finally consider extending the range of spin-spin interactions, $J_{i,j}\neq0$ for $|i-j|\le R$, as an alternative route to domain-wall binding, which does not suffer from incompatibility with the (explicit or emergent) $\mathbb{Z}_2$ symmetry (Sec.~\ref{sec_LR}).
The crucial feature that arises in these systems is the coexistence in the spectrum of both ``topologically charged'' excitations (kinks and antikinks) and ``neutral'' confined bound states.
During prethermal dynamics, confined excitations only generate vacuum fluctuations, resulting in long-lived coherent oscillations of the order parameter. Dynamical generation of unbound kinks and antikinks triggers the vacuum decay, and hence the decay of DTC spatiotemporal order.
We prove that this phenomenon is heavily suppressed by the nonlocal nature of topological excitations. 
In fact, we rigorously establish an exponential enhancement of the order-parameter lifetime,
i.e. $\gamma\sim\epsilon^{2R+1}$, under mild genericity assumptions on the couplings $J_{i,j}$ [Fig.~\ref{fig:intro}{\it (c)}].
In other words, the fastest process leading to DTC order melting occurs at a perturbative order that grows with the interaction range $R$.
Leveraging this result, we finally conjecture that for algebraically decaying interactions $J_{i,j} = J/|i-j|^\alpha$ the decay rate is asymptotically suppressed faster than any power of the perturbation, $\gamma\sim\epsilon^{ A \epsilon^{-1/(\alpha-2)}}$, in the parameter range $\alpha>2$ where no prethermal order is possible. (As $\alpha$ approaches $2$ from above, however, the lifetime $1/\gamma$ is eventually superseded by the heating timescale, see below.)
We present numerical simulations that not only confirm our theory predictions, but even point to a much more robust and extended stability than analytically understood.

The consequences of our findings are further discussed in the conclusive Sec.~\ref{sec_conclusions}:
\begin{itemize}
\item The stabilization mechanism identified in this work does {\it not} rely on Floquet-prethermal finite-temperature order, nor on eigenstate order. Rather, it relies on the long-term metastability of ``false vacua'', familiar from high energy physics~\cite{SchwingerMechanism,ColemanFalseVacuum}: atypical states with finite energy density,  that decay through rare macroscopic tunneling phenomena. This idea extends the theory of time crystals beyond  previously known mechanisms, circumventing some of their limitations (cf. Fig.~\ref{fig_timeline}). 
\item This work indicates a clear route to observing DTC behavior in a class of quantum simulation platforms of growing importance, namely Rydberg-dressed arrays of neutral atoms~\cite{RydbergDressed,Zeiher17coherent,RydbergDressed2}. 
These systems are characterized by a great degree of control  on the interaction range and strength.
Compared to conventional prethermal  and MBL DTCs, the milder requirements to observe confinement-stabilized DTCs come at the price of reducing the set of initial states exhibiting DTC response.
\item
Finally, as a byproduct on the theory side, our work clarifies a long-standing issue, namely the nature of the apparent anomalous persistence of the order parameter in the quench dynamics of quantum spin chains with algebraically decaying interactions $J_{i,j}=J/|i-j|^\alpha$, $\alpha>2$ previously observed in several numerical studies~\cite{ZunkovicPRL18_Merging,HalimehPRB17_PersistentOrder,GorshkovConfinement}. 
Our theory establishes that a strong enhancement of the order parameter lifetime {\it is to be generally expected}  in the parameter regime $2<\alpha\ll \infty$, and predicts its functional form as a function of the quench magnitude.  
This has the important consequence, not recognized before, that long-lived nonequilibrium order can be sustained by systems in the ``short-range'' regime $\alpha>2$, which cannot spontaneously form an ordered state in thermal equilibrium.
The meltdown of this metastable order is ultimately triggered at long times by the dynamical generation of deconfined topological excitations (domain walls).
\end{itemize}

\section{Order parameter decay in the kicked transverse-field Ising chain}
\label{sec:FIM}

Our starting point is the standard Ising Hamiltonian 
\be
\label{eq:H}
H_J = -J \sum_{j=1}^{L} Z_{j} Z_{j+1} ,
\ee
where $J>0$ is the ferromagnetic coupling, $X_{j}, Y_{j}, Z_{j}$ are the local spin-$1/2$ Pauli matrices for the $j$th spin, and periodic boundary conditions ($j+L \equiv j$) are assumed.
The Floquet dynamics is obtained by periodically intertwining the evolution governed by $H_J$ with kicks $K=\prod_j K_j$ at integer times $t_n=n=1,2,\dots$. The resulting single-cycle time-evolution (Floquet) operator reads
\be
\label{eq_Floquet}
{U} = K \, V_{J}, \qquad V_{J} =   e^{i J \sum_{j} Z_{j} Z_{j+1} } .
\ee

To observe the simplest realization of time-crystalline spatiotemporal order, the system is initially prepared in the fully polarized state with positive magnetization along the $\hat z$ direction, 
namely, one of the two degenerate ground states of $H_{J}$. 
This has a  product-state form ${|+\rangle \equiv |\cdots \up\up\up \cdots \rangle}$,
where $|\!\up\rangle$ ($|\!\down\rangle $)
denotes the eigenvector of the Pauli matrix $Z$ with eigenvalue $+1$ ($-1$).
The kick $K$ is taken to rotate each spin by an angle
$\pi $ around a transverse axis:
\footnote{The choice of period $1$ sets the energy scale: throughout this work, all Hamiltonian parameters will be considered dimensionless.}
\be
{K \longrightarrow K_{\pi/2}  =  e^{i \frac \pi 2 \sum_{j} X_{j}}.}
\ee
In this case, the time evolved state after $n$ kicks
$|\Phi(n)\rangle = {U}^{n}|+\rangle$ 
exhibits a sequence of perfect jumps between $|+\rangle$ and
the other ground state of $H_J$, namely $|-\rangle \equiv \ket{\cdots \downarrow\downarrow \downarrow \cdots}$.
The persistent nonvanishing value of the order parameter 
\be
m(n) = \braket{\Phi(n)|Z_j|\Phi(n)}
\label{eq:orderpar}
\ee
in both space and time, being equal to
$(-1)^n$,
gives rise to a trivial example of a macroscopic subharmonic response.
This behavior, however, relies on a fine-tuning of the kick strength, $g=\pi/2$.
The existence of a \textit{nonequilibrium phase of matter} exhibiting time-crystal behavior revolves around the stability of this spatiotemporal order to arbitrary (sufficiently weak) Floquet perturbations in the thermodynamic limit $L\to\infty$.

\begin{figure*}[t!]
  \includegraphics[width=\textwidth]{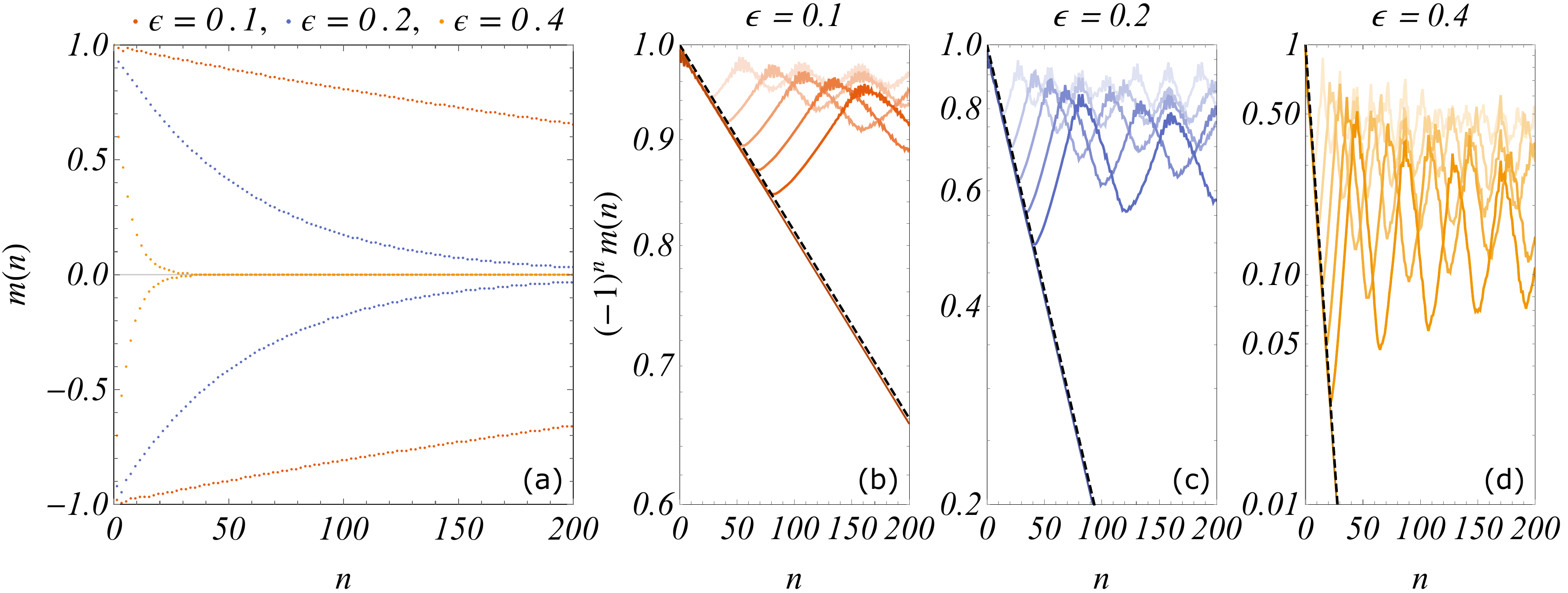}
  \caption{{\it (a)} Evolution of the order parameter under the Floquet Ising dynamics in Eq.~\eqref{eq_toggling},
    with $J = 1$ and different values of the kick strength $\epsilon$.
    Data have been obtained by analytically solving the dynamics of the model, as explained in Appendix~\ref{app:orderIsing}.
    {\it (b-d)} Absolute value of the order parameter in logarithmic scale: thick lines are the same data as
    in panel {\it (a)} and are for $L=\infty$. Shaded lines are the results of ED simulations for finite systems
    with different sizes $L=10,15,20,25,30$, from lighter to darker colors.
    Dashed black lines denote the asymptotic exponential decay $e^{-\gamma n}$, with $\gamma$ predicted by Eq.~\eqref{eq_tauexact}.}
\label{fig:Z_ising}
\end{figure*}

\subsection{Exact decay rate in the thermodynamic limit and finite-size effects}
\label{subsec_exactdecayintegrable}

The simplest perturbation to the above Floquet protocol consists in performing imperfect spin flips, i.e., \be
\label{eq_perturbedtransversekick}
K \longrightarrow K_{\pi/2+\epsilon}=e^{i(\frac \pi 2+\epsilon)\sum_j X_j}
,
\ee  
with $\epsilon \neq 0$. 
Since the perfect kick $K_{\pi/2} = i^{L} P$ can be factored out of $K$ and is proportional to the global $\mathbb{Z}_2$-spin flip operator $P = \prod_{j}X_j$,
the expectation value of the local order parameter~\eqref{eq:orderpar}
over $n$ periods can be expressed as
\be
\label{eq_toggling}
m(n) = (-1)^n
\langle + | [K_{\epsilon} V_{J}]^{-n} 
Z_{j}  
[K_{\epsilon} V_{J}]^n | + \rangle,
\ee
where we used the properties $P Z_{j} P = -Z_{j}$ and $[V_J,P]=0$,
while $K_{\epsilon}=\exp(i \epsilon \sum_j X_j)$.
Equation~\eqref{eq_toggling} expresses the fact that the absolute value of the magnetization evolves as if it were governed by the Floquet operator $K_{\epsilon} \, V_{J}$ with kick strength equal to $\epsilon$, where the perfect kick has been completely gauged away by switching to a toggling frame of reference, leading to the multiplicative factor with alternating sign $(-1)^n$.

The persistence of time-crystalline order 
is related to the preservation of a finite absolute value of the local order parameter $|m(n)|$ for large times $n\to\infty$.
This question has been addressed in previous works investigating finite-size chains. In particular, the analysis of Ref.~\cite{Angelakis_Ising} led to a positive answer based on finite-size scaling of the order parameter obtained by exact diagonalization (ED) of short chains $L\lesssim 20$.
This finding contradicts the generic expectations of the absence of long-range order in excited states of one-dimensional clean short-range interacting systems. 

Here, by computing the exact dynamics of the magnetization for an infinite chain using the integrability of the model, as detailed in Appendix~\ref{app:exactIsing}, we establish that the order parameter decays exponentially in time as $|m(n)|\sim e^{-\gamma n}$. The rate $\gamma$ is found to be
\be
\label{eq_tauexact}
\gamma =  - \int_{0}^{\pi}\frac{dp}{\pi} \partial_p\phi_{p} \ln  { |\cos(\Delta_p)|} ,
\ee
where the quasiparticle spectrum $\phi_{p}$ and the Bo\-go\-liu\-bov angle $\Delta_p$ resulting from the diagonalization of the quadratic Floquet Hamiltonian are defined by the equations $\cos(\phi_{p}) =  \cos(2J)\cos(2\epsilon) + \sin(2J)\sin(2\epsilon)\cos(p)$   and $
\cos(\Delta_{p}) = \cos(\theta_{p})\cos(p)+\sin(\theta_{p})\cos(2\epsilon)\sin(p)
 $, respectively. The explicit expression of $\theta_{p}$ is given in Appendix~\ref{app:diagIsing}.
By Taylor expanding the exact result~\eqref{eq_tauexact} for small perturbations $\epsilon$, we find that the rate $\gamma$ scales as 
\be
\label{eq_tauscaling}
\gamma = \frac{16}{3\pi\sin^{2}(2J)} |\epsilon|^{3} + O(|\epsilon|^5).
\ee

Figure~\ref{fig:Z_ising} reports the exact evolution of $m(n)$ for increasing values of $\epsilon$, for $J=1$ (colored data sets). In all cases, the resulting decay rate excellently reproduces the analytical result~\eqref{eq_tauexact} (dashed black lines). As it is evident in Fig.~\ref{fig:Z_ising}{\it (a)}, the decay remains quite slow even for moderate values of $\epsilon$. 
Furthermore, in Fig.~\ref{fig:Z_ising}{\it (b)-(d)}, we compare the thermodynamic-limit evolution with that of finite chains of length $L=10\div30$. {The plot illustrates the dramatic impact of finite-size effects. The apparent persistence of the order parameter in finite systems has been detected in previous works~\cite{Angelakis_Ising, Choudhury_Ising, PhysRevB.99.144304}. For the kicked Ising chain~\cite{Angelakis_Ising, Choudhury_Ising}, these strong finite-size effects can be attributed to a large overlap of the initial state with the magnetized ground state of the integrable Floquet Hamiltonian, in agreement with the numerical findings of Ref.~\cite{PizziPRB20_CleanDTC} and similarly to the static case~\cite{ce19,collura2019relaxation}.}

\subsection{Physical interpretation of the order-parameter decay as domain-wall spreading}
\label{subsec_physicalinterpretationDW}

The scaling in Eq.~\eqref{eq_tauscaling} with $\epsilon$ can be understood in in\-tui\-ti\-ve terms, considering that the system has exact quasiparticles which behave as noninteracting fermions. 
The dynamics in Eq.~\eqref{eq_toggling} is equivalent to a quench from the ground state of the classical Ising Hamiltonian~\eqref{eq:H}, evolving with a kicked Ising chain deep in the ferromagnetic phase $|\epsilon|\ll J$.
In this case, the free fermions can be interpreted as topologically protected excitations, i.e., domain walls (kinks and antikinks), interpolating between the two degenerate magnetized ground states.
To the lowest order in the kick strength $\epsilon$,
the quench creates a small density $\rho=\mathcal{O}(\epsilon^2)$ of spin flips, whose constituent pair of domain walls freely spread  along the chain with maximum velocity $v=\mathcal{O}(|\epsilon|)$.
The domain of reversed spins extending between a kink-antikink pair grows linearly in time until one of them meets another domain wall initially located far away. The decay rate of the order parameter can  be predicted in terms of this semiclassical picture~\cite{riegerigloi11}: we find
\be \gamma\sim\rho v = \mathcal{O}(|\epsilon|^3). \ee
Indeed, the exact result in Eq.~\eqref{eq_tauexact} obtained from the asymptotic expansion of the determinant of a large block Toeplitz matrix~\cite{PhysRevLett.106.227203, Calabrese_2012}, precisely takes the form of a product of quasiparticle group velocity $|\partial_p \phi_p|$ at momentum $|p|$ and (for small $\epsilon$) {number $\sin^2(\Delta_p) \simeq -\ln|\cos^2(\Delta_p)|$} of excited quasiparticle pairs  with momenta $(p,-p)$ in the initial state, averaged over momenta. This substantiates the intuitive interpretation above.

The exact result thus quantitatively confirms the intuitive model of order parameter meltdown by spreading domain-wall pairs, illustrated in Fig.~\ref{fig:intro}{\it (a)}.
More importantly, this picture highlights what makes time-crystalline order doomed to melt in one-dimensional systems:
Preventing domain-wall pairs from unbounded separation requires certain microscopic mechanisms, such as Anderson (many-body) localization induced by quenched disorder. In the rest of this article, we will investigate under which circumstances domain-wall confinement can provide a robust stabilization mechanism.

\section{Domain-wall confinement and deconfinement in the kicked mixed-field Ising chain}
\label{sec_mixedfield}

The previous section unambiguously illustrates how domain-wall spreading underlies the time-crystal melting in clean, locally interacting, spin chains.
A celebrated proposal to overcome this occurrence and protect long-range order out of equilibrium hinges upon disorder-induced localization: in such case, domain walls behave like particles moving in a random background, and spatial localization can arise from destructive interference, as first foreseen by Anderson~\cite{AndersonLocalization}, even in the presence of many-body interactions~\cite{BAA,HuseNandkishore,AbaninRMP}. This basic mechanism of localization-protected order~\cite{HusePRB13_LocalizationProtectedQuantumOrder,ChandranPRB14_MBLProtectedSPT}  has been proposed to stabilize time-crystalline behavior for arbitrarily long times~\cite{ElsePRL16,KhemaniPRL16}.
Here we explore a different mechanism to prevent domain-wall spreading, namely domain-wall confinement.

In this section we briefly review the current understanding of this multifaceted phenomenon (Sec.~\ref{subsec_confinementoverview}) and extend it to driven systems,
exemplified by the spin-chain dynamics~\eqref{eq_Floquet} subject to weak kicks $K \simeq \mathbb{1}$ about an arbitrary tilted axis.
Specifically, we
use the tools of many-body perturbation theory to reformulate the order parameter-dynamics in terms of the motion of effective domain walls (Secs.~\ref{subsec_effectiveDWdynamics}) and demonstrate domain-wall confinement (Sec.~\ref{subsec_twobody}).  
Finally, in Sec.~\ref{subsec_deconfinement}, we come back to our main discrete time-crystal problem with kicks close to perfect spin flips $K \simeq i^L P$, and understand the order-parameter melting, illustrated in Fig.~\ref{fig:intro}({\it b}).

\subsection{Confinement in quantum spin chains}
\label{subsec_confinementoverview}

For the benefit of readers who may not be familiar with domain-wall confinement in quantum spin chains, in this subsection we provide a brief overview. 

Particle confinement is a nonperturbative phenomenon arising in certain gauge theories, 
which {consists} in the absence of \textit{colored} asymptotic states: all stable excitations of the theory above the ground state are \textit{colorless} bound states of elementary particles \cite{Weinbergbook2}.
An intuitive picture of this phenomenon is given by the formation of a gauge-field string connecting a quark-antiquark pair, the energy cost of which provides an effective confining potential that grows linearly with the spatial separation between the two particles.
As a result, the quark and antiquark bind together into composite neutral particles called \textit{mesons}. When a large physical separation between them is enforced, the potential energy stored in the string becomes sufficient to produce another pair of particles out of the vacuum, which bind with the old particles to form two mesons, making the observation of isolated quarks impossible.

An analogous confinement phenomenon naturally arises for domain walls in quantum spin chains.
Its mechanism was proposed by McCoy and Wu in 1978~\cite{McCoyWuConfinement} and later studied in a variety of theoretical~\cite{DelfinoDecayIFF,ShankarConfinement,FonsecaZamMesonSpectrumContinuum,RutkevichMesonSpectrumContinuum,RutkevichConfinementXXZ,RutkevichMesonSpectrumLattice} and experimental~\cite{GrenierPRL14,WangSpinonConfinement,BeraEsslerSpinonConfinement} works. 
The core ingredient here is a first-order quantum phase transition, i.e., the explicit lifting of a spontaneously broken discrete symmetry. In the ferromagnetic quantum Ising chain ($H=-J\sum_jZ_jZ_{j+1} - g \sum_j X_j$, with $|g|<J$), this can be simply realized by  introducing a longitudinal field $-h\sum_j Z_j$, which generates an energy penalty for the reversed magnetic domain separating a pair of domain walls, analogous to a string tension. The energetic cost for separating the pair thus grows proportionally to the distance, giving rise to a linear confining potential that fully suppresses the spreading at arbitrarily high energies, binding  the pair of ``topologically charged'' particles (kink and antikink) into ``topologically neutral'' bound states, referred to as mesons by analogy with particle physics.
In certain quasi-one-dimensional magnetic insulators, similar effective longitudinal fields are provided at a mean-field level by inter-chain interactions; the resulting tower of mesonic excitations (spinon bound states) has been spectacularly observed with inelastic neutron scattering~\cite{GrenierPRL14,WangSpinonConfinement,BeraEsslerSpinonConfinement}.

Recently, it has been realized that confinement in quantum spin chains and in $(1+1)$-dimensional lattice gauge theories can be generically mapped onto each other~\cite{LeroseSuraceQuasilocalization} via elimination/introduction of matter degrees of freedom exploiting the local constraints posed by gauge invariance~\cite{SuraceRydberg,ZoharCiracMatterIntegration}.
This substantiates intuitive pictures of the nonequilibrium dynamics of spin chains in terms of prototypical phenomena in gauge theories, such as vacuum decay~\cite{RutkevichFalseVacuum,LeroseSuraceQuasilocalization,Milsted20_Collision,Lagnese21_FalseVacuumDecay}, string dynamics~\cite{MazzaTransport,LeroseSuraceQuasilocalization,Verdel19_ResonantSB,RobinsonNonthermalStatesShort}, and string inversions~\cite{SuraceRydberg}.

Dynamical signatures of domain-wall confinement have been recently attracting a growing interest, starting from 
Ref.~\cite{Kormos:2017aa}. The suppression of domain-wall spreading stabilizes the order parameter even out of equilibrium.
This stabilization also applies to the dynamics starting from the ``false vacuum'' magnetized \textit{against} the longitudinal field, which is a very atypical highly excited state without domain walls~\cite{MazzaTransport,LeroseSuraceQuasilocalization}. The basic explanation of this phenomenon is that domain-wall pairs excited on top of the false vacuum are also confined into ``anti-mesons''. Furthermore, domain-wall pair production out of the false vacuum is strongly suppressed, despite being energetically allowed and entropically favorable, as it requires a locally created virtual pair  to tunnel across a high-energy barrier. This effect, akin to the Schwinger mechanism in quantum electrodynamics~\cite{ItzyksonZuber}, results in an exponentially long lifetime of the order parameter~\cite{RutkevichFalseVacuum,LeroseSuraceQuasilocalization}.

\subsection{Effective domain-wall dynamics for weak kicks}
\label{subsec_effectiveDWdynamics}

Here we show that analogous considerations on domain-wall confinement carry over to \textit{driven} quantum Ising chains, under the assumption that the periodic kicks are weak. 
We consider the spin chain evolving with the Floquet operator in Eq.~\eqref{eq_Floquet}, with the simplest $\mathbb{Z}_2$-symmetry-breaking periodic kicks, i.e., a  tilted rotation axis away from the $x-y$ plane:
\begin{equation}
\label{eq_tiltedkick}
K \longrightarrow K_{\epsilon, h} = e^{i \sum_j(\epsilon X_j + h Z_j)} \, . 
\end{equation}
We define $\hat\epsilon=\max(\epsilon,h)$ to conveniently keep track of the global magnitude of the kick. 
For clarity, we stress that here, we are considering a weak kick $K_{\epsilon, h}$, with$\epsilon \ll 1$, in contrast with the almost perfect spin flip $K_{\pi/2 + \epsilon}$,  (compare with Eq.~\eqref{eq_perturbedtransversekick}) which is relevant for the stability of DTC.

The first problem that we encounter is related to the breaking of integrability for $h\neq0$: domain walls cease to exist
as exact quasiparticles throughout the many-body Floquet spectrum, seemingly obscuring the physical interpretation
of the magnetization dynamics. However, we can still make sense of an effective picture of domain-wall dynamics
for weak perturbations, using the rigorous theory of prethermalization of Ref.~\cite{DeRoeckVerreet}.
Namely, one can construct a sequence of {time-periodic}, quasi-local unitary transformations $\{e^{i \hat\epsilon^m S_m}\}_{m=1,\dots,p}$,
designed to remove from the time-dependent Hamiltonian all the terms that change the total number of domain walls 
\be
\label{eq_D1}
D_1 \equiv \sum_j D_{j,j+1}  \;, \quad D_{j,j+1} \equiv  \frac{1-Z_j Z_{j+1}}{2}.
\ee 
The resulting transformed Floquet operator 
\be
\label{eq_schiefferwolff}
{{U}}_p = e^{-i \hat\epsilon^p S_p} \cdots e^{-i\hat\epsilon S_1} \; K_{\epsilon,h} V_J \;  e^{i\hat\epsilon S_1} \cdots e^{i\hat\epsilon^p S_p}
\ee 
conserves $D_1$
up to terms of order $p+1$, i.e., 
$
\big[{{U}}_p,D_1\big]=\mathcal{O}(\hat\epsilon^{p+1})
$.
The perturbative series generated by this construction represents a nontrivial generalization of the well-known Schrieffer-Wolff transformation for static Hamiltonians~\cite{FrohlichSchriefferWolff,LossSchriefferWolff,McDonaldSchiefferWolffHubbard,AbaninRigorousPrethermalization}.
The generators $\{S_m\}$ take the form of sums of local operators, whose number and support size grow proportionally to the perturbative order $m$.

The construction of the transformation $e^{iS_{\le p}} \equiv e^{i \hat\epsilon S_1} \cdots e^{i \hat\epsilon^{p} S_{p}}$ for arbitrarily large $p$ aims at asymptotically producing an exactly domain-wall-conserving Floquet operator.
However, the resulting perturbative series is expected to have divergent (asymptotic) character in general, similarly to the static case~\cite{AbaninRigorousPrethermalization}, suggesting a late-time breakdown of the conservation of $D_1$ and an eventual heating to infinite-temperature.
Nevertheless, the transformed picture still contains extremely useful information on the transient dynamics.
The analysis of Ref.~\cite{DeRoeckVerreet} shows that, when $J$ is strongly incommensurate with the driving frequency $2\pi$,
\footnote{More precisely, $J/(2\pi)$ must be an irrational number in the Diophantine class, $\lvert J/2\pi - p/q \rvert > C/q^\tau$ for all $p,q\in\mathbb{Z}$.} 
the breakdown of $D_1$ conservation ---and hence heating--- must be extremely slow.
In fact, the optimal (with respect to suitable operator norms) truncation order depends on the magnitude of the perturbation as $p^*\sim C/\hat\epsilon^{3+\delta}$, where $\delta>0$ is arbitrary and $C=C(\delta)$ is a constant.
The very fact that $p^*$ scales up with the smallness of $\hat\epsilon$ itself yields a  \textit{nonperturbatively small} truncation error.
In turn, this translates into a quasi-exponentially long time window~\cite{DeRoeckVerreet}\footnote{Notice that time and frequency scales are dimensionless here, as the driving period is fixed at $1$.}
\be
\label{eq_timescale}
T_{\text{preth}} \ge \frac 1 {\hat\epsilon} \exp\bigg( \frac{c} {\hat\epsilon^{ 1/ (3+\delta)}} \bigg) 
\ee 
within which 
---for the purpose of computing the dynamics of local observables--- the nonequilibrium evolution of an initial state $\ket{\psi_0}$ can be approximated as
\begin{equation}
\label{eq_transformedpicture}
{U}^{n}\ket{\psi_0} \; \simeq \; e^{iS_{\le p^*}}  ({{U}'}_{p^*})^{n} e^{-iS_{\le p^*}} \ket{\psi_0},
\end{equation}
where ${{U}'}_{p^*} $ is the approximate Floquet operator obtained from ${{U}}_{p^*} $ by truncating terms beyond the order
$\hat\epsilon^{p^*}$. By construction, ${{U}'}_{p^*} $ exactly conserves the number of domain walls, $\big[{{U}'}_{p^*},D_1\big]=0$.

The physical consequence of this analysis is that  the ``bare'' (i.e., unperturbed) domain-wall occupation number $D_{j,j+1}$ on the bond $(j,j+1)$ 
acquires a perturbative quasi-local ``dressing'' $e^{-iS_{\le p^*}} D_{j,j+1} e^{iS_{\le p^*}}$ for small perturbations. 
The density of such dressed domain walls remains approximately conserved at least for the long timescale $T_{\text{preth}}$ in Eq.~\eqref{eq_timescale}. 
Note, however, that this heating timescale only represents a rigorous lower bound, and it is not expected to be tight in general.
In the integrable limit $h=0$, the underlying algebraic structure of the model produces cancellations to all orders which make this perturbative series convergent, leading to \textit{exact} dressed domain-wall quasiparticles. In this case heating is completely suppressed beyond the above timescale~$T_{\text{preth}}$. 
As soon as $h \neq 0$, instead, these emergent domain walls are expected to eventually decay after~$T_{\text{preth}}$ (see also Refs.~\cite{RobinsonNonthermalStatesShort,RobinsonNonthermalStatesLong}). 
In any case, as long as we deal with dynamical phenomena occurring in the long Floquet-prethermal time window $0\le t \le T_{\text{preth}}$, we can switch to the effective picture where states and observables are transformed via the unitary operator $e^{iS_{\le p^*}}$, and work therein with the effective domain-wall conserving Floquet operator ${{U}'}_{p^*} $, according to  Eq.~\eqref{eq_transformedpicture}.

As it results from the above discussion, we can analyze the evolution of the order parameter by switching to the transformed picture, 
\be
\label{eq_effectivedyn}
m(n)=\braket{\psi_0| U^{-n} Z_j U^n |\psi_0} \simeq  \braket{\psi_0'| ({U}'_{ p^*})^{-n} Z'_j ({U}'_{ p^*})^n |\psi_0'},
\ee
where $\ket{\psi_0'}= e^{-iS_{\le p^*}} \ket{\psi_0}$, $Z'_j = e^{-iS_{\le p^*}} Z_j e^{iS_{\le p^*}}$, and the approximation, due to truncating $U_{ p^*}$ to ${U}'_{ p^*}$, holds up to the long timescale in Eq.~\eqref{eq_timescale}. In this transformed picture, the number $D_1$ of domain walls is an exact quantum number, and the Hilbert space fractures into separate blocks labelled by $D_1$.

The perturbative construction introduced in Ref.~\cite{DeRoeckVerreet} to prove the theorem leading to Eq.~\eqref{eq_timescale} is hardly manageable for explicit low-order computations. In practice, we have found it more convenient to resort to a combination of the replica  resummation of the Baker-Campbell-Hausdorff (BCH) expansion of  Ref.~\cite{ProsenPolkovnikovPRL18_ReplicaBCHResummation} and a standard static Schrieffer-Wolff transformation (see, e.g., Ref.~\cite{AbaninRigorousPrethermalization}). In both approaches, the existence of a well-defined expansion requires incommensurability of the coupling $J$ with $2\pi$. This condition is necessary to ensure that the domain-wall number uniquely labels the unperturbed sectors of the Floquet operator, and thus remains a good quantum number throughout the perturbative construction. We remark that while the scheme of Ref.~\cite{DeRoeckVerreet} generally produces a time-dependent effective Hamiltonian, the combined BCH resummation and Schrieffer-Wolff transformation produce a static effective Hamiltonian, with presumably similar convergence properties.

In Appendix \ref{app_derivationFloquet}, we use the latter approach to derive the expression of $S$ and $U'$ to lowest order $p=1$, reported here:
\be
\label{eq_Utilde}
{U}'_1  = e^{+i \sum_j J \, Z_j Z_{j+1} + \epsilon \, (P^\uparrow_{j-1} X_{j} P^\downarrow_{j+1} + P^\downarrow_{j-1}  X_{j} P^\uparrow_{j+1})
 +  h \, Z_j } \, ,
\ee
\begin{multline}
  \label{eq_S1}
  S_{\le1} = -\frac{\epsilon}{2} \sum_j P_{j-1}^{\uparrow} [ X_j - \cot(2J) Y_j ] P_{j+1}^{\uparrow} +\\+  P_{j-1}^{\downarrow} [ X_j + \cot(2J) Y_j ] P_{j+1}^{\downarrow} \, .
\end{multline}
In Eqs.~\eqref{eq_Utilde},~\eqref{eq_S1}, the operators $P_j^{\uparrow,\downarrow}$ represent local projection operators onto the ``up'' and ``down'' states along $z$ of spin $j$.
The off-diagonal processes in ${U}'_1$ are given by the terms proportional to $ \epsilon$, which describe nearest-neighbor hopping of domain walls to the left or to the right. 
Processes which create or annihilate pairs of domain walls have been removed from the evolution operator ${U}_1$ through the unitary transformation $e^{i S_{\le1}}$; in this formulation, these processes only show up  in the expression of the transformed initial state $\ket{\psi_0 '}=e^{-i  S_{ \le1}}\ket{\psi_0}$.

The general structure of the expansion and further details are discussed in Appendix~\ref{app_derivationFloquet}.
Higher-order terms generate further small corrections in the effective Floquet Hamiltonian. In particular, at order $p$, one has terms in $U'_p$ that correspond to at most $p$ displacements of one domain wall or more adjacent domain walls to a neighboring bond. For example, at second order one has diagonal terms, plus nearest-neighbor hopping terms analogous to those in $U'_1$, plus new terms proportional to $P^\uparrow_{j-1} (S^+_{j} S^+_{j+1}+S^-_{j} S^-_{j+1}) P^\downarrow_{j+2}$ (next-nearest-neighbor hopping), $P^\downarrow_{j-1} (S^+_{j} S^-_{j+1}+S^-_{j} S^+_{j+1}) P^\downarrow_{j+2}$ (pair nearest-neighbor hopping), and analogous flipped combinations. 
These additional terms do not modify the conclusions below.
Likewise, the higher-order Schrieffer-Wolff generator $S_{\le p}$ flips at most $p$ neighboring spins.

\begin{figure*}[t!]
  \includegraphics[width=0.95\textwidth]{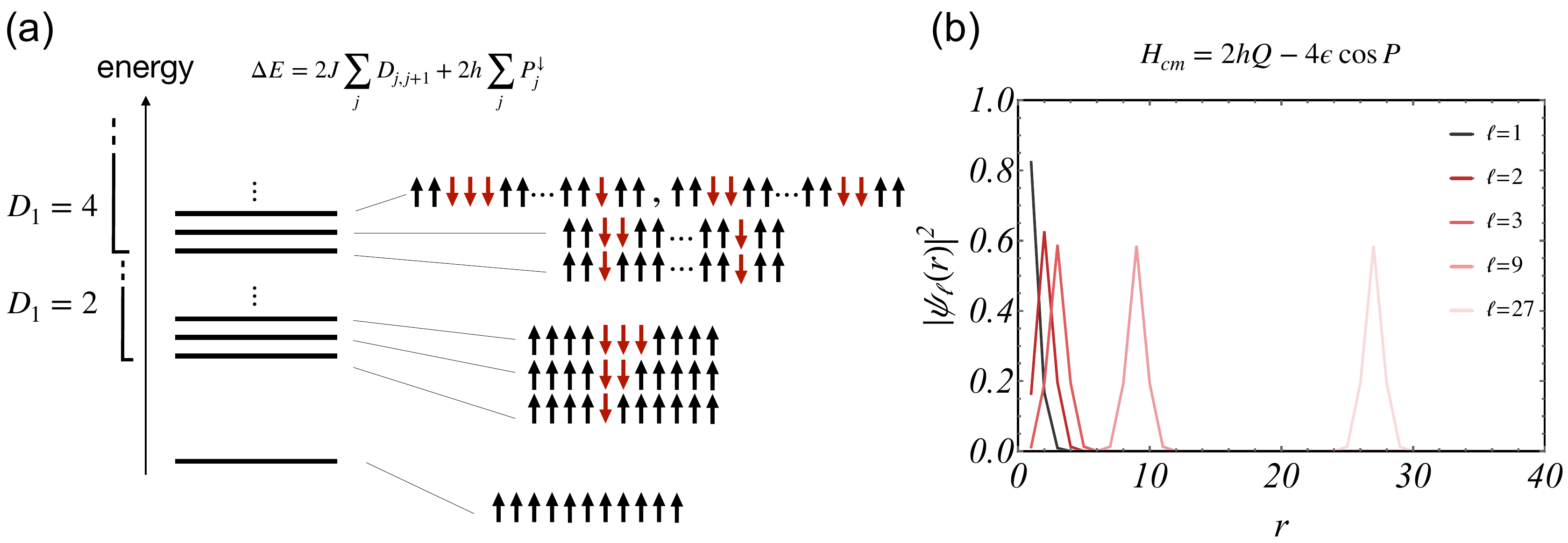}
  \caption{{\it (a)} Sketch of the low-energy spectrum of the Floquet operator in Eq.~\eqref{eq_Utilde} for $\epsilon\to0$.
    The vertical axis represents the relative eigenphase $\Delta E$ with respect to the polarized state $\ket{+}$.
    {\it (b)} A selection of mesonic eigenfunctions of the two-body problem, cf. Eq.~\eqref{eq_mesonwavefunction}, for $\epsilon/h=1.5$.}
  \label{fig:2body}
\end{figure*}

\subsection{Domain-wall confinement and order-parameter dynamics}

\label{subsec_twobody}

Armed with the rigorously established picture of effective domain-wall dynamics for weak kicks, we now discuss domain-wall confinement and its implications for the evolution of the order parameter.
Here, as in the previous subsection, we discuss the Floquet dynamics generated by weak kicks $U=K_{\epsilon,h} V_J$. In the next subsection we will go back to studying robustness of the DTC signal with $U=K_{\pi/2} K_{\epsilon,h} V_J$. 

Equation~\eqref{eq_effectivedyn} describes the evolution of the magnetization in the transformed domain-wall picture. 
The transformed initial state $\ket{+ '}$ consists of  a low density of order $\epsilon^2$ of flipped spins, 
\be
\label{eq_initialdensity}
\langle + |  e^{i  S_{\le 1}} \, D_{j,j+1} \, e^{-i  S_{\le 1}} |  + \rangle = \frac {\epsilon^2} {2 \sin^2 2J} + \mathcal{O}(\hat\epsilon^3) \, ,
\ee
as, to lowest order, this state is obtained by rotating the spins in $\ket{+}$ by an angle $\epsilon/\sin(2J)$ around a transverse axis
(note that each spin flip carries \textit{two} domain walls).
Furthermore, since $Z'_j = Z_j + \mathcal{O}(\epsilon)$, for the purpose of understanding the nature of the evolution of $m(n)$ (i.e., persistent or decaying), we can drop the correction to $Z_j$.

Since the initial state is composed of dilute tight pairs of domain walls, we can enlighten the resulting order-parameter dynamics by studying the two-body problem.
The intuitive picture of the evolution of $m(n)$ in terms of the motion of domain-wall pairs becomes asymptotically exact in such a low-density limit. 
The nature of this evolution (persistent or decaying order) depends on the effective Floquet operator ${U}'$ governing  the motion of domain walls. Equation~\eqref{eq_Utilde} describes domain walls of ``mass'' $2J$, hopping to neighboring lattice bonds with amplitude $\epsilon$, and experiencing a confining potential $V(r)=2hr$ which ties them to a neighbor at a distance~$r$.

The two-body problem is obtained by projecting the effective Hamiltonian ${H}_1$, defined by the exponent of ${U}'_1$ in Eq.~\eqref{eq_Utilde}, onto the sector spanned by two-particle basis states $\ket{j_1,j_2}$, where the integers $j_1<j_2$ label the positions of two domain walls along the chain.
The resulting two-body Hamiltonian is
\begin{multline}
\label{eq_H2body}
H_{\text{2-body}} = \sum_{j_1<j_2} V(j_2-j_1) \ket{j_1,j_2}  \bra{j_1,j_2}
\\  - {\epsilon}  \sum_{j_1<j_2} \big(\ket{j_1+1,j_2} + \ket{j_1,j_2+1}\big) \bra{j_1,j_2}  + \text{ H.c.}
\end{multline}
where the hard-core constraint $\ket{j_1=j_2} \equiv 0$ is understood.
The second term represents nearest-neighbor hops of the domain walls, whereas the first one acts as a linear {confining}  potential $V(r)= 2 h r$ as a function of the distance $r=j_2-j_1>0$.
While for $h=0$ one has a continuum of unbound traveling domain walls, as soon as $h\neq0$ the spectrum changes nonperturbatively to an infinite discrete tower of bound states.

Due to translational invariance of the initial state $\ket{+'}$, domain-wall pairs are only generated with vanishing
center-of-mass momentum $K=0$. The relative coordinate wavefunction $\psi(r)$ satisfies a Wannier-Stark equation with
a hard-wall boundary condition $\psi(0)=0$, yielding~\cite{FogedbyTwoKinkSolution,LeroseSuraceQuasilocalization}
the exact mesonic masses
\begin{equation}
\label{eq_mesonspectrum}
    {E}_\ell=2h \, \nu_\ell(2\epsilon/h) \equiv - 2h \times \text{ \{$\ell$-th zero of $x \mapsto \mathcal{J}_x(2\epsilon/h)$\} }
\end{equation}
 $\ell=1,2,\dots$,
and wavefunctions
\begin{equation}
\label{eq_mesonwavefunction}
    \psi_{\ell}(r) = \mathcal{J}_{r-\nu_\ell}(2\epsilon/h),
\end{equation}
where $\mathcal{J}$ is the standard Bessel function.
\footnote{Note that $k\in[-\pi/2,\pi/2)$, because $k$ and $k+\pi$ generate the same solution up to a phase: Since $\mathcal{J}_{\alpha}(-z)=e^{i\pi \alpha}\mathcal{J}_{\alpha}(z)$, when $k\mapsto k+\pi$ the wavefunction $\psi$ gets multiplied by $(-)^s e^{i\pi(r-\nu_\ell)}=e^{-i \pi \nu_\ell} (-)^{2 j_2}=e^{-i \pi \nu_\ell}$, i.e., a global phase.}
For $\epsilon\to0$, one finds the energy levels ${E}_\ell=2h \ell$, corresponding to a domain of $\ell$ reversed spins,
$\psi_\ell(r)=\delta_{\ell,r}$. 
Figure~\ref{fig:2body}{\it (a)} reports a sketch of the low-energy spectrum of the Floquet Hamiltonian in this limit.
For finite $\epsilon/h$, the eigenfunctions can still be adiabatically labelled by the integer $\ell$.
Panel {\it (b)} reports a selection of mesonic eigenfunctions $\psi_\ell(r)$ in the center-of-mass frame, for $\epsilon/h=1.5$. 
Boundary effects are visible for small $\ell \lesssim 2\epsilon/h$,  whereas for larger $\ell$ the wavefunctions become essentially Wannier-Stark localized orbitals, i.e., rigidly shifted copies of each other.

We can formulate a more intuitive analysis of  the two-body dynamics, 
which will turn out to be fruitful later to analyze  the time-crystalline behavior. 
To this aim, we introduce the canonically conjugated operators $Q,P$ defined by
\be
Q = \sum_{r} r \ket{r}\bra{r}, 
\qquad e^{iP} = \sum_{r} \ket{r+1}\bra{r} \, ,
\ee
which correspond to the position and the momentum in the center-of-mass frame, i.e., the distance between the two domain walls and their relative momentum;
one verifies $[Q,P]=i$.
\footnote{Alternatively, one can use a momentum-space representation $\ket{r} \mapsto \psi_r(k)=e^{ikr}$: in this description, one has $Q \mapsto -i \partial_k$ and $P \mapsto k$, whence the canonical algebra is manifest.}
Using these variables, the center-of-mass frame Hamiltonian becomes
\be
H_{\text{cm}} = 2 h Q - 4 {\epsilon} \cos P,
\ee
where the domain is $Q>0$ and a hard-wall boundary condition at $Q= 0$ is understood.
Classical trajectories are bounded in the $Q$ direction and are translationally invariant away from the boundary $Q=0$.
Indeed, the Heisenberg evolution equations can be integrated exactly in the bulk $Q\gg 2\epsilon/h$
(i.e., neglecting the boundary condition), which gives
\be
\label{eq_Heisenbergeqsol}
\left\{
\begin{split}
Q(t)  =&  Q(0) \\
& + \frac {{2\epsilon}}{h} \Big[  \sin 2ht \sin P(0) + (1-\cos 2ht) \cos P(0) \Big] , \\
P(t)  = & P(0) +2 h t \, .
\end{split}
\right .
\ee
Equation~\eqref{eq_Heisenbergeqsol} indicates that the relative momentum $P$ revolves freely around the Brillouin zone, whereas the relative coordinate $Q$ performs stable  \emph{Bloch oscillations}, remaining localized near the corresponding initial condition. In physical terms, the mutual confining potential creates an effective Wannier-Stark ladder which pins the distance between the two domain walls. 

\begin{figure}[t!]
  \includegraphics[width=0.45\textwidth]{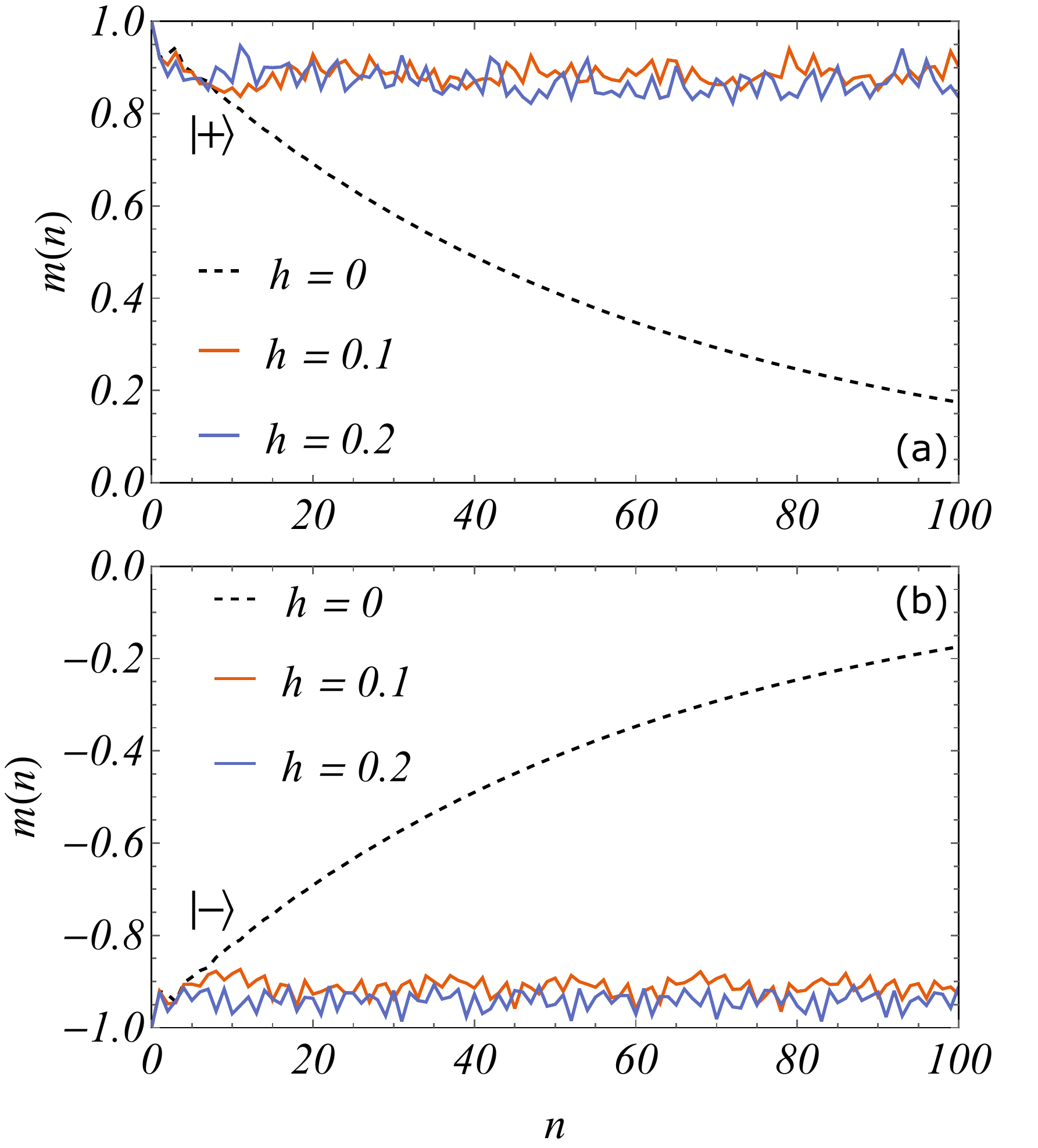}
  \caption{Floquet dynamics of the order parameter induced by the kick $K_{\epsilon,h}$,
    with $\epsilon = 0.2$ and different values of $h$, in an infinite chain $L=\infty$, obtained through iTEBD simulations. 
    The system is initially prepared either {\it(a)} in the ``vacuum'' $|+\rangle$, or {\it(b)} in the ``false vacuum'' $|-\rangle$.
    The order-parameter meltdown for $h=0$ (dashed black lines)
    gets strongly suppressed by the presence of a small $h \neq 0$ (full lines).}
  \label{fig:Ising_driven}
\end{figure}

As anticipated above, the solution of the two-body problem sheds light on the many-body dynamics.
Each of the dilute spin flips in the initial state $\ket{+'}$ overlaps significantly with the lightest mesonic wavefunctions.
\footnote{more specifically, the overlap is significant with the first $N \simeq \epsilon/h$ mesonic states}
The total magnetization thus exhibits a persistent oscillatory behavior, characterized by multiple frequencies associated with the ``masses'' of the mesonic bound states.
As long as the typical separation between distinct initial domain-wall pairs ---the inverse $\simeq \epsilon^{-2}$ of the density in Eq.~\eqref{eq_initialdensity}--- exceeds by far the size $\simeq 1+ \epsilon/h$ of the excited bound states, the evolution of the magnetization $m(n)$ can be described as resulting from the coherent superposition of independent small-amplitude meson oscillations, 
similarly to the undriven case discussed in Ref.~\cite{Kormos:2017aa}.
We note that inelastic meson scattering is expected to trigger asymptotic thermalization; however, the inelastic cross section drops rapidly for small $\epsilon$~\cite{Surace_2021}, making the transient out-of-equilibrium state extremely long lived~\cite{BanulsWeakThermalization,LinMotrunichOscillations}.

Analogous considerations can be made for the dynamics starting from the ``false vacuum'' state $\ket{-}$  (or, equivalently, taking $h\mapsto -h$). Two consecutive domain walls are subject to an \textit{anticonfining} potential, decreasing linearly with their separation. Due to lattice effects, however, the domain walls cannot accelerate arbitrarily towards large distances, as their hopping kinetic energy is bounded. Thus, they form a tower of bound states, formally analogous to the true ground-state excitations \eqref{eq_mesonwavefunction} discussed above. 
The dynamics of the order parameter thus follows a similar pattern, exhibiting a stable antimagnetization with small-amplitude multiple-frequency oscillations superimposed due to antimesons. 
 
The ultimate decay of the antimagnetization due to resonant domain-wall pair production is expected to occur at very long times, 
as the high energy cost $4J$ of creating two domain walls has to be compensated by a gain $2h r$ associated with a large ground-state bubble of size $r$ extending between them. To realize this,  the two locally created virtual domain walls have to tunnel through a distance $r^* \simeq 2J/h$, {which suggests a nonperturbative lifetime. 
The more rigorous estimation of Ref.~\cite{DeRoeckVerreet} leads to the quasi-exponentially long time in Eq.~\eqref{eq_timescale}.} This phenomenon is closely related to the Schwinger mechanism in quantum electrodynamics, as explained in Ref.~\cite{LeroseSuraceQuasilocalization}. 

The confinement scenario for the periodically kicked mixed-field Ising chain has been verified numerically, simulating the Floquet dynamics induced by $\mathbb{Z}_{2}$-symmetry-breaking periodic kicks $K_{\epsilon,h}$ by means of the infinite time-evolving block decimation (iTEBD) algorithm~\cite{TEBD}. 
Figure~\ref{fig:Ising_driven} reports the behavior of the order parameter as a function of the number of $n$ kicks. In fact, either starting from the vacuum $\ket{+}$ [panel {\it (a)}], or from the false-vacuum $\ket{-}$ [panel {\it (b)}], 
a small value of the longitudinal component $h$ is sufficient to induce a nonperturbative change in the order parameter dynamics: domain walls get confined into (anti)mesons, thus hindering the melting of the order parameter, which remains finite for exponentially long times (continuous colored lines).

\subsection{Deconfinement by driving and DTC lifetime}
\label{subsec_deconfinement}

In the last section we established that domain-wall confinement induced by a $\mathbb{Z}_2$-symmetry-breaking kick component stabilizes both the magnetization when quenching from the ``true vacuum'' and the antimagnetization when quenching from  the ``false vacuum'' (cf. Fig.~\ref{fig:Ising_driven}). 
We are now ready to come back to our main problem of time-crystalline order, and discuss how generic (non-$\mathbb{Z}_2$-symmetric) kick imperfections impact the order parameter lifetime determined in Sec.~\ref{sec:FIM}.

We consider kicks $K$ in Eq.~\eqref{eq_Floquet} of the form of imperfect spin flips:
\be
\label{eq_imperfectkick}
K \longrightarrow K_{\pi/2}K_{\epsilon,h} = i^L P e^{i \sum_j(\epsilon X_j + h Z_j)}.
\ee
To make progress, generalizing the approach of Sec.~\ref{sec:FIM}, we switch to the toggling frame, i.e., rewrite the \textit{two-cycle} Floquet operator reabsorbing the perfect kick:
\begin{eqnarray}
{U}^2 & = & (K_{\pi/2} K_{\epsilon,h} V_J ) \, ( K_{\pi/2} K_{\epsilon,h} V_J ) \nonumber \\ & \equiv & (-)^L \, ( K_{\epsilon,-h} V_J) \,  ( K_{\epsilon,h} V_J ) 
\label{eq_twocyclefloquet}
\end{eqnarray}
where, as in Eq.~\eqref{eq_toggling}, we have exploited the fact that $(-i)^L K_{\pi/2}=  P$ flips the $Z$ axis, leaving the $\mathbb{Z}_2$-symmetric interactions $V_J$ invariant. Here, however, the toggling frame makes the symmetry-breaking longitudinal component $h$ of the kick periodically flip in sign. The dynamics can thus be seen as generated by strong interactions and weak kicks only (without perfect flips), alternating the sign of the longitudinal component of the kick at each period.

The theory of Ref.~\cite{DeRoeckVerreet} discussed above for $U$, straightforwardly applies to $U^2$ as well: it guarantees the existence of a close-to-identity {time-periodic} unitary transformation $e^{i\bar S_{\le p}}$ such that, in the transformed frame, the two-cycle Floquet dynamics described by Eq.~\eqref{eq_twocyclefloquet} approximately conserves the number of domain walls over a long prethermal timescale analogous to Eq.~\eqref{eq_timescale}. 
In the previous section we have discussed the transformation $e^{iS_{\le p}}$ for $U=K_{\epsilon,h} V_J$ [see Eqs.~\eqref{eq_schiefferwolff} and~\eqref{eq_transformedpicture}]. 
Unfortunately, there is generally no simple relation between  $e^{iS_{\le p}}$ and $e^{i\bar S_{\le p}}$, because
the generators $S_{\le p}$ and the effective Floquet operator ${U}'_p$ depend on both $\epsilon$ and $h$. 
In particular, $h$ and  $-h$ produce generally \textit{different} operators $S_{\le p}$. 
This fact prevents us from straightforwardly combining the two transformations for $K_{\epsilon,h} V_J$ and $K_{\epsilon,-h} V_J$ into a single one for $U^2$.

However, the lowest-order result in Eq.~\eqref{eq_S1} shows that $S_{\le 1}$ is actually independent of $h$. 
This simplification allows us to directly combine the two transformations into a single one for $U^2$ right away: 
substituting into Eq.~\eqref{eq_twocyclefloquet}, we find that the resulting lowest-order transformation $e^{i \bar S_{\le 1}}$ coincides with $e^{i S_{\le 1}}$ in Eq.~\eqref{eq_S1}: 
\begin{eqnarray}
  U^2 &\simeq & (-)^L e^{i S_{\le 1}} {U}'_1(-h) \, {U}'_1(h) e^{- i  S_{\le 1}}  \nonumber \\
  & \equiv & (-)^L e^{i \bar S_{\le 1}} \big({U}^2\big)'_1 e^{- i \bar S_{\le 1}} 
 \label{eq_effectivefloquet2cycle}
\end{eqnarray}
where ${U}'_1(h)$ is expressed in Eq.~\eqref{eq_Utilde}. 
In contrast with ${U}'_1(\pm h)$,
$ \big({U}^2\big)'_1$ is not expressed as the exponential of a time-independent local Hamiltonian, but it results from two time steps 
where the longitudinal field switches between $h$ and $-h$.
On the other hand, the occurrence that the lowest-order transformation $\bar S_{\le 1}=S_{\le 1}$ is time-independent here relies on the special form~\eqref{eq_imperfectkick} chosen for the kick perturbation, which makes  the present derivation especially simple.

\begin{figure}[t!]
  \includegraphics[width=0.5\textwidth]{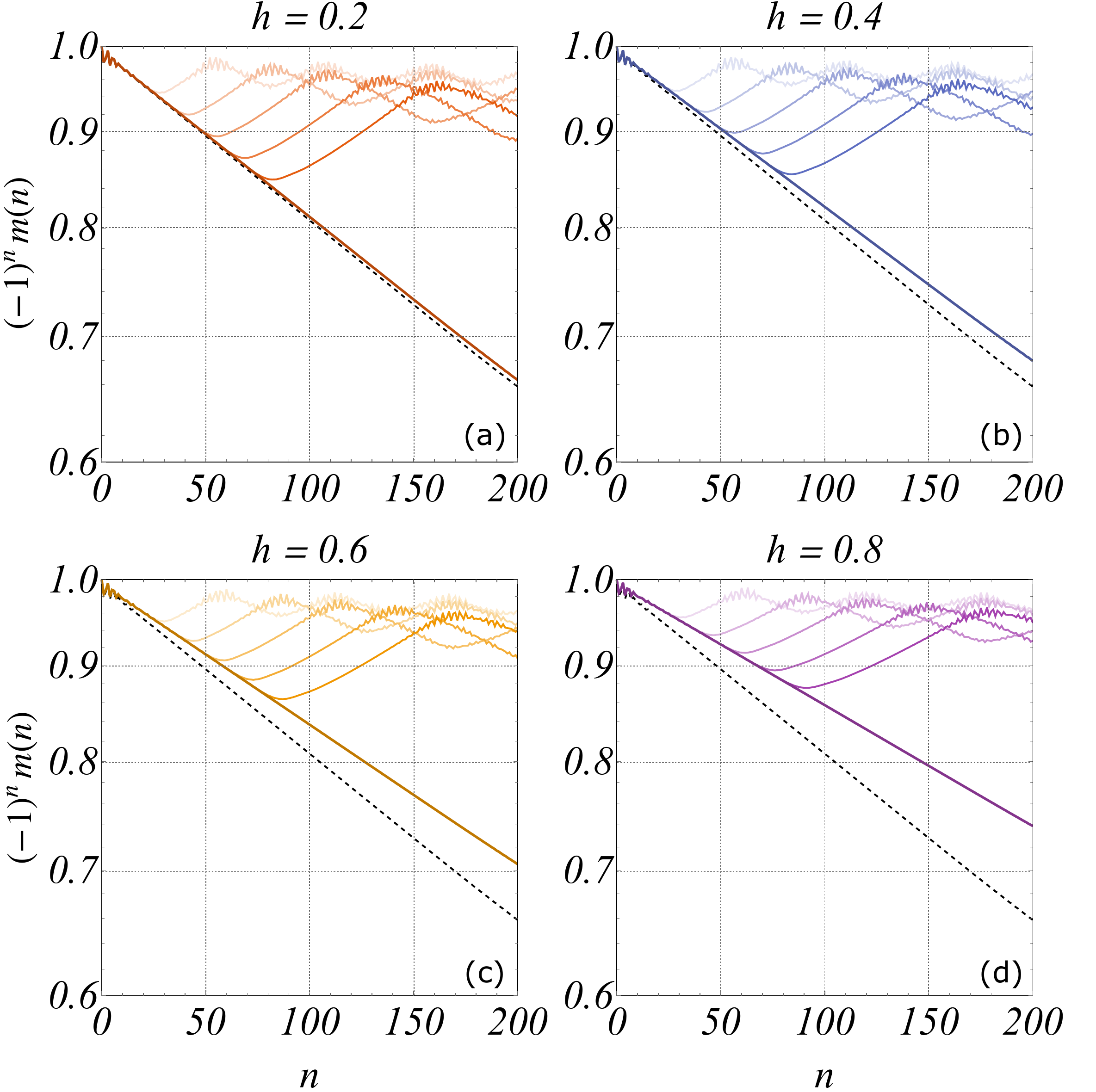}
  \caption{Log-linear plots of the order parameter time-evolution under the Floquet dynamics induced by
    $K_{\pi/2} K_{\epsilon,h} V_J$ for $\epsilon=0.1$ and different values of $h$.
    Thick lines are the iTEBD data, which are compared with ED results for finite system with different
    sizes $L= 10,15,20,25,30$ (shaded lines, from lighter to darker).
    Dashed black lines are the results for $h=0$ and are plotted for comparison.}
\label{fig:Z_piKick}
\end{figure}

Thus our problem amounts to study the driven dynamics of a dilute gas of domain-wall pairs.
To the level of approximation considered above, domain walls are
subject to the unitary dynamics expressed by Eq.~\eqref{eq_Utilde} where, crucially, the confining string tension $h$ regularly flips in sign at integer times, as dictated by Eq.~\eqref{eq_effectivefloquet2cycle}.
Generalizing the analysis of Sec.~\ref{subsec_twobody}, the nature of the evolution of the order parameter ---and hence the fate of time-crystal behavior in the presence of confinement--- will be essentially determined by the solution of the  two-body problem. 

Solving the two-body dynamics amounts to composing the evolution map in Eq.~\eqref{eq_Heisenbergeqsol} with $+h$ and $-h$. Even though domain walls are completely bound into mesons or antimesons within each individual period, we demonstrate that the periodic switching between the two leads to deconfinement, and thus meltdown of the system magnetization.
The exact two-cycle Floquet map restricted to the two-body space in the ``bulk'' (i.e., for $Q \gg 2 {\epsilon}/h$) is equivalent to the composition of two maps given by Eq.~\eqref{eq_Heisenbergeqsol} for $t=1$, with $+h$ and $-h$, respectively. The result of this composition is
\be
\left\{
\begin{split}
  Q(2t+2) & = Q(2t) + 4\frac {{\epsilon}}{h} \big[ \cos P(2t) - \cos (P(2t)+h) \big] , \\
  P(2t+2) & = P(2t) \, .
\end{split}
\right .
\ee
This two-cycle map is equivalent to one generated by an effective static Hamiltonian,
\be
\left\{
\begin{split}
Q(2t+2) & = e^{+i 2 \overline{H}_{\text{cm}}} \, Q(2t) \, e^{-i 2 \overline{H}_{\text{cm}}}, \\
P(2t+2) & = e^{+i 2 \overline{H}_{\text{cm}}} \, P(2t) \, e^{-i 2 \overline{H}_{\text{cm}}},
\end{split}
\right .
\ee
which reads
\be
\label{eq_barHcm}
\overline{H}_{\text{cm}} =\frac {{2\epsilon}}{h}   \big[ \sin P - \sin (P+h) \big]  ,
\ee
as can be readily verified.

Remarkably, $\overline{H}_{\text{cm}}$ is a pure-hopping Hamiltonian, without interaction potentials.
Its eigenstates are no longer bound states localized around a finite value of $Q$, but deconfined plane waves with a definite momentum $P$.
The periodic switching between $+h$ and $-h$ averages out the $\mathbb{Z}_2$-breaking confining potential, and effectively restores the symmetry, similarly to what happens in high-frequency-driven discrete time crystals~\cite{ElsePRX17_PrethermalDTC}.

\begin{figure}[t!]
  \includegraphics[width=0.5\textwidth]{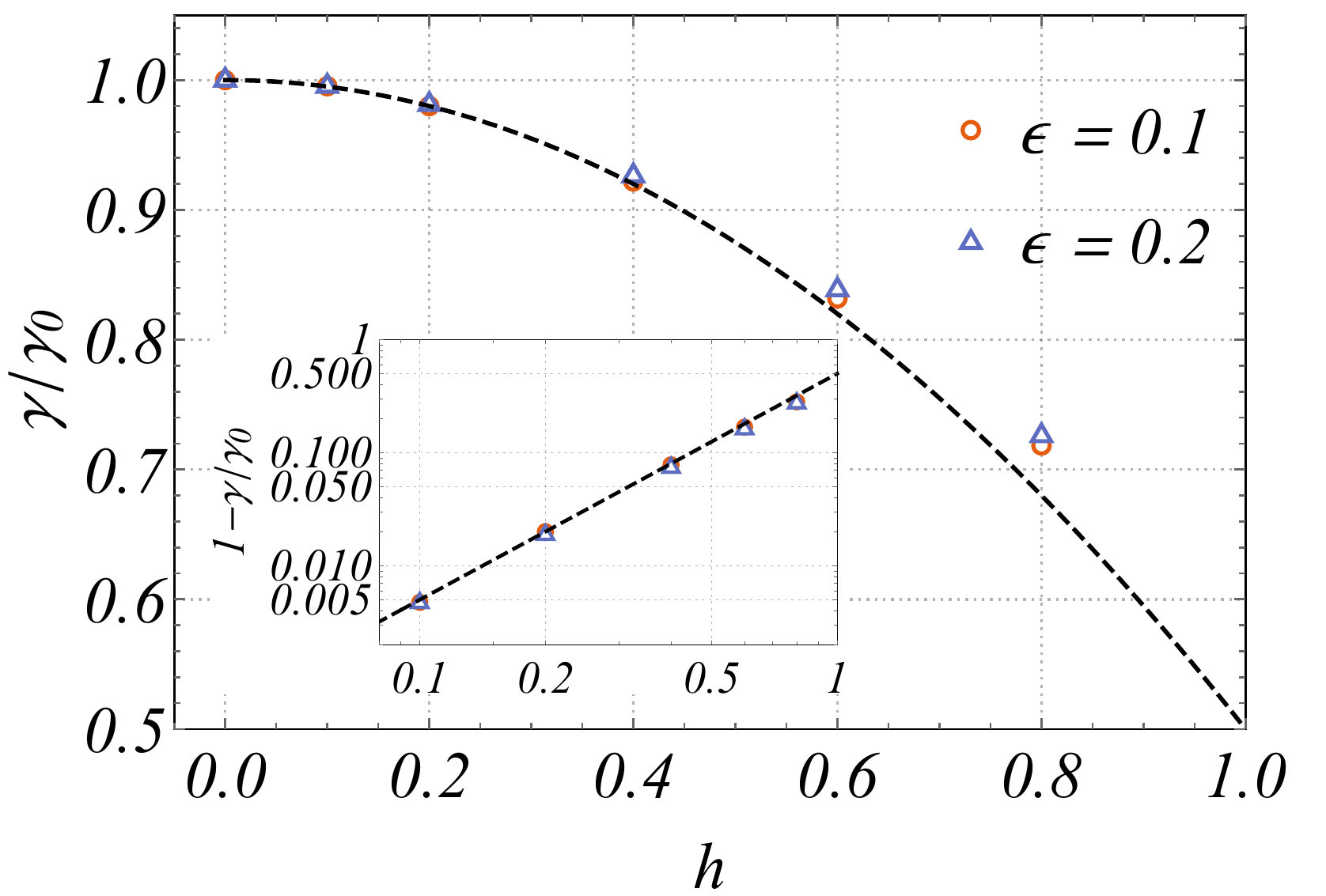}
  \caption{Scaling of the decay rate $\gamma=\gamma(\epsilon,h)$ of the order parameter
    (cf.~Fig.~\ref{fig:Z_piKick}) as a function of the longitudinal kick component~$h$ for different fixed values of~$\epsilon$.
    Rates have been rescaled by $\gamma_0=\gamma(\epsilon,h=0)$. 
    The dashed line is the result of the fit $\gamma/\gamma_0 = 1-c \, h^2$ for the relative correction,
    where $c \simeq 1/2$. The inset shows the same data in log-log scale.}
  \label{fig:gamma_h}
\end{figure}

A semiclassical description of the effective domain-wall dynamics is portrayed in Fig.~\ref{fig:intro}{\it (b)}, where it is highlighted how the periodic switching of the sign of the confining potential $h$ leads to an effective ballistic spreading of the reversed bubble delimited by two domain walls, with a renormalized maximum effective velocity $v_{\text{eff}}=v_{\text{eff}}(\epsilon,h)$.
From Eq.~\eqref{eq_barHcm}, we find the approximate dispersion relation $\overline{v}(k) = \frac {2\epsilon} h [ \sin k - \sin(k+h) ]$, whence  
\be
\frac{v_{\text{eff}}(\epsilon,h)}{ v_{\text{eff}}(\epsilon,0) }= 1-\mathcal{O}(h^2).
\ee
As we neglected all terms in $H_{\text{2-body}}$ beyond the first order  in $\epsilon$ and $h$ [cf. Eq.~\eqref{eq_H2body}], we cannot expect the correction to be quantitatively accurate. However, numerical simulations shown in Fig.~\ref{fig:Z_piKick} and~\ref{fig:gamma_h} clearly indicate a relative reduction of the decay rate $\gamma$ of the order parameter by a factor
\be
\frac{\gamma(\epsilon,h)}{ \gamma(\epsilon,0) } \simeq 1- \frac {h^2}{2},
\ee
which is compatible with a relative decrease  of order $h^2$ for both the density $\rho$ of spin flips in the initial state and the spreading velocity, according to the formula $\gamma \sim \rho v_{\text{eff}}$.

The bottom line of this analysis is that the symmetry-breaking perturbation, leading to confinement of excitations in the static case, does enhance the lifetime of time-crystalline behavior, although only by parametrically decreasing the prefactor $a=a(J,h)$ of the perturbative decay law $\gamma \sim a \epsilon^3$. 
The above arguments can be easily modified to account for generic symmetry-breaking perturbations of the perfect kick. 
Such perturbations, thus, do not qualitatively modify the picture of the order parameter meltdown found in the integrable case in Eq.~\eqref{eq_tauscaling}. 
Although we derived this  result in the lowest perturbative order, we expect that the $\mathbb{Z}_2$ symmetry is restored to all orders, similarly to what happens in MBL and high-frequency driven prethermal DTCs~\cite{VonKeyserlingkPRB16_AbsoluteStability,ElsePRX17_PrethermalDTC}.
Our simulations of Fig.~\ref{fig:Z_piKick} show no signature of slowdown of the computed exponential decay, clearly confirming the expectation.

We finally remark that, similarly to the integrable case, the order-parameter dynamics is strongly affected by the finite size of the chain: our ED data for $L=10\div30$, shown in Fig.~\ref{fig:Z_piKick} (shaded lines), display a deceptive persistence of the order parameter, as previously observed in different contexts~\cite{Davide-Rydberg}, whereas it eventually decays to zero for systems in the thermodynamic limit (thick straight lines). Note that both operators $K$ and $V$ can be exactly applied  to the many-body wavefunction, which allows us to efficiently simulate dynamics of $30$ spins with a reasonable amount of resources.

\subsection{Enhancement of DTC lifetime at dynamical freezing points}
\label{subsec_freezing}

The theoretical analysis above assumed the Ising coupling $J=\mathcal{O}(1)$ to be the dominant scale in the Floquet operator, and the transverse and longitudinal fields $\epsilon, h \ll 1$ to be weak perturbations to the perfect spin flip.
In this Subsection, building on the recent Refs.~\cite{haldar18onset,Haldar21dynamical,Luitz20prethermalization}, we point out that the DTC lifetime can be strongly enhanced if the kick perturbation is tuned to particular large [$\mathcal{O}(1)$] angles, referred to as {\it dynamical freezing points}.
The argument goes as follows.

We first of all rewrite our kicking protocol in Eq.~\eqref{eq_imperfectkick} in the equivalent form considered in Ref.~\cite{Luitz20prethermalization}.
The two-cycle Floquet operator $U^2=PKVPKV$ can be turned into the form $P\tilde K VP\tilde K V$ with a modified kick
\begin{equation}
  K = e^{i \sum_j(\epsilon X_j + h Z_j)} \quad \leadsto \quad
  \tilde K = e^{i \sum_j\tilde\epsilon X_j} e^{i \sum_j \tilde h Z_j} 
  \label{eq_euler}
\end{equation}
using the Euler angle decomposition of the former rotation. In fact,
there exist angles $\alpha,\beta,\gamma$, functions of $\epsilon$ and $h$, such that 
\begin{equation}
  e^{i (\epsilon X + h Z)} = e^{i\alpha Z} e^{i\beta X} e^{i\gamma Z}
\end{equation}
(to lighten the notation, we suppress $\sum_j (\dots)_j$ in all exponents in the rest of this Subsection).
Thus, we can rewrite 
\begin{equation}
  \begin{split}
    U^2 = & P e^{i\alpha Z} e^{i\beta X}
    e^{i\gamma Z} e^{iJZZ} P e^{i\alpha Z} e^{i\beta X}
    e^{i\gamma Z} e^{iJZZ} \\
    = &  e^{-i\alpha Z} e^{i\beta X}
    e^{-i\gamma Z} e^{iJZZ}  e^{i\alpha Z} e^{i\beta X}
    e^{i\gamma Z} e^{iJZZ} \\
    = & e^{-i\alpha Z} e^{i\beta X}
    e^{i(\alpha-\gamma) Z}
    e^{iJZZ}   e^{i\beta X}
    e^{i\gamma Z} e^{iJZZ}
  \end{split}
\end{equation}
and finally perform a unitary change of frame,
\begin{equation}
  e^{i\alpha Z} U^2 e^{-i\alpha Z} = e^{i\beta X} e^{i(\alpha-\gamma) Z}  e^{iJZZ} e^{i\beta X} e^{-i(\alpha-\gamma) Z} e^{iJZZ}
\end{equation}
to obtain our claim above, upon posing $\tilde\epsilon\equiv \beta$ and $\tilde h \equiv \alpha-\gamma$. 
Note that the initial state and the magnetization observable are unaffected by the last unitary transformation $e^{i\alpha Z}$, so the two Floquet operators are fully equivalent.
Thus, we will consider the equivalent form of the Floquet operator with the modified kick in Eq.~\eqref{eq_euler}.

The key observation of Ref.~\cite{Haldar21dynamical} -- exploited in Ref.~\cite{Luitz20prethermalization} in the context of DTC response in NMR experiments -- is that certain suitably chosen longitudinal field driving amplitudes proportional to the driving frequency may cause an averaging effect on the noncommuting transverse field perturbation, such that in the rotating frame co-moving with the strong drive the perturbation is heavily suppressed.
This idea is implemented in our scheme similarly to what done in Ref.~\cite{Luitz20prethermalization}: We  rewrite our two-cycle Floquet operator as
\begin{equation}
\begin{split}
    U^2=& e^{i\tilde\epsilon X} e^{i\tilde hZ} e^{iJZZ}
e^{i\tilde \epsilon X} e^{-i\tilde hZ} e^{iJZZ} \\ =&
e^{i\tilde\epsilon X} e^{iJZZ}
e^{i\tilde\epsilon [\cos(2\tilde h) X + \sin(2\tilde h) Y]} e^{iJZZ} \, .
\end{split}
\end{equation}
As in Refs.~\cite{Haldar21dynamical,Luitz20prethermalization}, here we consider the regime where $J$ and $\tilde\epsilon$ are small, so that we can apply a standard high-frequency expansion to the transformed Floquet operator $U^2$. In our case, this expansion is equivalent to the plain BCH expansion.
To lowest order, we obtain 
\begin{equation}
  \label{eq_gaugedaway}
  U^2 \simeq e^{i\tilde \epsilon [1+\cos(2\tilde h)] X + i\tilde\epsilon \sin(2\tilde h) Y+ i2JZZ} \, .
\end{equation}
The lowest-order dynamics are solvable by mapping to free fermions, similarly to our treatment in Sec.~\ref{sec:FIM}: rotating around the $Z$ axis by a suitable angle we realize that the exponent in Eq.~\eqref{eq_gaugedaway} describes a standard quantum Ising chain with coupling $2J$ and a transverse field generally proportional to $ \tilde\epsilon$. Thus, excitations are domain-wall-like for $\tilde\epsilon$ small enough. As we proved in Sec.~\ref{sec:FIM}, the magnetization of this driven spin chain will decay exponentially with a rate $\gamma\sim\tilde\epsilon^3$.
Nevertheless, we may cancel the perturbation to lowest order by setting $\tilde h=\pi(k+1/2)$ with $k$ integer: at these {\it dynamical freezing points} the two-cycle Floquet operator reads
\begin{equation}
  \label{eq_freeze}
  U^2 = e^{i\tilde\epsilon X} e^{iJZZ} e^{-i\tilde\epsilon  X} e^{iJZZ} \quad \simeq e^{i2JZZ}
\end{equation}
where the last equality holds to lowest order.
As it is evident, in this case the decay rate of the magnetization becomes of higher order.
Furthermore, unlike the general case, for these specific choices of $\tilde h$ the {\it exact} Floquet dynamics is solvable to all orders in $J,\tilde\epsilon$, as is evident from the first equality above: all four unitaries are mapped to quadratic fermions by the same Jordan-Wigner transformation. 
Computing the exact dispersion relation of the fermionic quasiparticles, we have verified that the excitations' bandwidth grows proportionally to $ \tilde\epsilon^2$ for small $\tilde\epsilon/J$ --
in other words, there is no accidental cancellation of $\mathcal O (\tilde\epsilon^2)$ corrections.

The argument above shows that our Floquet model with longitudinal field set to one of the dynamical freezing points $\tilde h=\pi(k+1/2)$  becomes an exactly solvable Floquet model, Eq.~\eqref{eq_freeze} above, similar to the one we considered in Sec.~\ref{sec:FIM}, but with the strength of the perturbation $\tilde\epsilon$ being replaced by $\tilde\epsilon^2$ due to a smart choice of the driving protocol. Applying our analysis of Sec.~\ref{sec:FIM}, we conclude that the magnetization decays exponentially in this model, with a strongly suppressed rate $\gamma \sim (\tilde\epsilon^2)^3=\tilde\epsilon^6$.
We note that this lifetime is (much) longer than that found in the XXZ-type spin chain considered in Ref.~\cite{Luitz20prethermalization}.
Lastly, we remark that if $\tilde h$ is slightly detuned from a dynamical freezing point, the Floquet dynamics may still be analyzed using the theory developed in the previous Subsections, showing that the expected decay rate remains qualitatively unaltered ($\gamma \sim \tilde\epsilon^6$).

\section{Stabilization of DTC response by interactions beyond nearest neighbors}
\label{sec_LR}

A weaker version of domain-wall confinement also arises in the ordered phase of spin chains with an interaction range extended over multiple sites, even in the absence of explicit symmetry-breaking fields. The basic mechanism was identified in Ref.~\cite{GorshkovConfinement}:
the separation of two domain walls involves an increase of the configurational energy, due to the increase in the number of frustrated bonds between pairs of spins beyond the nearest neighbors.
As discussed in Refs.~\cite{GorshkovConfinement,LeroseDWLR} and experimentally verified in Ref.~\cite{TanConfinementExperiment}, 
this gives rise to an effective attractive potential $v(r)$ between two domain walls.
The resulting physics is thus reminiscent of that generated by a longitudinal field. Here, however, the interaction tail tunes the shape and depth of the effective potential well, which, for the kind of interactions relevant to this work, grows sublinearly at large distances.

Below we show that this general domain-wall binding mechanism also arises in the Floquet context. 
Crucially, as it does not rely on explicitly breaking the symmetry, it does not suffer from time-averaging effects demonstrated in Sec.~\ref{sec_mixedfield}, opening the door to a true enhancement of the time-crystal order lifetime, as pictorially illustrated in Fig.~\ref{fig:intro}{\it (c)}.
As a core result of this section, we establish that an increase of the interaction range $R$ (i.e., $J_{i,j} \neq 0$ only if $|i-j|\le R$) leads to an \emph{exponential} enhancement of the order-parameter lifetime, $\gamma \sim a \, \epsilon^{2R+1}$. 
We further discuss the practical conditions for observing this confinement-stabilized DTC behavior, and the implications for experiments with Rydberg-dressed interacting atomic chains, recently realized in the lab~\cite{RydbergDressed, RydbergDressed2}.
Based on this result, we will finally argue that long-range algebraically decaying interactions with a generic exponent $\alpha$ (not necessarily smaller than $2$) stabilize the order parameter over timescales longer than any inverse power of the kick perturbation $\epsilon$.

This section is organized as follows. In Sec.~\ref{subsec_effectiveDWdynamicsLR}, we generalize the derivation of an effective domain-wall conserving Floquet Hamiltonian of Sec.~\ref{subsec_effectiveDWdynamics} to a chain with arbitrary Ising couplings $J_{i,j}$ beyond nearest neighbors.
Hence, in Sec.~\ref{subsec_DWbinding} we study the two-body problem, which is richer than the corresponding one studied in Sec.~\ref{subsec_twobody} due to the coexistence of bound states and deconfined continuum, and determine the conditions for domain-wall binding and their implications for the evolution of the order parameter.
Building on the intuition from the two-body problem, in Sec.~\ref{subsec_expsuppression} we switch to a more scrupulous analysis and prove that, in the asymptotic regime of weak kick perturbation $\epsilon$, the order-parameter decay rate gets heavily suppressed as $\gamma \sim a \, \epsilon^{2R+1}$.
This and related theory predictions are numerically verified in Sec.~\ref{subsec_numerics}. 
Finally, in Sec.~\ref{subs_LRlimit} we discuss the limit of long-range interactions and argue that the decay becomes nonperturbatively small.

\subsection{Effective domain-wall dynamics for weak kicks}
\label{subsec_effectiveDWdynamicsLR}

We generalize Eq.~\eqref{eq_Floquet} by considering a kicked Ising chain with arbitrary longer-range couplings, defined by the Floquet operator
\be
\label{eq_FloquetLR}
U = K V_{\mathbf{J}}
, \qquad
V_{\mathbf{J}} = e^{i \sum_{j=1}^L \sum_{r=1}^R J_r Z_j Z_{j+r}} 
\ee
with periodic boundary conditions.
We have denoted by
$\mathbf{J}\equiv (J_1,J_2,\dots,J_R)$ the array of coupling strengths at increasing distances, with
$\pi/2 > J_1 \ge J_2 \ge \dots\ge  J_R  > 0 $, and we implicitly assumed $R < L/2$.
We consider the range $R$ fixed and independent of the system size $L$ 
(the long-range limit $R\propto L$ will be discussed in Sec.~\ref{subs_LRlimit}).
For simplicity, we take the kick $K$ as in Eq.~\eqref{eq_perturbedtransversekick}, i.e.,
\be
K \longrightarrow K_{\pi/2+\epsilon} = e^{i(\pi/2+\epsilon)\sum_j X_j} = i^L P e^{i\epsilon\sum_j X_j} .
\ee
As shown in the previous Sec.~\ref{sec_mixedfield}, an explicit symmetry-breaking component $h$ of the kick is not expected to qualitatively enhance the time-crystal lifetime.
Thus, to lighten our arguments, we will discard it.
We will later numerically verify that, indeed, those terms do not impact the lifetime of the DTC response.
Note that, despite $h=0$, the couplings beyond nearest neighbors break integrability.
As in Eq.~\eqref{eq_toggling}, transforming to the toggling frame gives
\be
m(n) = (-1)^n
\langle +  | [K_{\epsilon} V_{\mathbf{J}}]^{-n} 
Z_{j}  
[K_{\epsilon} V_{\mathbf{J}}]^n | + \rangle,
\ee
so we focus on weak kicks $K_{\epsilon}$ close to the identity.

Along the lines of the first part of this paper, we want to build intuition on the evolution of the order parameter in terms of the dynamics of domain walls.
The construction of the effective domain-wall conserving Floquet Hamiltonian of Sec.~\ref{subsec_effectiveDWdynamics} can be straightforwardly generalized to the present case of the Floquet operator
$U=K_\epsilon V_{\mathbf{J}}$
with interactions beyond nearest neighbors.
To this aim, we first work in the regime of ``weak confinement'' $J_{2},\dots,J_{R} \ll J_1$.
As the dominant scale $J_1$ in the problem couples to the number of domain walls $D_1$ [defined in Eq.~\eqref{eq_D1}],  we can set up a perturbation theory similar to that of Sec.~\ref{subsec_effectiveDWdynamics}, where we had $h\ll J$. 

The derivation closely parallels that of Sec.~\ref{subsec_effectiveDWdynamics}, based on the general theory
of Ref.~\cite{DeRoeckVerreet}. In this case the perturbative parameter is
$\hat \epsilon = \max (\epsilon, J_{2},\dots,J_R)$.
The rigorous bounds of Ref.~\cite{DeRoeckVerreet} ensure that the density of perturbatively dressed domain walls
remains accurately conserved over the long prethermal timescale in Eq.~\eqref{eq_timescale},
where the numerical constant $c$ is here adjusted to account for the longer range $R$ of the perturbation operator.

Similarly to the case discussed in Sec.~\ref{subsec_effectiveDWdynamics}, for low-order explicit computations it is more practical to
follow a different scheme from Ref.~\cite{DeRoeckVerreet} and aim for a {static} effective Floquet Hamiltonian $H^F$ 
by combining the two exponentials of the product $K_\epsilon V_{\mathbf{J}}$ 
using the replica resummation of the BCH expansion~\cite{ProsenPolkovnikovPRL18_ReplicaBCHResummation}, order by order in the kick imperfection $\epsilon$,
and hence perform a conventional static Schrieffer-Wolff transformation on $H^F$, order by order in $\hat \epsilon$.
The presence of arbitrary Ising couplings here requires a nontrivial generalization of the calculation in Ref.~\cite{ProsenPolkovnikovPRL18_ReplicaBCHResummation}.
The structure of the resulting expansion is worked out in Appendix~\ref{app_derivationFloquet}.

The lowest-order result reported here is a simple generalization of Eqs.~\eqref{eq_Utilde},~\eqref{eq_S1}:
\be
\label{eq_HFLR}
H^F_{1,1} = - \sum_{j,r\ge 2} J_r \, Z_j Z_{j+r} - \epsilon \, 
(P^\uparrow_{j-1} X_{j} P^\downarrow_{j+1} + P^\downarrow_{j-1}  X_{j} P^\uparrow_{j+1}),
\ee
\begin{multline}
\label{eq_S1LR}
  S_{\le1} = -\frac{\epsilon}{2} \sum_j P_{j-1}^{\uparrow} [X_j - \cot(2J_1) Y_j ] P_{j+1}^{\uparrow} +\\
  +  P_{j-1}^{\downarrow} [X_j + \cot(2J_1) Y_j ] P_{j+1}^{\downarrow} \, .
\end{multline}
Due to the arbitrary range of the perturbation, the appearance of higher order terms in the effective Hamiltonian and in the Schrieffer-Wolff generator is more cumbersome than that discussed in Sec.~\ref{subsec_effectiveDWdynamics}. Appendix~\ref{app_derivationFloquet} presents the general hierarchical structure of the expansion and explicitly reports the second-order result, for illustration. 
Terms of order $p$ in $H^F$ contain at most $p$ spin-flip operators $S_j^\pm\equiv \frac 1 2( X_j\pm i Y_j)$ separated by a distance at most $R$, i.e., they contain strings of the form $S^{\mu_1}_{j_1} \cdots S^{\mu_q}_{j_q}$ with  $\mu_1,\dots,\mu_q=\pm$, $q\le p$, and $j_1<\dots<j_q$ with $j_{n+1}-j_n \le R$ for all $n$. {Such a product 
  of spin-flip operators (which we occasionally refer to as a quasilocal ``cluster'') is  multiplied by complicated products of diagonal $Z_j$ operators -- with coefficients depending on the couplings $\{J_r\}$ -- located not farther away than $R$ sites from the cluster}.
Finally, projectors $P^{\uparrow}_j,P^{\downarrow}_j$ applied to spins adjacent to the position of the spin-flip operators ensure that the spin flips only move domain walls without creating or annihilating them.
Likewise, the Schrieffer-Wolff generator at order $p$ contains clusters of at most $p$ spin-flip operators, with the same locality properties as discussed above.

\subsection{Domain-wall binding}
\label{subsec_DWbinding}

Similarly to Sec.~\ref{subsec_twobody}, we now analyze the dynamics of the order parameter in terms of the motion of conserved domain walls in the Schrieffer-Wolff-transformed picture. 
The initial state in the transformed picture $\ket{+'} = e^{-i S_{\le p^*}} \ket{+}$ can be viewed as a dilute gas of quasilocal clusters of $p=1,2,\dots,p^*$ flipped spins, the respective density being suppressed as $\epsilon^{2p}$.
To lowest order $p=1$, these are isolated spin flips, i.e., adjacent pairs of domain walls. Thus, to this order of approximation, the evolution of the order parameter can be understood starting from the two-body problem. 

Let us begin with qualitative considerations.
Interactions beyond the nearest neighbors lead to the presence of a discrete set of bound states, coexisting with a continuum of unbound domain walls for larger energy.
Furthermore, interactions between distant spins also favor the formation of more structured ``molecular'' bound states out of larger clusters of domain walls.
The rich nonequilibrium dynamics of the system, including the anomalously slow decay of the order parameter, results from the coexistence of topological and nontopological excitations in the spectrum (i.e., unbound domain walls and bound pairs) which we now turn to quantitatively analyze. 

\begin{figure*}[t!]
  \includegraphics[width=0.95\textwidth]{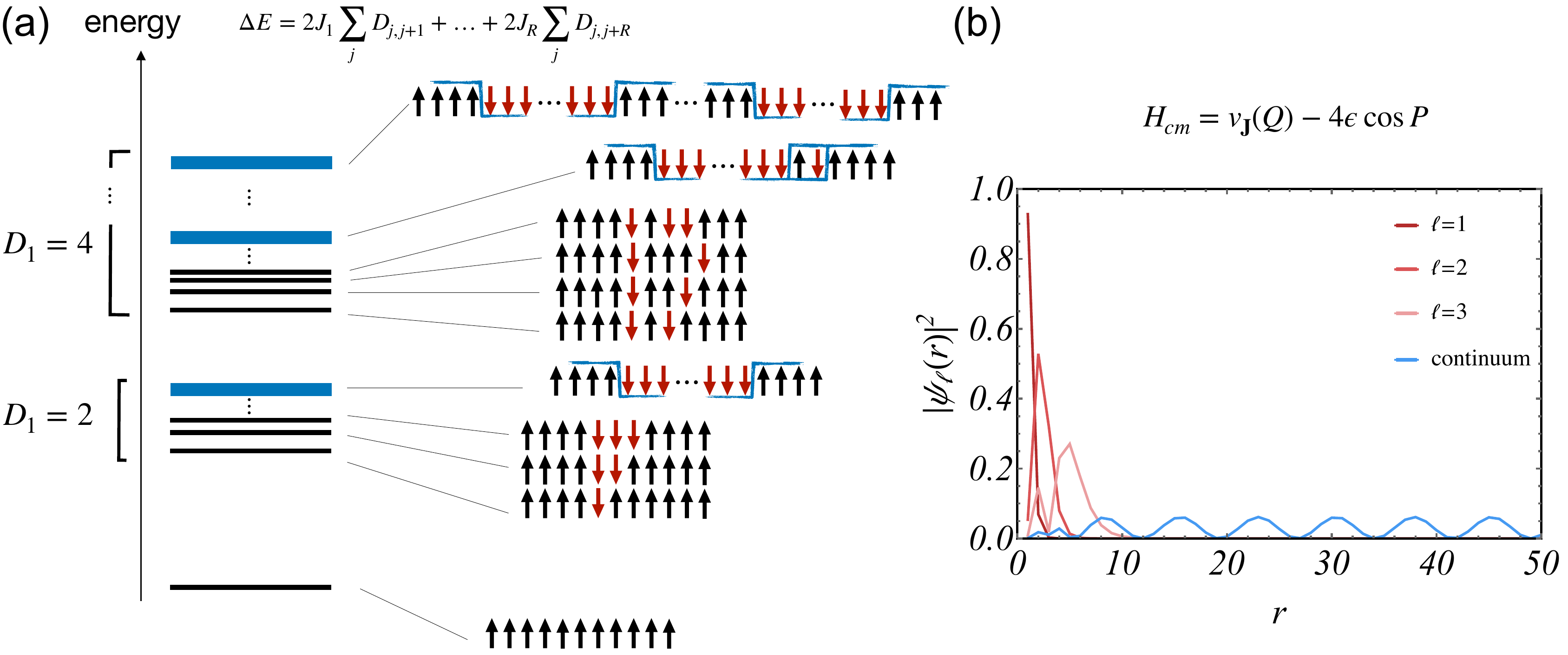}
  \caption{{\it (a)} Sketch of the eigenstates of the Floquet Hamiltonian $H^F$ for $\epsilon\to0$.
    The vertical axis represents the excitation energy $\Delta E$ above the Floquet ground state.
    {\it (b)} A selection of eigenstates of the two-body problem~\eqref{eq_2bodyLR} in the center-of-mass frame,
    for $J_r=1/r^\alpha$ for  $r\le R=10$, $\alpha=3$, and $\epsilon=0.1139$.}
  \label{fig:2bodyLR}
\end{figure*}

The effective domain-wall conserving Floquet Hamiltonian~\eqref{eq_HFLR} projected onto the two-particle sector gives the following first-quantized two-body problem, analogous to Eq.~\eqref{eq_H2body}:
\begin{multline}
  \label{eq_H2bodyLR}
  H_{\text{2-body}} =   \sum_{j_1<j_2} v_{\mathbf{J}}(j_2-j_1) \ket{j_1,j_2}  \bra{j_1,j_2}
  \\  - {\epsilon}  \sum_{j_1<j_2} \big(\ket{j_1+1,j_2} + \ket{j_1,j_2+1}\big) \bra{j_1,j_2}  + \text{ H.c.},
\end{multline}
where 
\be
v_{\mathbf{J}}(r) = 
\left\{
\begin{split}
  4 \sum_{d=1}^r d J_d + 4r \sum_{d=r+1}^R J_d, & \qquad \text{if } r<R , \\
  4 \sum_{d=1}^R d J_d \equiv v_{\mathbf{J}}(\infty), &  \qquad \text{if } r\ge R .
\end{split}
\right.
\ee
The two-body potential $v_{\mathbf{J}}(r)$ grows as a function of the distance up to $r=R$, then flattens out.
Thus, the potential well hosts a finite number of bound states, which grows to $R-1$ upon decreasing $\epsilon\to0$.
In field-theoretical language, this discrete set of  energy levels forms the mass spectrum of nontopological particles.
Above this, a continuum of scattering states appears, built out of two unbound
domain walls;
in field-theoretical language, the spectrum contains stable topologically charged particles, i.e., kinks and antikinks. The topological nature of these excitations stems from the fact that they can be locally created or destroyed in globally neutral pairs only, not individually.

The bound (nontopological) and unbound (topological) excitations can be distinguished by being labelled by a real-space or momentum-space quantum number.
To understand this, let us transform to the center-of-mass frame $\Psi(j_1,j_2)=e^{iK(j_1+j_2)}\psi(j_2-j_1)$ and set $K=0$ due to translational invariance of the nonequilibrium initial state (cf.~Sec.~\ref{subsec_twobody}).
The reduced wavefunction $\psi$ satisfies the Schr\"{o}dinger equation
\be
\label{eq_2bodyLR}
v_{\mathbf{J}}(r) \psi(r) - 2\epsilon [\psi(r+1)+\psi(r-1) ] = E \psi(r)
\ee
in the domain $r>0$, subject to the boundary condition $\psi(0)\equiv 0$.
This equation defines the center-of-mass frame Hamiltonian $H_{\text{cm}}$.
For $\epsilon\to0$, the eigenfunctions in the center-of-mass frame $\psi_\ell(r)=\delta_{\ell,r}$ correspond to contiguous reversed domains of $\ell$ spins, with eigenvalues $E_\ell=v_{\mathbf{J}}(\ell)$. 
In this limit, the discrete label $\ell=1,\dots,R-1$ thus has the physical meaning of distance between the two domain walls.
Due to the discreteness of the spectrum, this labelling can be adiabatically continued to finite $\epsilon$, where eigenfunctions feature quantum fluctuations of the physical distance. 
On the other hand, the degenerate levels $E_{R}=E_{R+1}=\dots=v_{\mathbf{J}}(\infty)$ split into a continuous band $E_k=v_{\mathbf{J}}(\infty)-4\epsilon\cos k$, with eigenfunctions labelled by the relative momentum $k$ of the two domain walls.
The binding potential $v_{\mathbf{J}}$ only affects these eigenfunctions via the relative scattering phase $e^{i\delta_k}$ between incoming and outgoing waves.
We can  estimate the stability range of the most excited [($\ell=R-1$)-th] bound state by the condition that the continuum band $E_k$ does not overlap the discrete energy level $E_{R-1}$. This yields the range $\epsilon \lesssim J_R$; for larger $\epsilon$, the bound state hybridizes with the unbound continuum.
Energy levels and eigenfunctions are further renormalized by higher-order processes of order $\hat\epsilon^2$, $\hat\epsilon^3$,\dots to be added to $H_{\text{2-body}}$, not included in Eq.~\eqref{eq_H2bodyLR}. These additional terms consist of diagonal terms and longer hops of domain walls by at most $2$, $3$, \dots lattice sites, respectively. For small enough $\epsilon$, the resulting quantitative corrections do not alter the qualitative structure of the spectrum discussed here.

Figure~\ref{fig:2bodyLR}{\it (a)} reports a sketch of the eigenstates of the Floquet Hamiltonian for $\epsilon\to0$.
Unlike in Fig.~\ref{fig:2body}, both discrete ``mesonic'' bound states and unbound ``domain-wall-like'' states appear.
The latter are highlighted by blue lines in the pictorial sketch, and their energy levels are marked by thick blue bars, indicating that they form continuous bands of width $\propto\epsilon$.
Panel {\it (b)} reports a selection of eigenfunctions of the two-body problem~\eqref{eq_2bodyLR} in the center-of-mass frame, for algebraically decaying coupling $J_r=1/r^\alpha$ truncated to the range $r\le R=10$, $\alpha=3$, and $\epsilon=0.1139$.
In this case, due to small couplings beyond the nearest neighbors and comparatively large hopping $\epsilon$, the number of bound states (red curves) is only $3$. As $\epsilon\to0$, it grows to $R-1=9$. Higher excited eigenstates are unbound plane waves (blue curves).

The two-body problem already gives us important hints on the nonequilibrium evolution of the order parameter.
As a matter of fact, confined domain walls only produce a weak oscillatory behavior of $m(n)$ with frequencies related to the excitation energies (masses) of the bound states.
Within the prethermal time window $0\le t \le T_{\text{preth}}$ in Eq.~\eqref{eq_timescale}, the order parameter decay is only ascribed to the dynamical production of unbound domain walls.
As discussed above, domain-wall-like excitations exist as higher energy excitations.
While in thermal equilibrium such domain-wall excitations are finitely populated, 
imperfect kicks $K_\epsilon$
will only excite the domain-wall continuum very weakly, precisely with amplitude $~\epsilon^R$, as generating unbound domain walls requires to flip a critical contiguous domain of at least $R$ spins.
This key insight suggests that the destruction of long-range order in the Floquet-prethermal Gibbs ensemble $e^{-\beta H^F}/Z$ is a very slow process, leaving large room for nonequilibrium time-crystalline behavior.

\subsection{Exponential suppression of the order-parameter decay rate}
\label{subsec_expsuppression}

Generalizing the reasoning of Sec.~\ref{subsec_physicalinterpretationDW},
the solution of the two-body problem suggests that the decay rate is severely suppressed:
The density of critical domains in the initial state is of order $\rho \sim (\epsilon^{R})^2$, and the spreading velocity of their constituents domain walls is $v \sim |\epsilon|$, leading to the estimate 
\be
\label{eq_decayLR}
\gamma \sim \rho v  = \mathcal{O}(\epsilon^{2R+1}).
\ee
For $R=1$, domain walls are the only stable excitations in the spectrum, and we recover the exact result $\gamma \sim \epsilon^3$ of Sec.~\ref{subsec_exactdecayintegrable}.
For $R>1$,  the appearance of bound states  is expected to significantly slow down the order parameter decay.

The argument based on the two-body problem 
is, however, too naive, as it completely neglects all multi-body processes and interactions between confined domain walls.
In fact, not only adjacent domain walls attract each other via the two-body potential $v_{\mathbf{J}}(r)$, but also mesonic bound states themselves are subject to an effective attraction when their distance is less than $R$.
\footnote{For comparison, note that mesonic bound states in the presence of a longitudinal field $h$, instead, interact only via local contact interactions, as the linear ``Coulomb'' potential is completely screened in one dimension.}
Such attractive forces can be thought of as residual interactions between bound states of elementary particles, physically analogous to nuclear forces that keep together protons and neutrons (i.e., bound states of quarks), or to molecular forces that keep together atoms (i.e., bound states of electrons and nuclei).
The meson-meson attractive potential can be defined as 
\be
w_{\mathbf{J},\ell_1,\ell_2}(r) = E_{\mathbf{J}}(\ell_1,r,\ell_2) - v_{\mathbf{J}}(\ell_1) - v_{\mathbf{J}}(\ell_2) <0 ,
\ee
where $E_{\mathbf{J}}(\ell_1,r,\ell_2)$ is the configurational energy of two reversed domains of length $\ell_1$ and $\ell_2$, separated by $r$ spins in between.
One finds
\begin{eqnarray}
w_{\mathbf{J},\ell_1,\ell_2}(r) & = & -4 \sum_{1\le i \le r_1}  \; \sum_{\ell_1+r+1 \le j \le \ell_1+r+\ell_2} J_{|i-j|}  \nonumber \\
& = & -4 \sum_{s={r+1}}^{\min(\ell_1+\ell_2+r-1,R)} n_s J_s ,
\end{eqnarray}
with $n_s \equiv \min(s-r,\ell_1,\ell_2,r+\ell_1+\ell_2-s)$.
Considering fixed  $\ell_1$ and $\ell_2$, 
this two-body potential $w_{\mathbf{J},\ell_1,\ell_2}(r)$ grows monotonically as a function of the distance $r$, from $w(r=1)<0$ to $w(r \ge R) \equiv 0$.
For instance, two isolated spin flips ($\ell_1=\ell_2=1$) experience an attractive potential proportional to the bare coupling, $w_{\mathbf{J},1,1}(r) = - 4J_r$. 

\begin{figure}[t!]
  \includegraphics[width=0.35\textwidth]{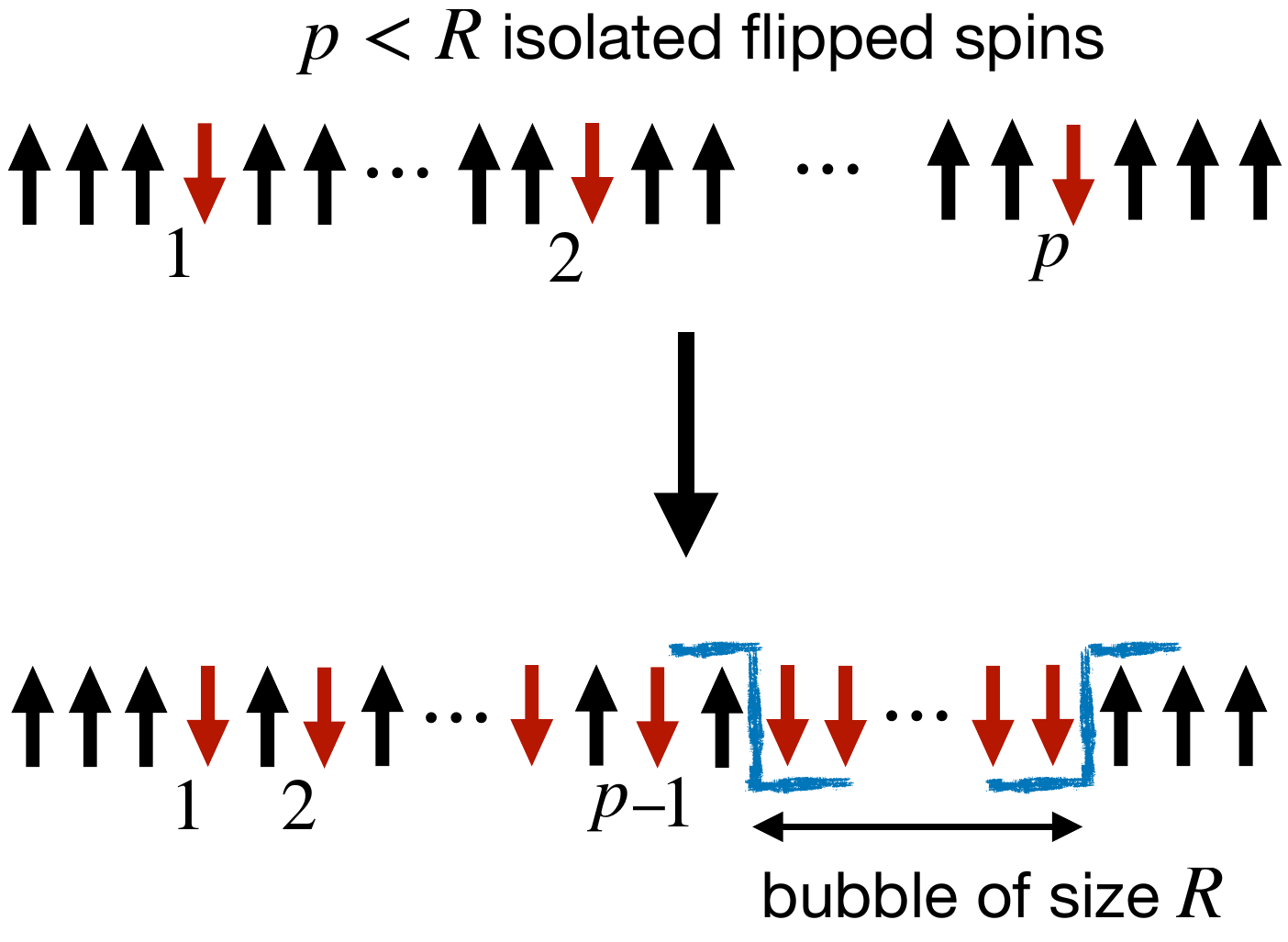}
  \caption{A quasilocal cluster of few ($p<R$) spin flips (top configuration) could agglomerate tightly and convert the accumulated extra interaction energy into the formation of a reversed bubble of size $R$ (bottom configuration), thus releasing a free domain wall, while conserving the total number of domain walls.
    While energetically allowed, such a process can be shown to be necessarily off-resonant under generic incommensurability assumptions on the couplings $\{J_r\}$ (see main text).
  }
  \label{fig:meltingprocess1}
\end{figure}

It is easy to imagine a process where a cluster of few flipped spins or small domains, initially distant from each other, agglomerate more tightly, making the gained reciprocal meson-meson interaction energy available to progressively enlarge a domain and finally release a domain-wall from the attraction of the rest of the cluster at distance $>R$, such that it can freely travel away and melt the order parameter.
{In Fig.~\ref{fig:meltingprocess1} we sketch such an example, where a finite ($R$-independent) cluster of initial spin flips
  possesses sufficient energy to release a traveling domain wall. }
It is a-priori conceivable that such processes could trigger a comparatively fast decay of the order parameter
at low perturbative order. Remarkably, however, it is possible to rigorously exclude such a scenario,
and prove that the fastest decay process occurs at order $R$.

To establish this result, we need to modify the perturbative Schrieffer-Wolff transformation of Sec.~\ref{subsec_effectiveDWdynamics} to explicitly account for the fact that all couplings $J_1,\dots,J_R$ are large compared to $\epsilon$ when the asymptotic regime $\epsilon\to0$ is considered. 
The unperturbed Floquet operator $V_{\mathbf{J}}=e^{i\sum_{j,r} J_r Z_j Z_{j+r}}$ defines highly degenerate sectors of the many-body Hilbert space, identified by the energy levels
\be
E(n_1,\dots,n_R) = E_{GS}+ 2\sum_r n_r J_r,
\ee
where $E_{GS}=-L\sum_r J_r$ is the unperturbed ground-state energy of the fully polarized state $\ket{+}$, and the non-negative integer $n_r \in \mathbb{N}$ has the meaning of total number of frustrated bonds at distance $r$.
Under the assumption of strong incommensurability of the couplings $\{ J_r \}$ in Eq.~\eqref{eq_diophantine} ---necessary to derive a well defined static Floquet Hamiltonian, as shown in Appendix~\ref{app_replica}--- each degenerate sector is in one-to-one correspondence with the set $\{n_r\}$.
For $\epsilon\to0$, transitions between such sectors are energetically suppressed and can be adiabatically eliminated.
In other words, we can dress the effective Hamiltonian within each sector, order by order in $\epsilon$, to account for all resonant processes occurring via virtual transitions between different sectors.
The strong incommensurability condition on the couplings guarantees that each such transition is accompanied by a finite energy denominator.
The construction can thus be formally carried out to all orders in $\epsilon$~\cite{DeRoeckVerreet}.

Let us illustrate how this procedure works within our time-independent approach.
Starting from the Floquet 
operator $U=K_\epsilon V_{\mathbf{J}}$, we combine the two exponentials into a Floquet Hamiltonian $H^F_p$ by generalizing the replica calculation of Ref.~\cite{ProsenPolkovnikovPRL18_ReplicaBCHResummation} as detailed in Appendix~\ref{app_replica}. Hence,
 we  seek a modified Schrieffer-Wolff unitary transformation $e^{i \epsilon^m \mathcal{S}_m}$,
iteratively for $m=1,\dots,p$,
which eliminates from $H^F_p$ all terms of order $\epsilon^m$ that violate the conservation of  
the any of the operators
\be
D_r= \sum_j \frac{1- Z_j Z_{j+r}}{2},
\ee
$r=1,\dots,R$, respectively coupled to $J_r$ in the unperturbed Floquet Hamiltonian.
 These operators are thus approximate conservation laws, meaning that their eigenvalues $\{n_r\}$ are good quantum numbers to label the eigenstates of the resulting truncated Schrieffer-Wolff-transformed Floquet Hamiltonian $ H^{F}_{R,p}$:
 \be
H^F_p = e^{i \mathcal{S}_{\le p}} H^{F}_{R,p} e^{-i \mathcal{S}_{\le p}} + \mathcal{O}(\epsilon^{p+1}),
\ee
\be
[H^{F}_{R,p},D_r]=0 \qquad \text{for all } r=1,\dots,R .
\ee
where $e^{i \mathcal{S}_{\le p}} \equiv  e^{i \epsilon \mathcal{S}_{1}} \cdots e^{i \epsilon^p \mathcal{S}_{ p}}$.
We remark that this scheme should be distinguished from the perturbation theory of Sec.~\ref{subsec_effectiveDWdynamicsLR}, aimed at the conservation of the quantity $D_1$ only for the effective Floquet Hamiltonian $H^{F}_{1,p}$; 
on the contrary, the effective Floquet Hamiltonian $H^{F}_{R,p}$ obtained here conserves all quantities $D_r$, $r=1,\dots,R$.
To emphasize this distinction, here we use a calligraphic notation for the generator $\mathcal{S}$ of the perturbative scheme in powers of the kick $\epsilon$, to avoid confusion with the previous generator $S$ of the  perturbative scheme in powers of $\hat\epsilon$.

As usual in the many-body context (cf. Sec.~\ref{subsec_effectiveDWdynamics}), the proliferation of possible processes as well as the decrease of energy denominators at high perturbative orders is expected to lead to a divergent (asymptotic) character of the series, corresponding to a mutual hybridization of the many-body energy bands arising from the splitting of the highly-degenerate unperturbed levels.
However, the perturbative series  provides valuable information on the slowness of the dynamical delocalization in Hilbert space. 
The timescale $T_{\text{preth}} $ over which the approximate conserved quantities $\{D'_r \equiv e^{-i \mathcal{S}_{\le p^*}} D_r e^{i \mathcal{S}_{\le p^*}}\}_{r=1}^R$ appreciably depart from their initial value can be estimated by finding the optimal truncation order $p^*$ of the series.
\footnote{Optimality is defined by the condition that the remainder is minimized with respect to a suitable operator norm.}
Within the time-dependent construction of Ref.~\cite{DeRoeckVerreet},  $p^*$ is found to depend on the magnitude of the perturbation as  $C \epsilon^{-1/ (2R+1+\delta)}$ where $\delta>0$ is arbitrary and $C=C(\delta)$ is a constant. This result leads to a stretched exponential lower bound on the prethermal timescale, which generalizes Eq.~\eqref{eq_timescale}:
\be
\label{eq_timescaleLRR}
T_{\text{preth}} \ge \frac 1 \epsilon \exp\bigg( \frac{c} {\epsilon^{1/(2R+1+\delta)}} \bigg) .
\ee
Within the long Floquet-prethermal window $0\le t \le T_{\text{preth}}$, the dynamics is guaranteed to take place within the Hilbert space sectors defined by the set of eigenvalues $n_r$ of the dressed operators $D'_r$, i.e., the (dressed) total numbers of frustrated bonds at distance $r$. 
Since the time-independent approach followed here (replica resummation $+$ standard static Schrieffer-Wolff) basically produces the same set of conserved quantities as the time-dependent scheme of Ref.~\cite{DeRoeckVerreet},
it is  natural to assume that this alternative scheme yields similar nonperturbatively long heating timescales (see also the relative discussion in Refs.~\cite{DeRoeckVerreet, ProsenPolkovnikovPRL18_ReplicaBCHResummation}).

Since ${H}_{R,p^*}^F$ conserves all operators $\{D_r\}$, an initial configuration can only evolve within the corresponding resonant subspace with fixed number of frustrated bonds $n_r$ at distance $r$.
The effective Hamiltonian ${H}_{R,p^*}^F$ is thus much more constrained than $H_{1,p^*}^F$ in Eq.~\eqref{eq_HFLR}. 
The transformed-picture initial state $e^{-i\mathcal{S}_{\le p^*}} \ket{+}$ is a low-density superposition of isolated quasilocal clusters of flipped spins.   
At $p$th order in $\epsilon$, these clusters may comprise at most $p$ flipped spins.

We are now in a position to clearly formulate our question on the order parameter evolution:
Can an initial cluster of $p<R$ flipped spins evolving via ${H}_{R,p^*}^F$ resonantly excite unbound domain walls?
Under the single assumption that the array $(J_1,\dots,J_R,2\pi)$ is strongly incommensurate
[as specified by Eq.~\eqref{eq_diophantine} in Appendix~\ref{app_replica}], we prove that such a configuration
with $p<R$ is never resonant in energy with order-melting configurations, i.e.,
configurations possessing a contiguous domain of $R$ reversed spins.

\begin{figure}[t!]
  \includegraphics[width=0.4\textwidth]{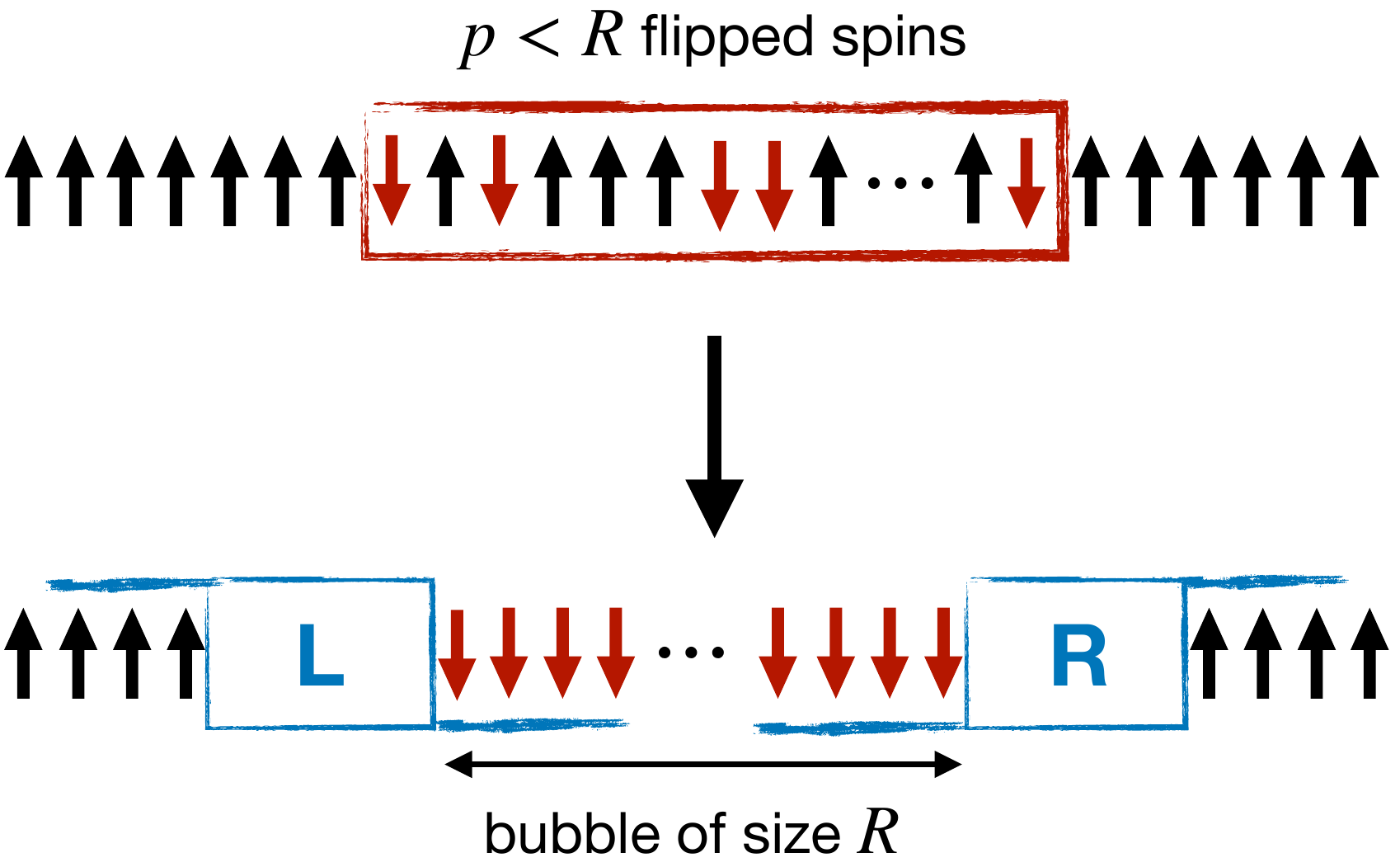}
  \caption{A candidate low-order  process leading to the meltdown of the order parameter.
    In the main text, we prove that energy resonance between the two configurations is not possible
    under a generic assumption of strong incommensurability of the couplings.}
\label{fig:meltingprocess}
\end{figure}

Referring to the illustration in Fig.~\ref{fig:meltingprocess}, the claim is that the top and bottom configurations are necessarily separated by an energy mismatch.
By the incommensurability condition, the existence of this energy mismatch is equivalent to the occurrence that $n_r \neq n^L_r + n^R_r$ for some $1\le r \le R$
(note that the process in Fig.~\ref{fig:meltingprocess1} is a particular case of Fig.~\ref{fig:meltingprocess}).

The proof follows from the
\begin{displayquote}
\textbf{Lemma.} A domain-wall-like configuration has $n_r \ge r$.
\end{displayquote}
To show this, we consider a  domain-wall-like configuration, with all spins pointing up (down) for $j\le j_L$ and down (up) for $j>j_R$ (where $j_R \ge j_L$).
We focus on the $r$ sublattices $\{i+jr | j\in\mathbb{Z}\}$, labelled by $i=1,\dots,R$.
Each sublattice exhibits a domain-wall-like configuration, and hence, it has at least one frustrated nearest-neighbor bond.
Each such bond maps to a bond at distance $r$ in the original lattice.
Thus, the original configuration has at least $r$ frustrated bonds at distance $r$, QED.

Using this lemma, it is easy to prove the main claim: considering again Fig.~\ref{fig:meltingprocess}, the bottom configuration is energetically equivalent to two isolated domain-wall-like configurations, defined by the content of the regions denoted L and R, as the reversed bubble in between can be made infinitely large without frustrating any further bonds.
By the lemma, $n^L_r \ge r$ and $n^R_r\ge r$.
In particular, $n^L_R + n^R_R \ge 2R$.
On the other hand, the top configuration in Fig.~\ref{fig:meltingprocess} has $p$ flipped spins, each of which can frustrate at most two bonds at distance $r$, for all $r$. In particular, $n_R \le 2p$.
Hence, clearly, the resonance condition $n_R = n^L_R + n^R_R$ is impossible to satisfy when $p<R$, which concludes the proof, QED.

The result proved here implies that processes involving {\it at least} $R$ spin flips are responsible for the order parameter decay.
In particular, we already identified above the fastest such process, arising from terms
\be
 \propto \epsilon^R \sum_j P^\uparrow_{-R+1} \cdots P^\uparrow_{j} \, S^-_{j+1} \cdots S^-_{j+R} \, P^\uparrow_{j+R+1} \cdots P^\uparrow_{j+2R} 
\ee
in $\mathcal{S}_{\le p^*}$ [recall $S_j^\pm\equiv \frac 1 2( X_j\pm i Y_j)$]. These terms give rise to isolated domains of $R$ contiguous reversed spins, with density $\propto \epsilon^{2R}$, appearing in the transformed-picture initial state $\ket{+'}=e^{i\mathcal{S}_{\le p^*}}\ket{+}$ at order $R$. The two domain walls separated by $R$ sites are free to hop away from each other with amplitude $\epsilon$, thus spreading the reversed domain and melting the system magnetization at a rate $\gamma \sim \epsilon^{2R+1}$, cf. Eq.~\eqref{eq_decayLR}.
Figure~\ref{fig_timeline}(a) reports a sketch of the timeline derived here; we reiterate that the heating timescale is only a lower bound, and might be non-tight in general.

We finally observe that tuning the longitudinal kick component to a dynamical freezing point as outlined in Sec.~\ref{subsec_freezing}, the decay rate of the order parameter may be further strongly suppressed, as the magnitude $\epsilon$ of the perturbation gets effectively replaced by $\epsilon^2$. Our theory shows that this results in a decay rate $\gamma\sim\epsilon^{4R+2}$. 
As interactions beyond the nearest neighbors are quite generic in experiments, our results indicate that, upon tuning the driving parameters to a dynamical freezing point, DTC response might appear perpetually stabilized for all practical purposes.

\subsection{Numerical simulations for $R>1$}
\label{subsec_numerics}

\begin{figure}[t!]
  \includegraphics[width=0.5\textwidth]{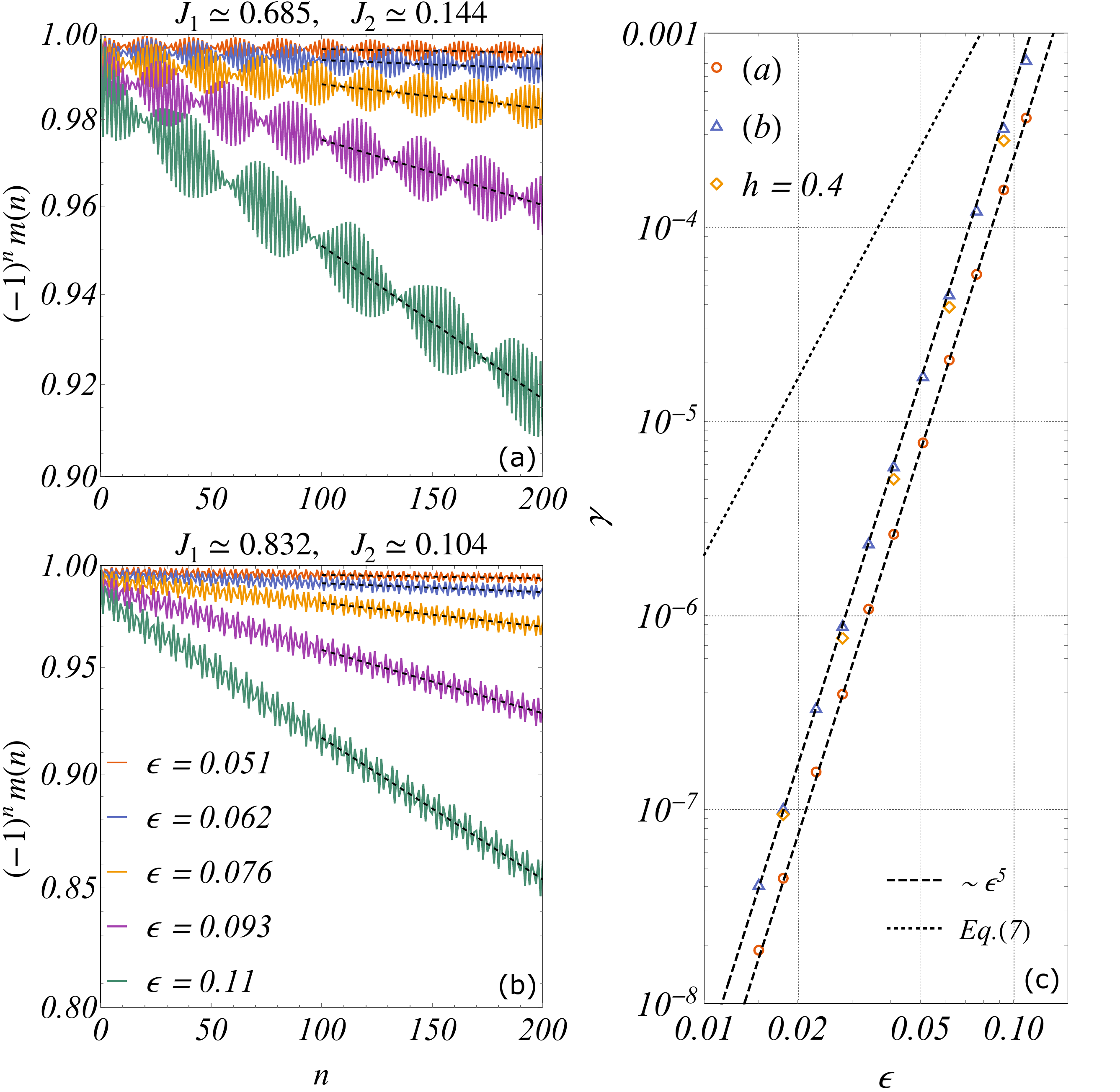}
  \caption{{\it (a-b)}  Log-linear plot of the order-parameter time evolution under the Floquet dynamics
    in the quantum Ising chain with next-to-nearest-neighbor interactions, obtained by means of iTEBD simulations.
    We set $J_{1} = 1 /\zeta(\alpha)$ and $J_{2} = (1/2)^{\alpha} / \zeta(\alpha)$, with $\alpha = 2.25$ {\it (a)} and $3$ {\it (b)}. 
    Dashed black lines are exponential fits. 
    {\it (c)} Scaling of the decay rate $\gamma$ as function of the kick strength $\epsilon$,
    where symbols correspond to the data extracted from panels {\it (a)} and {\it (b)}.
    The agreement with the predicted $\epsilon^5$ law is perfect.
    The third dataset shows the result for the same parameters as {\it (b)}, with the addition of a longitudinal
    component $h=0.4$ in the kick. As it is evident, the decay rate is essentially unaltered.}
\label{fig:J1_J2_gamma}
\end{figure}

\begin{figure}[t!]
  \includegraphics[width=0.5\textwidth]{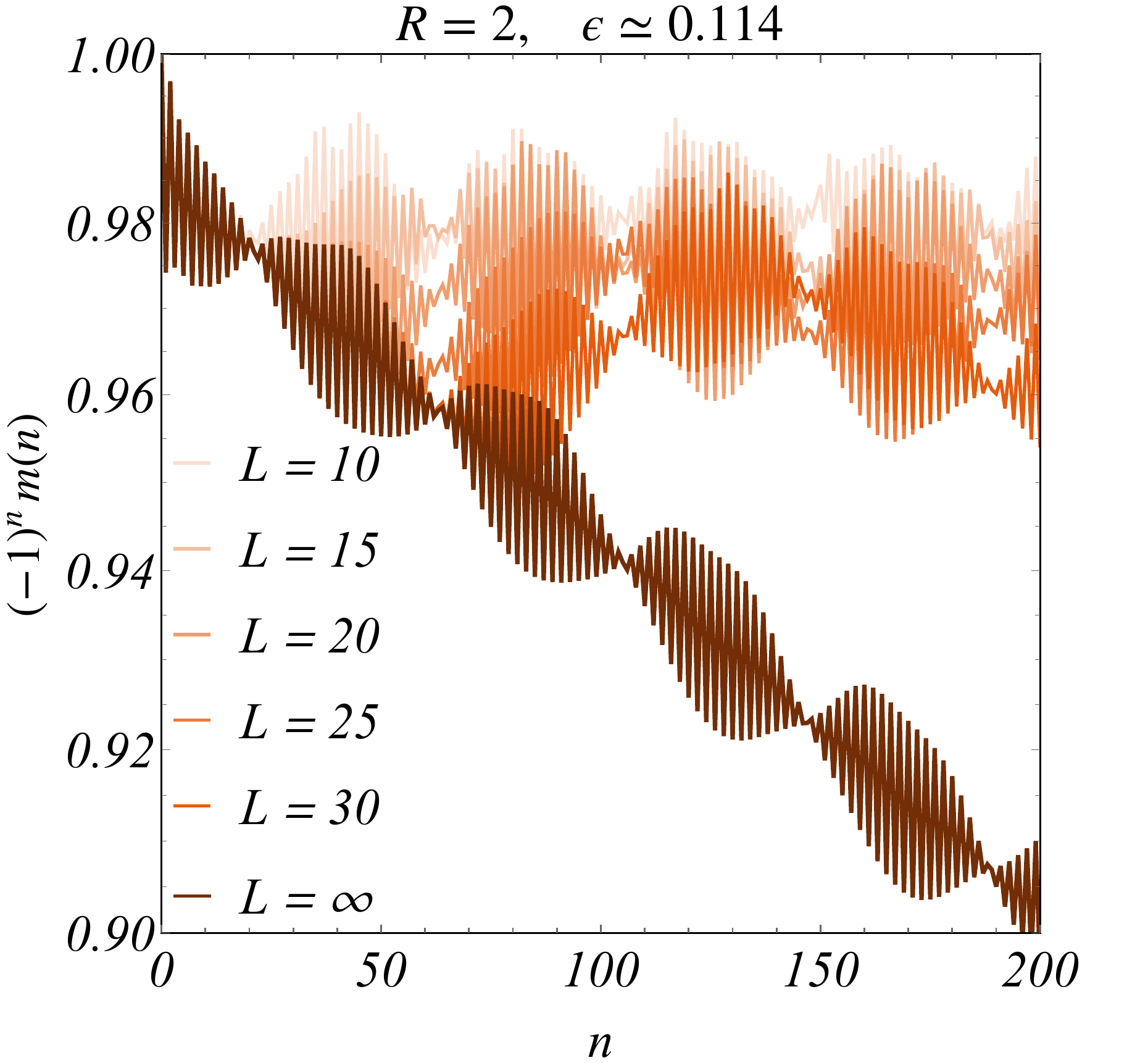}
  \caption{Same as in Fig.~\ref{fig:J1_J2_gamma}{\it (a)}, for $\epsilon = 0.11390625$. 
    The iTEBD data (thick dark line) are compared with ED for finite system with different sizes $L = 10,15,20,25,30$
    (shaded red lines, from lighter to darker).}
\label{fig:J1_J2_finitesize}
\end{figure}

The above theoretical analysis is supported by extensive numerical simulations that we performed for $R=2$ (namely, for the Ising chain with next-to-nearest-neighbor interactions). In our numerics, the couplings have been fixed as $J_1 = 1/\zeta(\alpha)$ and
$J_2 = (1/2)^{\alpha}/\zeta(\alpha)$, 
where $\zeta(x)=\sum_{r=1}^\infty 1/r^x$ denotes the Riemann zeta function.
We fix either $\alpha =2.25$ or $3$ (thus larger than $2$), to be consistent with the next section concerning long-range systems.
This choice is largely arbitrary at this level; we anticipate that it allows a direct comparison with the data presented in the next section.
{Note that the incommensurability of the couplings $\{J_r\}$ would be guaranteed for irrational $\alpha$'s, but we take rational values to further test of the robustness of our analytical predictions.}

We found that the Floquet evolution of the absolute value of the order parameter is always compatible with the theoretically expected exponential decay for all kick imperfections $\epsilon \neq 0$. 
To validate our theory prediction on the suppression of the decay rate $\gamma$  compared to the deconfined case of nearest-neighbor interactions, we performed thermodynamic-limit iTEBD simulations for a sequence of small values of $\epsilon$.
Results are reported in panels {\it (a)} and {\it (b)} of Fig.~\ref{fig:J1_J2_gamma}. 
Note that the curvature of the exponential decay is hardly visible on the accessible timescale for such small values of $\epsilon$; however, $\gamma$ can be accurately extracted.
The resulting scaling of $\gamma$ vs $\epsilon$ is reported in panel {\it (c)}, which clearly shows how the numerical data points follow the theory prediction (dashed black lines). 
The nearest-neighbor analytical result, given by Eq.~\eqref{eq_tauexact} and valid for  $R=1$, is also reported for comparison (dotted black line).
We finally report in panel {\it (c)} additional results in the presence of a sizeable longitudinal kick component $h=0.4$. As expected from our analysis of Sec.~\ref{sec_mixedfield}, the decay rate is essentially unaltered.

In Fig.~\ref{fig:J1_J2_finitesize} we compare the thermodynamic data with the finite-size ED results. 
(Note that for small $\epsilon$ the entanglement entropy growth is slow enough to push iTEBD simulations to unusually large numbers of driving periods~\footnote{For the iTEBD numerical simulations we let the auxiliary dimension $\chi$ growing up to $1024$; notwithstanding, only few demanding simulations saturated this maximum bond, with an accumulated truncation error which remains always under control.}.)
Also in this case, the absolute value of the order parameter gets stuck to a nonzero value for $L < \infty$. However, both iTEBD data and finite-size results show the same frequency of oscillations.

\begin{figure}[t!]
\includegraphics[width=0.5\textwidth]{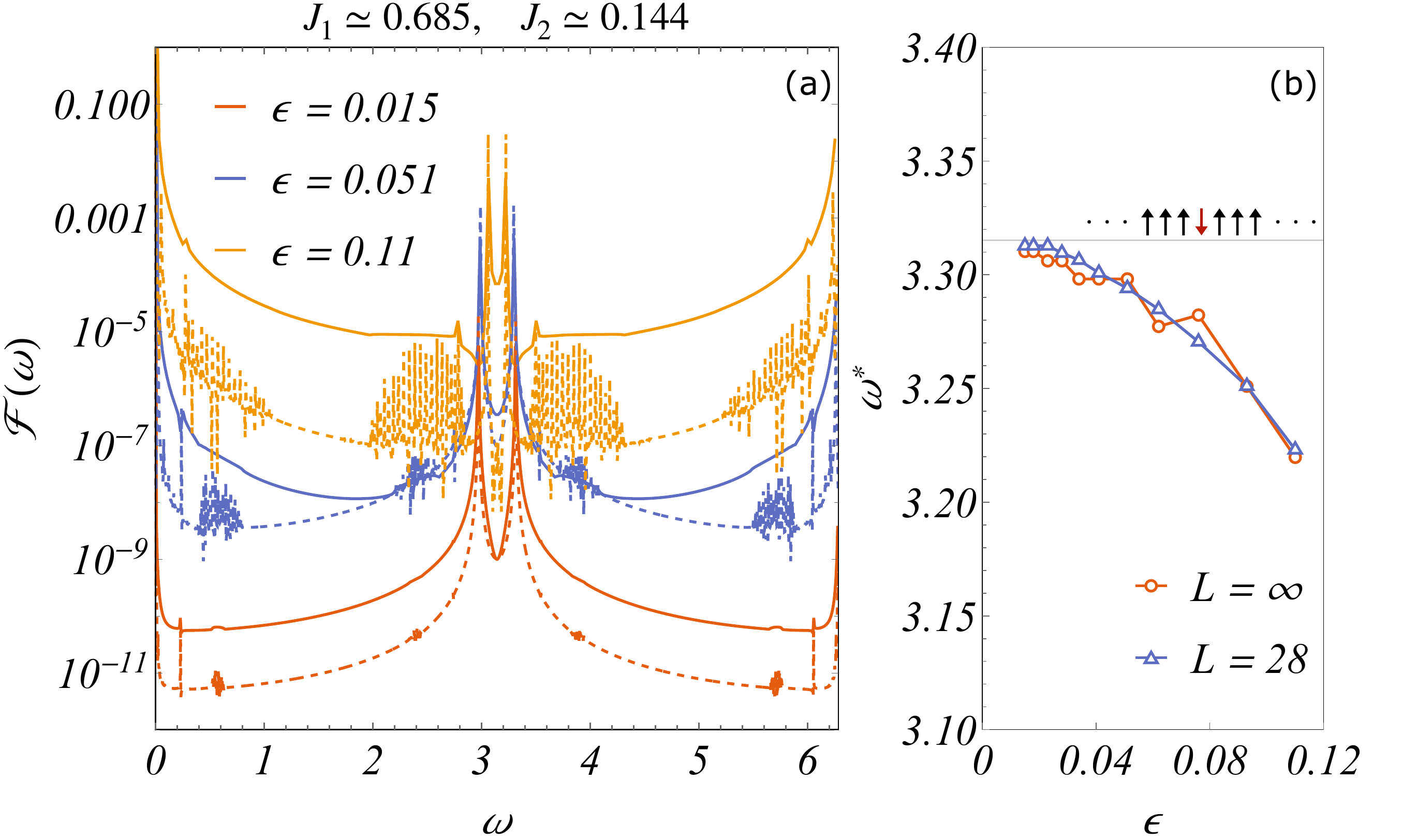}
\caption{{\it (a)} Discrete Fourier transform of the time series (up to 800 kicks) in Fig.~\ref{fig:J1_J2_gamma} for representative values of the kick strength $\epsilon$ (full lines). Dashed lines are the analogous data obtained via ED for $L=28$ and longer time series (up to 2000 kicks).
{\it (b)} Position of the main peak (at $\omega >\pi$) as a function of $\epsilon$,
which is expected to match the ``classical" single spin-flip excitation value for $\epsilon\to 0$ (see main text for details).}
\label{fig:J1_J2_Fourier}
\end{figure}

The frequencies of such oscillations can be extracted from the Fourier power spectrum of the time series for $|m(n)|$ (Fig.~\ref{fig:J1_J2_Fourier}). We observe that data in the thermodynamic limit and at finite size are consistent, and manifest the same main peaks symmetrical around $\omega=\pi$ [panel {\it (a)}]. 
 (Note that the finite-size spectrum shows some spurious frequencies due to time-recurrence effects.)
Further confirmation of the validity of our perturbative analysis is provided by the scaling of the position of the main peak, which is approaching the ``classical'' value of the single spin-flip excitation $4J_1 + 4J_2$ for $\epsilon \to 0$ [panel {\it (b)}].

\subsection{Long-range limit}
\label{subs_LRlimit}

The results of the previous section demonstrate that the order parameter decay rate is suppressed as $\gamma \sim a_R \epsilon^{2R+1}$, as $\epsilon\to0$. 
In this last section, we discuss the long-range limit $R\to\infty$.
In particular, we focus on the experimentally relevant case of algebraically decaying interactions $J_r=J/r^\alpha$. 
This model is relevant to the dynamics of effective qubits in  quantum simulators based on trapped ions (tunable $0<\alpha<3$)~\cite{Blatt2012,Richerme2014} and Rydberg atoms ($\alpha=6$)~\cite{browaeys2020review,BernienRydberg}.
In these setups, nontrivial quantum dynamics are generated by additional magnetic fields acting on the spins, whose spatiotemporal variations can be efficiently controlled in the experiment.
We thus hereafter interpret $\mathbf{J}=\{J_1,\dots,J_R\}$ as a finite-range truncation of the parent sequence of couplings $\{J_r=J/r^\alpha\}_
{r=1}^\infty$.
[Note that for a generic (irrational) value of $\alpha$, these truncated sequences are expected to satisfy the strong incommensurability condition in Eq.~\eqref{eq_diophantine}.]
Within this perspective, it is interesting to shift our viewpoint to the functional dependency of the rate $\gamma$ on $R$.

Extracting the scaling of the prefactor $a_R$ would involve keeping track of the magnitude of the subset of processes of order $\epsilon^R$ which trigger the order parameter decay in our combined replica $+$ Schrieffer-Wolff transformations. 
This is in principle straightforward but practically unfeasible, due to the rapid growth of the complexity of high-order perturbative computations. 
However, as a crude conservative estimate, we can bound $a_R$ from above by the total magnitude of all terms of order $\epsilon^R$.
This type of bounds are worked out in the related analysis of Ref.~\cite{DeRoeckVerreet}, as well as in many previous works on rigorous prethermalization theory~\cite{MoriPRL16,AbaninPRB17_EffectiveHamiltonians,AbaninRigorousPrethermalization,ElsePRX17_PrethermalDTC,LinMotrunichExplicitQuasiconserved}, to obtain estimates of the thermalization timescales, like Eq.~\eqref{eq_timescaleLRR}. 
The ubiquitous scenario resulting from these works is that the total magnitude (measured by a relevant operator norm) of all terms perturbatively generated at order $p$ first decreases exponentially with $p$, before plateauing at $p=p^*$ and finally diverging rapidly. 
In the case of interest here, Ref.~\cite{DeRoeckVerreet} finds $p^* \sim \epsilon^{-1/(2R+1+\delta)}$.
Since any finite range $R$ is largely superseded by $p^*$ for small enough perturbation $\epsilon$, 
the exponential suppression $\epsilon^{2R+1}$ dominates over the prefactor $a_R$, and the decrease of $\gamma$ upon increasing $R$ is effectively exponential.

However, taking the limit $R\to\infty$ is subtle, as it does not commute with the asymptotic perturbative limit $\epsilon\to0$:
Setting heuristically $R=\infty$, the heating bound in Eq.~\eqref{eq_timescaleLRR} trivializes.
Taken literally, this occurrence suggests that a fast violation of the effective conservation laws of $\{D'_r\}$
has to be expected for the long-range interacting system, which could in principle lead to a fast order parameter meltdown.
In particular, this would preclude any meaningful extrapolation of the results of the previous section to long-range interactions.

\begin{figure*}[t!]
  \includegraphics[width=0.9\textwidth]{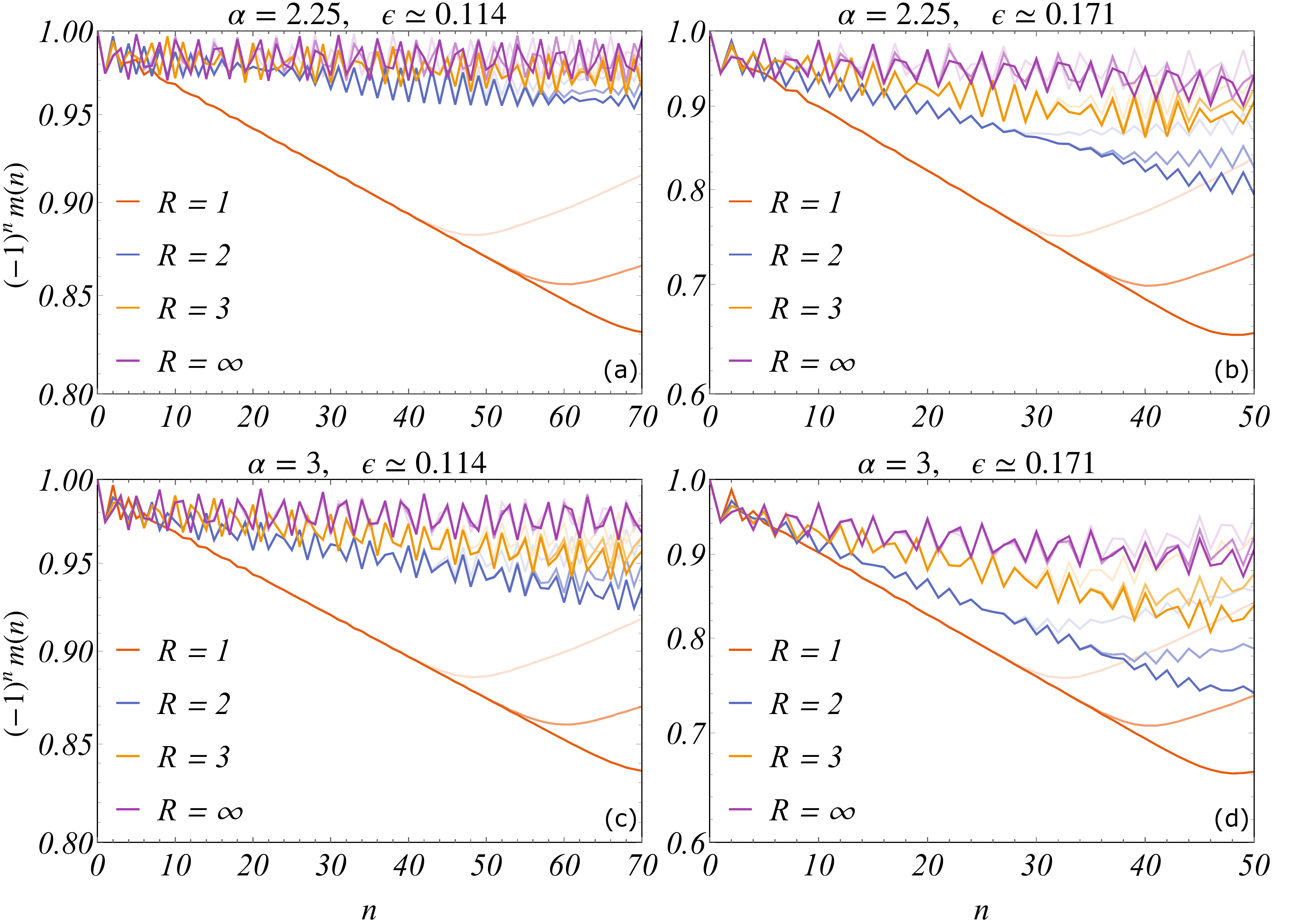}
  \caption{Order parameter decay for increasing interaction range $R$ of the Ising couplings,
    from nearest-neighbor ($R=1$) to long-range ($R=\infty$) interactions, as indicated in the legends.
    Data come from ED simulations for chains of length $L = 20,25,30$ (shaded lines from lighter to darker).
    The various panels represent different values of the decay exponent $\alpha$ and of the kick strength $\epsilon$.
    Note that a strong stabilization of the DTC response occurs even for $J_{r>1} \ll \epsilon$, i.e.,
    well beyond the strict perimeter of our analytical theory.}
  \label{fig:longrange_finitesize}
\end{figure*}

Numerical simulations, however, suggest the opposite behavior: we find that increasing the truncation radius $R$ of long-range interactions $J_r=J/r^\alpha$ to the maximum $R=L/2$, for arbitrary $\alpha$, leads to a dramatic increase of the order parameter lifetime, as clearly visible from the data shown in Fig.~\ref{fig:longrange_finitesize} for $\alpha=2.25$ [panels {\it (a)-(b)}] and $\alpha=3$ [panels {\it (c)-(d)}].
In these simulations, we have chosen  Kac rescaled interactions, i.e., $J=1/\zeta(\alpha)$ (cf. Fig.~\ref{fig:J1_J2_gamma}), so that results for different $\alpha$ can be fairly compared.
The reported data point at a robust stabilization of the DTC signal, well beyond our analytical theory of Sec.~\ref{subsec_expsuppression}: in fact, one can see that the kick strength taken there,
$\epsilon = 0.114$
and $0.171$, correspond to quite big rotations of the spins at each kick, by angles $\approx 13^\circ$ and $\approx 20^\circ$, respectively.
These perturbations are actually much larger than the considered couplings beyond nearest neighbors. 
In spite of this, the order parameter decay is extremely suppressed upon increasing $R$, and hardly visible in the long-range limit; moreover, this occurrence is not a finite-size effect.

To resolve this apparent contradiction, we observe that the bound~\eqref{eq_timescaleLRR} on prethermalization is unnecessarily pretentious for our purposes: it expresses the expected timescale of quasiconservation of a large number of operators $\{D'_r\}_{r=1}^R$, \emph{uniformly} in the many-body spectrum. 
As we take $R=L/2$, energy levels become infinitely dense in the thermodynamic limit away from the band edges, thanks to the strong incommensurability condition which prevents them from being degenerate. 
In fact, the preservation of an extensive number of commuting operators $\{D'_r\}_{r=1}^{L/2}$ would make the system effectively many-body localized, contrary to conventional delocalization scenarios for translationally invariant models~\cite{DeRoeckHuveneersScenario,Schiulaz1}.
The slow dynamics of highly excited states resulting from this long-range limit is thus nonstandard, and the actual heating  timescales (or thermalization timescales for time-independent systems) are presently unclear, even for static (undriven) systems~\cite{ZunkovicPRL18_Merging,HalimehPRB17_PersistentOrder,GorshkovConfinement,LeroseDWLR}.

On the other hand, the evolution of the order parameter relevant to this work takes place in a particular corner of the many-body Hilbert space, corresponding to the low-energy sector of an approximate Floquet Hamiltonian.
While the perturbative series might be severely divergent at low orders in the long-range limit when measured by uniform operator norms, the same bounds are far too loose when the construction is restricted to a low-energy sector with dilute excitations, relevant for the purpose of this work.
\footnote{To draw a suggestive analogy, the standard Schrieffer-Wolff transformation is believed to be convergent in the low-energy sector, and divergent (asymptotic) when considered uniformly in the many-body Hilbert space. This is related, on the one hand, to the effectiveness of adiabatic theorems for gapped ground states and, on the other hand, to the belief in eigenstate thermalization for arbitrarily weak generic perturbations of integrable systems.
}
In other words, while low-order perturbative transitions from highly excited configurations with extensively many frustrated bonds are very likely to hit resonances, this becomes extremely unlikely when the initial state is the polarized state $\ket{+}$ considered in this work, since quasilocal clusters of flipped spins generated by the weak kicks are far away from each other and hence can hardly cooperate to produce resonant transitions.
This argument suggests that the lower bound on the timescale $T_{\text{preth}}$ in Eq.~\eqref{eq_timescaleLRR} is far too conservative for the dynamics originating from the state $\ket{+}$, for arbitrary $R$. 
A tighter bound would be needed to correctly account for the decrease of the density of states at low energy.
In the same regime, a long-lasting suppression of heating has to be expected in the limit of long-range interactions $R\to\infty$, as well.
We reiterate that, even for finite-range interactions, our numerical simulations indicate a stability well beyond the analytical theory presented above -- cf. Figs.~\ref{fig:intro},~\ref{fig:longrange_finitesize}.
Along the lines of the discussion above, our intuition is that while plenty of resonances are bound to occur when the range is long, in the dilute quench dynamical setup of our work many of these resonances are inconsequential, as the only ``dangerous'' processes for the order parameter meltdown are those that cause an inner rearrangement of a cluster of domain walls involving the relocation of the leftmost (or rightmost) of them to distance $R$ from the next. It is likely that such specific resonances form a much smaller set that is further dynamically obstructed compared to the naive expectation from brute-force norm bounds.

Setting up a generalized Schrieffer-Wolff perturbative scheme aimed at estimating the timescales involved in the intricate slow dynamics of the long-range interacting chain appears as a formidable problem, which we leave to future investigations.
Here, building on the insight of Sec.~\ref{subsec_expsuppression} and above, we formulate the conjecture that the timescale of the fastest process leading to the order parameter decay can be estimated in terms of the initial density of bubbles of reversed spins, whose walls are free to spread away from each other.
Remarkably, such plausible scenario leads to a decay rate $\gamma$ \emph{beyond perturbation theory}, i.e., smaller than any power of $\epsilon$.

As suggested in the discussion above, even though the elimination of domain-wall nonconserving processes is formally valid throughout the entire many-body spectrum only for $\alpha \gg 1$, in the low-energy sector one can naively perform several perturbative steps to preserve $D_1$ (and a few other operators $D_r$, as well) for arbitrary $\alpha$.
Thus, we reconsider the two-body problem of Eq.~\eqref{eq_H2bodyLR}, and set $J_r=J/r^\alpha$ and $R=\infty$.
We obtain
\begin{multline}
\label{eq_H2bodyLRinfty}
H_{\text{2-body}} = \sum_{j_1<j_2} v_{\alpha}(j_2-j_1) \ket{j_1,j_2}  \bra{j_1,j_2}
\\ + {\epsilon} \sum_{j_1<j_2} \! \big(\ket{j_1+1,j_2} + \ket{j_1,j_2+1}\big) \bra{j_1,j_2} + \text{H.c.},
\end{multline}
where 
\be
v_{\alpha}(r) = 
4 \sum_{d=1}^r d J/d^\alpha + 4r \sum_{d = r+1}^\infty J/d^\alpha \, .
\ee
The potential grows from $v_\alpha(r=1)=4J \zeta(\alpha)$ to $v_\alpha(r=\infty)=4J \zeta(\alpha-1)$ (for $\alpha>2$) or $\infty$ (for $\alpha\le 2$). The asymptotic behavior at large $r$ is 
\be
\label{eq_potentialLR}
v_{\alpha}(r) \underset{r\to\infty}{\thicksim}  
4J \times 
\left\{
\begin{array}{ll}
\displaystyle \frac{r^{2-\alpha}}{(2-\alpha)(\alpha-1)} & \text{for } \alpha <2, \vspace*{1mm} \\
\log r & \text{for } \alpha =2, \\
\displaystyle \zeta(\alpha-1)-\frac{r^{-(\alpha-2)}}{(\alpha-2)(\alpha-1)} & \text{for } \alpha > 2. 
\end{array}
\right.
\ee

Let us now discuss the result in Eq.~\eqref{eq_potentialLR}.
For $\alpha\le2$, the binding potential is confining at large distances. Hence, the hierarchy of domain-wall bound states exhausts the excitation spectrum~\cite{GorshkovConfinement}. 
The infinite energetic cost of isolated domain walls for $\alpha\le2$ underlies the persistence of long-range order at finite temperature~\cite{Dyson1969,thouless1969long}, circumventing the standard Landau argument against the existence of thermal phase transitions in one dimension.
Indeed, at a prethermal level, the effective Floquet Hamiltonian resulting from the resummation of the BCH expansion supports long-range order at finite temperature for small but finite $\epsilon$.
%
\begin{figure}[t!]
\includegraphics[width=0.48\textwidth]{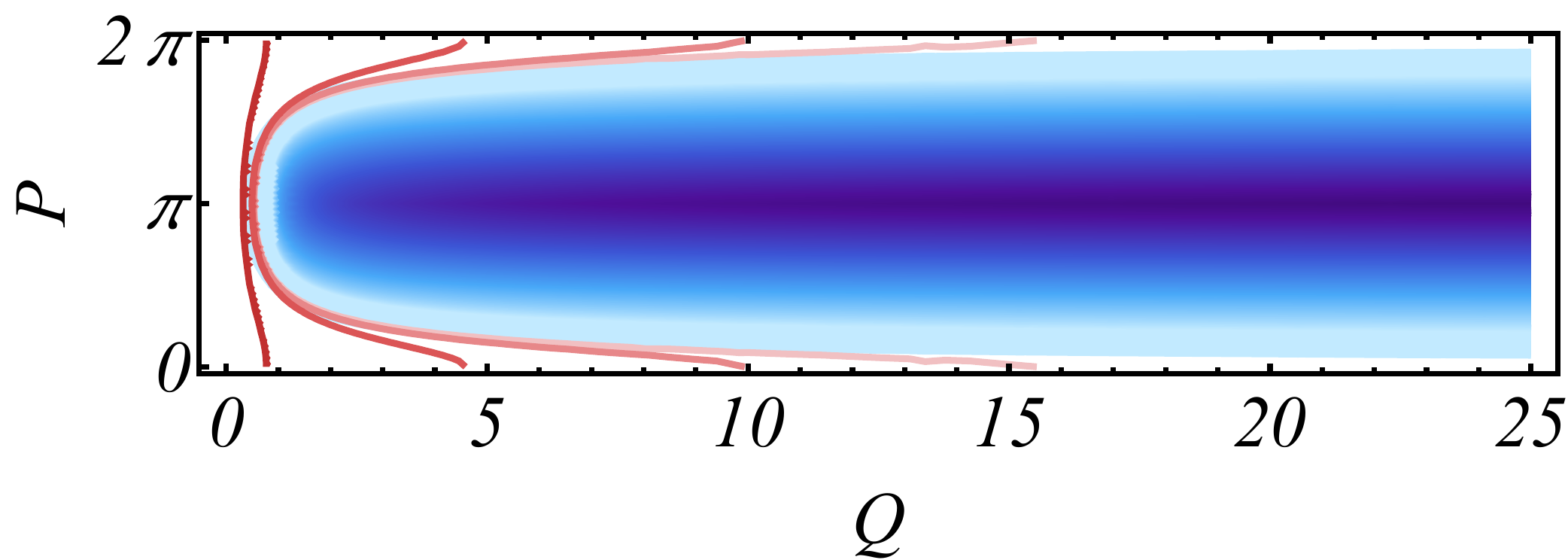}
\caption{Semiclassical energy eigenstates of the two-body problem in Eq.~\eqref{eq_H2bodyLRinfty} with $\alpha=3$ and $\epsilon=0.1139$, represented by classical trajectories in phase space $(Q,P)\in[0,\infty)\times [0,2\pi)$ governed by the Hamiltonian $H_{\text{2-body}}(Q,P)= v_\alpha(Q)-4\epsilon \cos P$, encircling an area equal to a multiple of Planck's constant $h=2\pi \hbar$ (Bohr-Sommerfeld quantization rule). Here we take $\hbar=1$, so that $h \equiv 2\pi$.
The continuous function $v_\alpha(Q)$ has been taken as in Eq.~\eqref{eq_potentialLR}.
For this choice of parameters, the center-of-mass potential well hosts four bound states, marked by red trajectories, bounded in $Q$.
All higher excited eigenstates (blue trajectories) are unbound plane waves, and form a continuum labelled by the asymptotic momentum $P$ for $Q\to\infty$.
}
\label{fig:semiclassical}
\end{figure}
%
On the other hand, the two-body potential is bounded at large distances for $\alpha>2$. 
In this case,
freely travelling domain-wall states appear, similarly to finite-range systems.
Due to their nonlocal nature, isolated domain-wall-like excitations cannot be locally created or destroyed; thus, they give rise to topologically protected quasiparticles, which form a continuum in the excitation spectrum above a  discrete sequence of nontopological bound states.
Unlike the case $R<\infty$, however, the potential only flattens out asymptotically for $r\to\infty$.
Consequently,
the number $\mathcal{N}_{\alpha}$ of such bound states critically depends  on the hopping amplitude $\epsilon$. 
The finite statistical weight of domain-wall-like quasiparticles in thermal equilibrium is what prevents long-range order at finite temperature for $\alpha>2$.

\begin{figure}[t!]
  \includegraphics[width=0.5\textwidth]{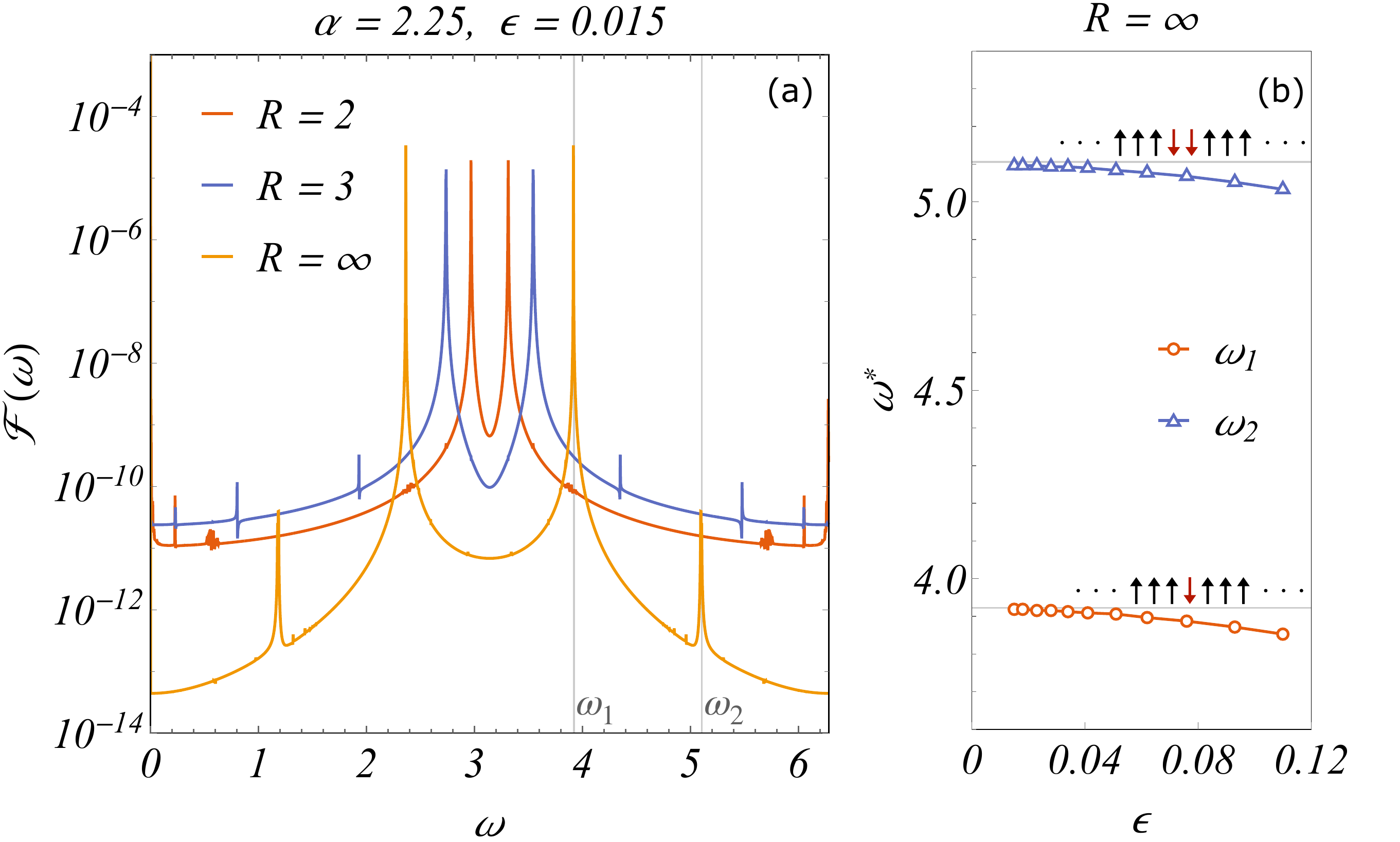}
  \caption{{\it (a)} Discrete Fourier transform of the order parameter time series for $\epsilon = 0.015$, $\alpha =2.25$
    and different interaction ranges $R$. Data have been obtained via ED with $L=28$ and time series up to 2000 kicks.
    {\it (b)} The position of the first two peaks (at $\omega >\pi$) as a function of $\epsilon$ in the long-range case $R=\infty$,
    which are expected to match the “classical” single or double spin-flip excitation value, for $\epsilon=0$
    (see main text for details).}
  \label{fig:longrange_Fourier}
\end{figure}

Figure~\ref{fig:semiclassical} reports an illustration of the two-body spectrum, obtained within a semiclassical approximation (which becomes quantitatively accurate in the continuum limit, i.e., for highly excited states).
Here we set $\alpha=3$ and $\epsilon=0.1139$.
Quantized trajectories undergo a transition between spatially localized (red) and delocalized (blue), representing a discrete sequence of nontopological confined bound states below a continuum of topological unbound domain walls.
For this choice of parameters, $\mathcal{N}_\alpha(\epsilon)=4$; however, this number grows as $\epsilon\to0$.
{A clear signature of these bound states is further given by the presence of pronounced peaks in the power spectrum of the absolute value of the order parameter time series. In Figure~\ref{fig:longrange_Fourier} we show the Fourier transform of finite-size data for $\alpha=2.25$. In the long-range limit ($R=\infty$)
the two main peaks correspond to the lower-energy bound states, namely the single and the double spin-flip excitation. In the classical limit ($\epsilon\to 0$) their energies are given respectively by
$\omega_{1} = 4\sum_{r=1}^{L/2} r^{-\alpha}/\zeta(\alpha)$ and
$\omega_{2} = \omega_{1} + 4\sum_{r=2}^{L/2} r^{-\alpha}/\zeta(\alpha)$.
}
On the contrary, the main signatures in the order parameter evolution of the presence of unbound domain walls in the spectrum is the (very slow) overall decay of the signal in Fig.~\ref{fig:longrange_finitesize}.

Generalizing our argument of Sec.~\ref{subsec_DWbinding}, the number of bound states $\mathcal{N}_{\alpha}$ can be estimated starting from the observation that isolated domain walls can freely hop to neighboring sites with amplitude $\epsilon$, which gives the dispersion law $E_k = \frac 1 2 v_{\alpha}(\infty) - 2 \epsilon \cos k$. The unperturbed bound-state wavefunctions for $\epsilon\to0$, $\psi_\ell(r)=\delta_{\ell,r}$, $\ell\in\mathbb{N}$, are precluded from hybridizing with the domain-wall continuum when their unperturbed energy $E_\ell=v_\alpha(\ell)$ is below the ``ionization threshold'' $v_\alpha(\infty)-4\epsilon$. The equation
\be
v_\alpha(\ell) = v_\alpha(\infty)-4\epsilon
\ee
thus identifies the highest stable bound state $\ell\equiv\mathcal{N}_{\alpha}$. Using the asymptotic expansion in Eq.~\eqref{eq_potentialLR}, we obtain
\be
\mathcal{N}_{\alpha} \sim (c_\alpha \epsilon/J)^{-1/(\alpha-2)}
\ee
where $c_\alpha=4(\alpha-2)(\alpha-1)$.

This result expresses how the number of bound states diverges as $\epsilon\to0$ for all $\alpha>2$.
Accordingly, the physical size of a critical reversed bubble triggering the order parameter meltdown grows unbounded in the asymptotic regime of weak perturbation.
In other words, our conjecture that the formation of a critical bubble is the fastest process leading to the decay of the order parameter, suggests that this decay has a {\it nonperturbative} origin in the long-range interacting spin chain,
\be
\label{eq_gammaLRinfty}
\gamma \sim \epsilon^{2 \mathcal{N}_\alpha+1} \sim \epsilon^{ A \epsilon^{-1/(\alpha-2)}
} \, ,
\ee
where we have defined $ A \equiv 2(c_\alpha/J)^{-1/(\alpha-2)}$.

We note that as $\alpha$ approaches $2$ from above, the lifetime $\gamma^{-1}$ from Eq.~\eqref{eq_gammaLRinfty} diverges. 
The lack of an appropriate heating bound in this regime (as discussed above) prevents us from estimating the location of a presumable crossover region $\alpha\approx \alpha^*\ge2$ between a vacuum-decay driven ($\alpha\gtrsim\alpha^*$) and a heating-driven ($\alpha\lesssim\alpha^*$) order-parameter decay.
In any case, we reiterate that the order-parameter lifetime is expected to be nonperturbatively long for all $\alpha$'s.

An explicit comparison between the various timescales can be drawn in the high-frequency driving limit of our Floquet model, i.e., taking 
\begin{equation}
    \label{eq_FloquetLRHF}
U 
\; = \; 
 i^L P \;
e^{i\tau \sum_j   \epsilon X_j }
\; e^{i \tau \sum_{j} \sum_{r=1}^\infty 
(J/r^\alpha)
Z_j Z_{j+r}  }
\end{equation}
with $\tau$ small.
In this case, rigorous bounds on the heating timescale  are available for all values of $\alpha$.
We refer to Fig.~\ref{fig_timeline}{\it (b)} for a comprehensive illustration.
For $1<\alpha\le\infty$, an exponential lower bound $T_{\mathrm{preth}} \ge \exp(C/\tau)$ applies uniformly
in $\alpha$~\cite{KUWAHARA201696} (assuming Kac rescaling of $J$ as above).
Within this long time window, heating is prevented by the quasiconservation of an effective Hamiltonian emerging in the toggling frame, of the form 
\begin{equation}
H^{{F}}= -\sum_j \bigg(\sum_{r=1}^\infty  J  \frac {Z_j Z_{j+r}} {r^\alpha} + \epsilon X_j \bigg) \; + \tau (\dots) + \tau^2 (\dots) + \dots
\end{equation}
For $\alpha>2$, this Floquet Hamiltonian governs the evolution of local observables~\cite{MachadoPRX20}. 
The arguments of this Section apply to the dynamics generated by $H^F$, leading to the long lifetime $1/\gamma$ of the confinement-stabilized DTC order parameter in Eq.~\eqref{eq_gammaLRinfty}.  Knowledge of the heating timescale allows us to keep track of a presumable crossover value $\alpha^*=\alpha^*(\epsilon,\tau)>2$ between this and the vacuum-decay timescale.
Upon decreasing $\epsilon\to0$, the value of  $\alpha^*$ is pushed to $\infty$; conversely, as $\tau\to0$, it is pushed towards $2$.
(We reiterate that the heating timescale is only a lower bound, and might be non-tight in general.)
For $1<\alpha\le 2$, the Floquet Hamiltonian supports long-range order at finite temperature. Thus, the mechanism of Floquet prethermalization suffices to stabilize DTC order. Its lifetime is only limited by heating processes~\cite{ElsePRX17_PrethermalDTC,MachadoPRX20}. 
Finally, for $0\le \alpha\le 1$, dynamics starting from a fully polarized initial state (such as our $\ket{+}$) is governed by an emergent semiclassical description over a timescale that diverges with system size~\cite{mori2018prethermalization,LerosePappalardiPRR20}, and the underlying asymptotic ``decoupling'' between the few collective degrees of freedom coupled to the drive and the extensive set of microscopic degrees of freedom, prevents the system from absorbing energy and heating up~\cite{LeroseKapitza,Bhakuni21Suppression}. Because of this nontrivial mechanism, the emergent DTC behavior has mean-field character in this regime~\cite{RussomannoPRB17_LMGDTC,pizzi2021higher}.

To summarize, while the Floquet-prethermalized state features a nonvanishing density of travelling domain walls precluding
large-scale spatiotemporal order, the dynamical production of these excitations by kick imperfections
goes through extremely slow vacuum-decay processes, delaying the Floquet prethermalization itself, ultimately paving
the way for a long window of genuinely nonequilibrium time-crystalline behavior compatible with numerical simulations.
Our argument that the fastest order-melting process is the fractionalization of a critical-size reversed bubble into a deconfined kink-antikink pair leads to the nonperturbatively long lifetime in Eq.~\eqref{eq_gammaLRinfty}.

\section{Conclusions}
\label{sec_conclusions}

In this paper, we have established a framework to understand and compute the order-parameter evolution in periodically driven quantum spin chains, hinging upon the effective dynamics of emergent domain-wall excitations. 
Within this framework, we have analyzed the impact of domain-wall confinement on the order-parameter decay, and established that a slight increase of the interaction range can result in a dramatic extension of the lifetime of DTC response, despite the absence of a long-range ordered Floquet-prethermal states.
The results of this paper delimit and characterize the theory of time crystals for disorder-free, finite-range interacting quantum spin chains driven at arbitrary frequencies, extending the current state of the art.

Observing confinement-stabilized DTC response does not require disorder, high-frequency drives, or fat-tailed interactions, which may not be easily accessible in many experimental setups.  
A naturally suited platform for implementation of the physics discussed here is given by Rydberg-dressed spin chains~\cite{RydbergDressed,Zeiher17coherent,RydbergDressed2}.
The tunable-range Ising interactions realized in these systems, described in our notations as
\be
J_r = \frac{J}{1+(r/r_c)^6},
\ee
 have an almost flat core for $r\lesssim r_c$, and cross over to a quick decay in the intermediate range $r\approx r_c$.
The value of the effective range $r_c$ can be efficiently tuned in the experiment, making this setup ideal to observe the confinement-stabilized DTC response identified in this work.
Indeed, necessary ingredients such as preparation of fully polarized states, application of global pulses, long coherence times and monitoring of the collective magnetization, have already been demonstrated in these experiments.

An important remark, however, is that the set of initial states which give rise to such a response is more limited than for the prethermal DTC.
The flexibility in perturbing the initial state by arbitrary local operations is set by the same parameter scale $\epsilon$ that controls the duration of the signal. 
This is, in a sense, reminiscent of DTC behavior associated with quantum many-body scars~\cite{maskara21dtc,bluvstein2021controlling}.
More generally, this work provides further evidence that increasing the range of interactions may generate nonthermal behavior in certain regimes~\cite{KastnerPRL11_Diverging,mori2018prethermalization,neyenhuis2017observation,LeroseLong,DefenuPNAS,LeroseDWLR,GorshkovConfinement,LerosePappalardiPRR20}, and may help realizing  genuinely nonequilibrium phases~\cite{LeroseKapitza,MachadoPRX20}.

 The results reported here clarify some confusion on the role of the system size in clean short-range interacting ``time crystals'', consistent with the numerical analysis of Ref.~\cite{PizziPRB20_CleanDTC}.
Furthermore, a crucial byproduct result of this work is the clarification of the nature of the apparent anomalous  persistence of the order parameter observed in several numerical investigations of long-range quantum Ising chains with $\alpha\gtrsim 2$ after a global quench of the transverse field from a ferromagnetic ground state~\cite{ZunkovicPRL18_Merging,HalimehPRB17_PersistentOrder,GorshkovConfinement}.
Due to pronounced finite-size effects and  severe slowdown of the dynamics in this regime, purely numerical calculations face significant challenges. 
The theory developed in this paper for the more general Floquet setting (Sec.~\ref{sec_LR}) applies equally well to these quench dynamics (of course, incommensurability of the couplings with $2\pi$ need not be assumed in this case).
This work provided solid analytical evidence that a strong enhancement of the order parameter lifetime {is to be generally expected  in the parameter regime $2<\alpha\ll \infty$}, and predicted its functional form as a function of the quench magnitude.  
This has the important consequence, seemingly not recognized before, that a long-lived nonequilibrium order can be sustained by systems that cannot exhibit long-range order in equilibrium, via the suppression of the dynamical creation of deconfined topological excitations (domain walls), ultimately responsible for melting the metastable nonequilibrium order. This occurrence is related to familiar macroscopic tunnelling phenomena in high-energy physics, such as the Schwinger mechanism~\cite{SchwingerMechanism} and Coleman's false vacuum decay~\cite{ColemanFalseVacuum}, where atypical excited states may only decay (and hence thermalize) through slow nonperturbative processes.

On a technical level, a few points remain open and are left to future work, including a
more rigorous derivation of Eq.~\eqref{eq_gammaLRinfty}, and a better understanding of the role of resonances in the unperturbed spectrum.

\begin{acknowledgments}
  We thank D. A. Abanin and W. De Roeck for discussions at the early stages of this project, and M. Heyl, M. Knap, R. Moessner, A. Pizzi, F. Surace for interesting comments.
  We gratefully acknowledge M. Heyl for drawing our attention on the implementation of Rydberg-dressed spin lattices in Ref.~\cite{RydbergDressed}.
  A.L. acknowledges support by the Swiss National Science Foundation.
\end{acknowledgments}

\appendix

\section{Exact kicked dynamics in the integrable Ising chain}
\label{app:exactIsing}
\label{app:diagIsing}
\label{app:orderIsing}

We start by analyzing the high-frequency regime  $J, \epsilon \ll 1$, which allows us to approximate
\be\label{eq:KU_approx}
 K_{\epsilon} V_{J} \simeq  e^{i J \sum_{j} [Z_j Z_{j+1} + (\epsilon/J) X_{j}]} \, ,
\ee  
using the BCH formula truncated to the lowest order.
The resulting stroboscopic dynamics is equivalent to that of a static transverse-field Ising Hamiltonian   
after a quench of the field from $0$ to $\epsilon/J$.
Exact calculations of the large-distance behavior of the two-point function in the thermodynamic limit have shown that the order parameter does relax to zero exponentially at late time (here $|\epsilon/J| <1$)~\cite{Calabrese_2012},
\be
|m(n)|
\simeq \left[\frac{1+\sqrt{1-(\epsilon/J)^2}}{2}\right]^{1/2} e^{-n/\tau_I},
\label{eq:highfreq}
\ee
with
$
\tau_{I}^{-1} =
J \left[ \frac{4}{3\pi} |\epsilon/J|^3 + O(|\epsilon/J|^5) \right], 
$
confirming the fact that the relaxation time can be very long, depending on the specific driving parameters.
Even though the formula~\eqref{eq:highfreq} is not expected to be accurate away from the high-frequency regime [i.e., for arbitrary $J=\mathcal{O}(1)$],
we still do expect the real relaxation time to scale as $\sim |\epsilon|^{-3}$ for $|\epsilon| \ll 1$.

Next, we proceed by analytically solving the dynamics induced by the Floquet operator $K_{\epsilon} V_{J}$.
To this aim, we exploit the fact that both $K_{\epsilon}$ and $V_{J}$
are Gaussian operators in terms of the spinless fermions $c^{(\dagger)}_{j}$,
introduced by the following Jordan-Wigner transformation
\be
Z_{j} = \prod_{i=1}^{j-1}(1-2 c^{\dag}_{i}c_{i})(c_{j}+c^{\dag}_{j}),\quad X_{j} = 1 - 2 c^{\dag}_{j}c_{j},
\ee
where $\{c_{i},c^{\dag}_{j}\} = \delta_{ij}$
and $\{c_{i},c_{j}\} = 0$.
In terms of such fermions, the evolution operators read
\be
K_{\epsilon}  =  e^{ i \epsilon \sum_{j} (c_{j}c^{\dag}_{j} - c^{\dag}_{j}c_{j}) }, \qquad
V_{J}  =  e^{  i J \sum_{j} (c^{\dag}_{j} - c_{j}) (c^{\dag}_{j+1} + c_{j+1}) },
\ee
whose product can be diagonalized by first going to the Fourier space
$
c_{j} = \frac{1}{\sqrt{L}} \sum_{p} e^{-i p j} \eta_{p},
$
where the sum runs over momenta that are quantized depending on the specific sector,
namely $p = 2\pi n /L + (\pi/L)(1+P)/2$ with $n = -L/2,\dots, L/2-1$.  
In terms of positive momenta, both operators can be recast 
in the following forms
\be\label{eq:UK_p}
K_{\epsilon} = \prod_{p>0}e^{\boldsymbol{\eta}^{\dag}_{p}  \mathbb{K}  \boldsymbol{\eta}_{p}},\qquad
V_{J} =  \prod_{p>0}e^{\boldsymbol{\eta}^{\dag}_{p}  \mathbb{V}  \boldsymbol{\eta}_{p}}, 
\ee
where we defined the row vector $\boldsymbol{\eta}^{\dag}_{p} \equiv (\eta^{\dag}_{p}, \,\eta_{-p})$,
and the $2\times 2$ matrices
\be
\mathbb{K} = -2 i \epsilon \, Z, \qquad \mathbb{V} =  2 i J \, [\cos(p) Z - \sin(p) Y].
\label{eq:KVmatr}
\ee
For different momenta, the operators entering  Eq.~\eqref{eq:UK_p}
commute between each other, therefore we only need to combine, for each $p>0$, 
two local Gaussian operators by exploiting the identity~\cite{PhysRevB.87.245107}
\be
e^{\boldsymbol{\eta}^{\dag}_{p}  \mathbb{K}  \boldsymbol{\eta}_{p}} \,
e^{\boldsymbol{\eta}^{\dag}_{p}  \mathbb{V}  \boldsymbol{\eta}_{p}}
=
e^{\boldsymbol{\eta}^{\dag}_{p}  \mathbb{H}  \boldsymbol{\eta}_{p}},
\qquad e^{\mathbb{H}} = e^{\mathbb{K}}e^{\mathbb{V}}.
\ee
The product of the two matrix exponents can be parametrized as an SU(2) rotation,
such that one has
\be
\mathbb{H} = i \phi_{p} \, \hat r_{p} \cdot \boldsymbol{\sigma},
\ee
where $\boldsymbol{\sigma} = \{X,Y,Z\}$ is the vector of the Pauli matrices, and
$\phi_{p}$ and $\hat r_{p}$ are defined below [cf.~Eqs.~\eqref{phip_def} and~\eqref{eq:r}].
The matrix $\mathbb{H}$ is then diagonalized by the following unitary transformation
\be
\mathbb{U}_{p} = e^{-i \epsilon Z} e^{-i \theta_p X /2},
\ee
with $\theta_p$ defined below in Eqs.~\eqref{thetap_def}.
The orthonormal eigenvectors in $\mathbb{U}_p$ 
can be used to construct the following fermionic operators 
\be\label{eq:xi}
{\bm \xi}^{\dag}_p \equiv
(\xi^{\dag}_{p},\,\xi_{-p})
=
{\bm \eta}^{\dag}_p  \mathbb{U}_p,
\ee
in terms of which the Floquet operator is diagonal and reads
\be\label{eq:KU_exact}
K_{\epsilon} V_{J}= 
e^{i\sum_{p>0}\phi_p {\bm \xi}^{\dag}_p Z {\bm \xi}_p} .
\ee

The dynamics induced by the Floquet operator in Eq.~\eqref{eq:KU_exact}
can be understood as a quench originated from the vacuum state in the
$P=+1$ sector. In practice, we may solve the dynamics, 
by connecting the pre- and post-quench diagonal fermions.
The initial diagonal fermions ${\bm \xi}^{0}_{p}$ are those which diagonalize 
the Ising Hamiltonian in Eq.~(\ref{eq:H}), and they are related to the model-independent
momentum-space fermions ${\bm \eta}_{p}$ via the Bogoliubov transformation
${\bm \xi}^{0}_{p} = e^{i p X /2} {\bm \eta}_{p}$.
The pre- and post-quench diagonal fermions are thus related
by the unitary transformation
$
{\bm \xi}_p = e^{i \theta_p X /2} e^{i \epsilon Z} e^{-i p X /2} {\bm \xi}^{0}_p
$.

To use the asymptotic prediction of the order-parameter decay 
in Ref.~\cite{Calabrese_2012}, we need to reshape the aforementioned 
unitary transformation connecting pre- and post-quench 
diagonal fermions into
$e^{-i \alpha_p Z} e^{i \Delta_{p} X/2} e^{i \alpha^{0}_{p} Z }$,
so that we can absorb the irrelevant phases into a redefinition of
new diagonal fermions $\tilde {\bm \xi}_{p} \equiv e^{i \alpha_p Z}  {\bm \xi}_{p}$
and $\tilde {\bm \xi}^{0}_{p} \equiv e^{i \alpha^{0}_p Z}  {\bm \xi}^{0}_{p}$.
These finally identify a single Bogoliubov rotation with angle $\Delta_p$,
such that
\be
\cos(\Delta_{p}) = \cos(\theta_{p})\cos(p)+\sin(\theta_{p})\cos(2\epsilon)\sin(p).
\ee
The Floquet dynamics 
therefore induces an unavoidable exponential decay of the magnetization, 
with exact decay rate given by
\bea\label{eq:gamma}
\gamma & = &  -\int_{0}^{\pi}\frac{dp}{\pi} \partial_p\phi_{p} \ln |\cos(\Delta_p)| .
\eea
A Taylor expansion of Eq.~\eqref{eq:gamma} for small values of $\epsilon$ produces the asymptotics in Eq.~\eqref{eq_tauscaling}.\\

\subsection{The exact Floquet operator}

The Floquet operator for the integrable kicked Ising model has been constructed from the matrices
$\mathbb{K}$ and $\mathbb{V}$ defined in Eq.~\eqref{eq:KVmatr}
so that the product of the two matrix exponent reads
\begin{widetext}
\be\label{eq:eKeU}
e^{\mathbb{K}}e^{\mathbb{V}} = 
\begin{bmatrix}
e^{-2i\epsilon} [\cos(2J) +i \sin(2J)\cos(p)]& 
-e^{-2i\epsilon} \sin(2J)\sin(p)\\
e^{2i\epsilon} \sin(2J)\sin(p) & 
e^{2i\epsilon} [\cos(2J) -i \sin(2J)\cos(p)]
\end{bmatrix}.
\ee
\end{widetext}
This is a unitary matrix with $\det[e^{\mathbb{K}}e^{\mathbb{V}} ] = 1$
and eigenvalues $\{e^{i\phi_{p}}, e^{-i\phi_{p}}\}$ such that
$\tr[e^{\mathbb{K}}e^{\mathbb{V}} ] = 2\cos(\phi_{p})$ with
\be
\cos(\phi_{p}) =  \cos(2J)\cos(2\epsilon) + \sin(2J)\sin(2\epsilon)\cos(p).
\label{phip_def}
\ee
The matrix in Eq.~\eqref{eq:eKeU} can be thus parametrized as an SU(2) rotation 
\be
e^{\mathbb{K}}e^{\mathbb{V}} = e^{i\phi_{p} \, \hat r_{p}\cdot \boldsymbol{\sigma}}
= \cos(\phi_{p}) + i \sin(\phi_{p}) \hat r_{p}\cdot \boldsymbol{\sigma},
\ee
where we introduced the unit vector $\hat r_{p} = \tr[e^{\mathbb{K}}e^{\mathbb{U}}\boldsymbol{\sigma}]/[2i\sin(\phi_{p})]$,
whose components are 
\be\label{eq:r}
\hat r_{p} = \frac{1}{\sin(\phi_{p})}
\begin{bmatrix}
\sin(2J)\sin(2\epsilon)\sin(p)\\
-\sin(2J)\cos(2\epsilon)\sin(p)\\
\sin(2J)\cos(2\epsilon)\cos(p)-\cos(2J)\sin(2\epsilon)
\end{bmatrix}.
\ee
We can parametrize $\hat r_{p}$ with the angles $\{\theta_p, \xi_p\}$
such that (notice that is not the standard polar-azimuth parametrization)
\be\label{eq:r_theta_xi}
\hat r_{p} = 
\begin{bmatrix}
\sin(\theta_p)\sin(\xi_p)\\
-\sin(\theta_p)\cos(\xi_p)\\
\cos(\theta_p)
\end{bmatrix}.
\ee
Comparing the definition in Eq.~\eqref{eq:r_theta_xi} with the results in Eq.~\eqref{eq:r},
we may easily identify $\xi_{p} = 2\epsilon$ (for $-\pi/2< 2\epsilon < \pi/2$) independent of the momentum,
and
\begin{subequations}
\label{thetap_def}
\bea
\sin(\theta_p) &=& \frac{\sin(2J)\sin(p)}{\sin(\phi_p)},\\
\cos(\theta_p) &=& \frac{\sin(2J)\cos(2\epsilon)\cos(p)\!-\!\cos(2J)\sin(2\epsilon)}{\sin(\phi_p)}, \hspace*{1cm}
\eea
\end{subequations}
so that the effective Hamiltonian $\mathbb{H} = i\phi_p \, \hat r_p \cdot \boldsymbol{\sigma}$
takes the form
\be
\mathbb{H} = i\phi_p 
\begin{bmatrix}
\cos(\theta_p) & i \sin(\theta_p)e^{-2i\epsilon}\\
-i \sin(\theta_p)e^{2i\epsilon} & -\cos(\theta_p)\\
\end{bmatrix},
\ee
and is diagonalized by the unitary transformation
\be\label{eq:Vp}
\mathbb{U}_{p} = 
\begin{bmatrix}
\cos(\theta_p/2) e^{-i\epsilon}& -i \sin(\theta_p/2) e^{-i\epsilon}\\
-i \sin(\theta_p/2) e^{i\epsilon} & \cos(\theta_p/2) e^{i\epsilon} \\
\end{bmatrix},
\ee
which can be rewritten as the product of two SU(2) rotations
\be\label{eq:Vp2}
\mathbb{U}_{p} = e^{-i \epsilon Z} e^{-i \theta_p X /2} .
\ee
Let us notice that, for $\epsilon = 0$, the transformation in Eq.~\eqref{eq:Vp2}
reduces to a single rotation around $\hat x$ with angle $\theta_p = p$,
inducing the usual Bogoliubov transformation diagonalizing the Ising Hamiltonian
in Eq.~\eqref{eq:H}.

\subsection{The order-parameter dynamics}

The free-fermion techniques outlined above  
allow the computation of the discrete time evolution of the two-point function
$\langle Z_{0}(n) Z_{\ell}(n) \rangle$, starting from
a $\mathbb{Z}_{2}$-symmetric state. 
Here we assume to work in the $P=+1$ sector, 
thus preparing the initial state to be $\ket{0_+} \equiv (\ket{+} + \ket{-})/\sqrt{2}$,
namely the vacuum of the $\xi^{0}_{p}$ fermions.
From the large-distance behavior of the two-point function we have
$|m(n)|^{2} = \lim_{\ell\to\infty} \langle Z_{0}(n) Z_{\ell}(n) \rangle$.

The order-parameter correlation function can be
computed using the Majorana fermions
\be\label{eq:majorana}
a_{2j} = i (c^{\dag}_j - c_{j}) , \qquad a_{2j-1} = c^{\dag}_{j} + c_{j},
\ee
which are Hermitian operators satisfying $\{a_{i},a_{j}\} = 2\delta_{ij}$.
Indeed, the expectation value of the two-point function involves the evaluation of
the Pfaffian of a skew-symmetric real matrix, namely
$\langle Z_{0}(n) Z_{\ell}(n) \rangle =  {\rm pf}  [ \mathbb{M}(n) ]$,
where
$
\mathbb{M}_{ij}(n)  =  -i\langle a_{i}(n) a_{j}(n) \rangle + i\delta_{ij}
$
with $\{i,j\} \in \{0,1,\dots, 2\ell -1 \}$. 
Actually, since we only need the modulus of the two-point function, 
we may just compute the determinant, since $\det[\mathbb{M}(n)] =  {\rm pf}  [ \mathbb{M}(n) ]^2$.

Therefore, the building block is the fermionic two-point function,
whose time evolution can be computed by re-writing 
the Majorana operators in Eq.~(\ref{eq:majorana}) in terms of the new
diagonal operators in Eq.~(\ref{eq:xi}), whose time-evolution is a trivial phase factor.
In particular, by using the fact that, in the thermodynamic limit,
\be
{\bm a}_j 
=\int_{-\pi}^{\pi} \frac{dp}{2\pi} e^{-i p j}
\mathbb{A} \,
{\bm \eta}_{p}, \quad
\mathbb{A}=
\begin{bmatrix}
1 & 1\\
-i & i
\end{bmatrix},
\ee
where ${\bm a}^{\dag}_j \equiv (a_{2j-1}, a_{2j})$,
we easily obtain after $n$ periods
\be\label{eq:corr_evolution}
\langle {\bm a}_j (n)  {\bm a}^{\dag}_l (n) \rangle
 =
\int_{-\pi}^{\pi} \frac{dp}{2\pi} \, e^{-ip(j-l)} \,
\mathbb{W}_{p}(n) \,
\langle {\bm\eta}_p  {\bm\eta}^{\dag}_p\rangle \,
\mathbb{W}^{\dag}_{p}(n),
\ee
where
$
\mathbb{W}_{p}(n) = \mathbb{A} \, \mathbb{U}_{p}\, e^{i n \phi_p \,Z} \, \mathbb{U}^{\dag}_{p}
$,
and the initial correlation function is easily computed as 
$
\langle {\bm\eta}_p  {\bm\eta}^{\dag}_p\rangle = 
e^{-i p X /2} \langle {\bm\xi}^{0}_{p}  {\bm\xi}^{0\dag}_p\rangle  e^{i p X /2}
$,
with
$
\langle {\bm\xi}^{0}_{p}  {\bm\xi}^{0\dag}_p\rangle = (1+Z)/2
$.

\section{Derivation of the effective Floquet Hamiltonian}
\label{app_derivationFloquet}

In this section, we present the explicit construction of the Floquet Hamiltonian which conserves at any order the number of domain walls $D_1$ in \eqref{eq_D1}. 
As explained in the main text, we follow a different procedure from Ref.~\cite{DeRoeckVerreet}, which allows us to obtain explicit low-order formulas more easily. 
For the sake of generality, we consider a kicked Ising chain with
variable-range interactions as in Eq.~\eqref{eq_FloquetLR}
and mixed-field kick as in Eq.~\eqref{eq_tiltedkick}. 
For convenience, we parametrize $\epsilon=\eta\cos\theta$, $h=\eta\sin\theta$ in Eq.~\eqref{eq_HKH0}.
It is useful here to introduce the associated Hamiltonians
\begin{equation}
\label{eq_HKH0}
     H_K = -\sum_j (\cos\theta X_j + \sin\theta Z_j) \; , \; H_0 = -\sum_{j} \sum_{r = 1}^R J_{r} Z_j Z_{j+r} \,.
\end{equation}
The procedure is based on two steps:
\begin{enumerate}
    \item using the replica resummation  of Ref.~\cite{ProsenPolkovnikovPRL18_ReplicaBCHResummation}, 
    we combine the kick generator $K = e^{- i H_K}$ with the interactions $V_{\mathbf{J}} = e^{-i H_0}$ into an approximate Floquet Hamiltonian,
\be
\label{eq_HF}
K \, V_{\mathbf{J}} \simeq e^{-i H^F};
\ee
\item we apply a static Schrieffer-Wolff transformation to $H^F$ to cancel order-by-order all the terms which do not commute with $D_1$.
\end{enumerate}
The two procedures are detailed in the next two subsections, respectively.

\subsection{Replica resummation for the kicked variable-range and mixed-field Ising chain \label{app_replica}}

 To keep the calculation as general as possible, we actually consider a generic 2-body interactions in the $\hat{z}$-direction
\begin{equation}
    H_K = - \sum_{i<j} J_{i,j} Z_i Z_j \;.
\end{equation}
The translationally invariant case in Eq.~\eqref{eq_HF} can be readily recovered at the end setting $J_{i,j} = J_{r = j-i}$.

The approach of Ref.~\cite{ProsenPolkovnikovPRL18_ReplicaBCHResummation} allows to obtain $H^F$ defined in Eq.~\eqref{eq_HF} perturbatively in small $\eta$.
\be
H^F = H_0 + \eta H_1 + \eta^2 H_2 + \dots  \; .
\ee
While this series is expected to be generically divergent, signalling the nonexistence of a physically meaningful conserved energy, truncations of the series do have a crucial physical meaning as approximate conserved energy for long times (see Ref.~\cite{ProsenPolkovnikovPRL18_ReplicaBCHResummation} and the discussion in the main text).

The $p$-th coefficient $H_p$ of the replica expansion, $p\ge1$, is obtained by writing the logarithm
\be
H^F = i \log e^{-i \eta H_K} e^{-i H_0}
\ee
as a limit $\log x = \lim_{n\to0} \frac 1 n (x^n-1)$, hence taking the $p$-th derivative with respect to $\eta$ at $\eta=0$ before analytically continuing the result for non-integer $n$ and taking the replica limit $n\to0$,
\be
H_p \!=\! i \lim_{n\to0} \frac 1 n  \frac 1 {p!}\frac{\partial^p}{\partial \eta^p} \!\bigg\rvert_{\eta=0}
\underbrace{
(e^{-i \eta H_K} e^{-i H_0}) \!\cdots\! (e^{-i \eta H_K} e^{-i H_0})}
_{n \text{ times}} .
\ee

The building block for explicit computations is the operator
\be
\widetilde{H}_m =
e^{im H_0} H_K e^{-im H_0}.
\ee
In the model under consideration, one obtains
\begin{eqnarray}
\widetilde{H}_m & = &
- \cos \theta \sum_j
\bigg[
\cos\Big(
2m \zeta_j
\Big) X_j \nonumber \\
&& -
\sin\Big(
2m \zeta_j
\Big) Y_j
\bigg]
- \sin \theta \sum_j Z_j
. \quad
\end{eqnarray}
where
$
\zeta_j 
=
\sum_{i (\neq j)} J_{i , j} Z_i$ is the effective field acting on spin $j$.
The first term of the replica expansion is then computed as
\begin{eqnarray}
  \label{eq_H1replica}
  H_1 & = &\lim_{n\to0} \frac 1 n \sum_{m=0}^{n-1} \widetilde{H}_m \nonumber \\
  & = & - \! \cos\theta \sum_j \! \Big( \zeta_j \cot \zeta_j \, X_j + \zeta_j \, Y_j \Big) - \! \sin \theta \sum_j \! Z_j , \qquad
\end{eqnarray}
We see that first-order terms contain at most one local spin-flip operator ($X_j$ or $Y_j$). Compared to the bare kick generator, such spin-flip terms are 
``decorated'' by diagonal operators.
The range of these decorations is inherited from that of the interaction couplings $J_{i,j}$ in $H_0$. In particular, the amplitude of a local spin-flip process is only influenced by the configuration of the spins in a neighborhood of radius $R$.
In the case of the conventional quantum Ising chain ($R=1$), Eq.~\eqref{eq_H1replica} reduces to the expression first obtained in Ref.~\cite{ProsenPolkovnikovPRL18_ReplicaBCHResummation}. Equation~\eqref{eq_H1replica} constitutes a generalization to Ising chains with arbitrary couplings.
In the high-frequency limit $J_{i,j} \to0$, we correctly retrieve the standard BCH expression at lowest order, $ H_1 \equiv H_K $.
Importantly, divergences appear in the replica coefficients of particular off-diagonal (``resonant'') transitions. 
This brings in the need for an incommensurability assumption on the nonvanishing interaction couplings, as discussed below. 

The complexity of the computation increases rapidly with the perturbative order.
To illuminate the general structure of the replica series expansion,
it is instructive to report here the result for $p=2$.
The second term is computed as~\cite{ProsenPolkovnikovPRL18_ReplicaBCHResummation}
\be
H_2 = \frac i 2 \lim_{n\to0} \frac 1 n \; \sum_{0\le m_1 < m_2 < n} \big[\widetilde{H}_{m_2},\widetilde{H}_{m_1}\big].
\ee
Calculation gives
\begin{widetext}
\be
\label{eq_H2replica}
\begin{split}
H_2 = &
+ \cos^2\theta \sum_j \lambda_j Z_j \\
&+
\cos^2\theta
\sum_{j_1,j_2} \Big(
\mu^{XX}_{j_1,j_2} X_{j_1} X_{j_2}
+
\mu^{YY}_{j_1,j_2} Y_{j_1} Y_{j_2}
+
\mu^{XY}_{j_1,j_2} X_{j_1} Y_{j_2}
+ \mu^{YX}_{j_1,j_2} Y_{j_1} X_{j_2} \Big) \\
&+ 
\cos\theta\sin\theta
\sum_j \Big( \nu^X_{j} X_j
+ \nu^Y_{j} Y_j \Big).
\end{split}
\ee
The coefficients $\lambda$, $\mu$, $\nu$ are diagonal operators obtained via the following analytical replica limits:
\be
\label{eq_2ndreplicacoeff}
\begin{split}
    \lambda_j &=
    \lim_{n\to0} \frac 1 n  
\sum_{0\le m_1 < m_2 < n}
- \frac 1 2
\sin[2(m_1-m_2)\zeta_j]
     \\
    \mu^{XX}_{j_1,j_2} &= \lim_{n\to0} \frac 1 n \sum_{0\le m_1 < m_2 < n} 
    \sin(2m_2 \zeta_{j_2})
    \sin(2 m_1 \zeta'_{j_1,j_2})
    \sin(2m_1 J_{j_1,j_2})-[m_1 \leftrightarrow m_2]
    \\
    \mu^{YY}_{j_1,j_2} &= \lim_{n\to0} \frac 1 n \sum_{0\le m_1 < m_2 < n} 
    \cos(2m_2 \zeta_{j_2})
    \cos(2m_1 \zeta'_{j_1,j_2})
    \sin(2m_1 J_{j_1,j_2})-[m_1 \leftrightarrow m_2]
    \\
    \mu^{XY}_{j_1,j_2} &= \lim_{n\to0} \frac 1 n \sum_{0\le m_1 < m_2 < n} 
    \cos(2m_2 \zeta_{j_2})
    \sin(2m_1 \zeta'_{j_1,j_2})
    \sin(2m_1 J_{j_1,j_2})-[m_1 \leftrightarrow m_2]
    \\
    \mu^{YX}_{j_1,j_2} &= \lim_{n\to0} \frac 1 n \sum_{0\le m_1 < m_2 < n} 
    \sin(2m_2 \zeta_{j_2})
    \cos(2m_1 \zeta'_{j_1,j_2})
    \sin(2m_1 J_{j_1,j_2})-[m_1 \leftrightarrow m_2]
    \\
    \nu^{X}_{j} &= \lim_{n\to0} \frac 1 n \sum_{0\le m_1 < m_2 < n} 
    \sin(2 m_1 \zeta_{j})
    - \sin(2 m_2 \zeta_{j})
    \\
    \nu^{Y}_{j} &= \lim_{n\to0} \frac 1 n \sum_{0\le m_1 < m_2 < n} 
    \cos(2 m_1 \zeta_{j})
    - \cos(2 m_2 \zeta_{j})
    \\
\end{split}
\ee
\end{widetext}
In these equations, $\zeta'_{j_1,j_2} \equiv \sum_{i(\neq j_1,j_2)}J_{i,j_1}Z_i = \zeta_{j_1} - J_{j_1,j_2} Z_{j_2}$.
The evaluation of the replica limits in Eq.~\eqref{eq_2ndreplicacoeff} can be easily performed with the help of algebraic manipulators such as Wolfram Mathematica.
In particular, finite replica limits exist in all cases, and $\mu^{YX}_{j_1,j_2} \equiv 0$.
A crucial property  is that
\be
\mu^{\alpha\beta}_{j_1,j_2} \; =
\; \mathcal{O}(J_{j_1,j_2}) \qquad \text{as } J_{j_1,j_2}\to0 
\ee
for all $\alpha$, $\beta$ = $X$, $Y$.
This implies that for interactions with finite range $R$, pairs of spin flips in $H_2$ can only occur at maximum distance $R$. 
Furthermore, for long-range interactions, large-distance pairs are suppressed as rapidly as the interaction tail.
We note that, compared to the first order term, a larger (finite) set of resonances appear at second-order.

The second-order processes described by Eq.~\eqref{eq_H2replica} include diagonal terms ($\propto \epsilon^2$, first line), single local spin-flips ($\propto h \epsilon$, third line), and pairs of spin-flips ($\propto \epsilon^2$, second line).
As in the first order term~\eqref{eq_H1replica}, spin-flip operators ($X_j$, $Y_j$, $X_{j_1}X_{j_2}$, $X_{j_1}Y_{j_2}$, $Y_{j_1}Y_{j_2}$)  are ``decorated'' by diagonal operators determined by the configuration of spins in a neighborhood of radius $R$ from the location of the flips.
Furthermore, double spin-flips come in quasilocal pairs, i.e., spins are flipped at most $R$ sites away from each other.

Higher-order terms in the replica expansion have increasingly complex coefficients, but crucially exhibit a hierarchical structure in terms of the maximum number and the quasilocality of off-diagonal spin-flip processes. 
This property follows from expressing the $p$-th order term as $p$ nested commutators of the building block $\widetilde{H}_m$ before taking the replica limit~\cite{ProsenPolkovnikovPRL18_ReplicaBCHResummation}.
In particular, terms of order $\epsilon^p$ in the Floquet Hamiltonian feature a quasilocal product of at most $p$ spin-flip local operators ($X$ or $Y$), each one located at most $R$ sites away from the next, ``dressed'' by diagonal coefficients involving operators $Z$ and couplings $J_r$.

As already noted  above, these coefficients may have poles when particular integer combinations of the couplings $\sum_r d_r J_r$, $d_r\in\mathbb{Z}$, equal a multiple of $2\pi$. Such divergences stem from corresponding sequence of perturbative transitions generated by the kick $K_\epsilon$ hitting a resonance in the unperturbed spectrum of $V_{\mathbf{J}}$.
These resonances make the perturbative series ill-defined. 
To resolve this issue, we must assume a condition of strong incommensurability of the couplings, such that integer combinations of $\{J_r\}$ and $2\pi$ remain sufficiently removed from zero unless correspondingly large integers are chosen. More precisely, we assume the following Diophantine condition: there exist $x,\tau>0$, such that for all nonzero integer arrays $\mathbf{n} \equiv (n_0,n_1,\dots,n_R) \in \mathbb{Z}^{R+1}$,
\be
\label{eq_diophantine}
\bigg\lvert\sum_{r=1}^R n_r J_r - 2\pi n_0 \bigg\rvert > \frac x {||\mathbf{n}||^\tau} \, ,
\ee
where $||\mathbf{n}|| \equiv \max (n_0,n_1,\dots,n_R) $.
It is not hard to show that almost all choices of the couplings $J_r$ satisfy this strong incommensurability condition for some $x>0$ when $\tau>R$ (see, e.g., the aforementioned Ref.~\cite{DeRoeckVerreet}). This condition guarantees that the resummation of the BCH series is formally well defined to all orders in $\epsilon$, and that the coefficients do not grow too wildly with the perturbative order.

\subsection{Derivation of the domain-wall conserving effective Hamiltonian}
\label{app_SW}

In this subsection we apply the Schrieffer-Wolf transformation to the Floquet Hamiltonian in Eq.~\eqref{eq_HF}, to obtain a new operator which conserves the number of domain walls $D_1$. For simplicity, we focus on the short range case $R = 1$ and we restrict to the first order. In this case, the effective fields reduce to $\zeta_j = J (Z_{j+1} + Z_{j-1})$ and the Floquet Hamiltonian up to first order takes the form (cf. Ref.~\cite{ProsenPolkovnikovPRL18_ReplicaBCHResummation})
\begin{equation}
    \label{eq_FloHSRFO}
    H_{\leq 1} = - J \sum_j Z_j Z_{j+1} + \eta H_1 
\end{equation}
where 
\begin{multline}
H_1 = 
- \cos\theta \bigg[
\Big(J \cot 2J + \frac 1 2  \Big) \sum_j X_j \\
  + \Big(J \cot 2J - \frac 1 2  \Big) \sum_j Z_{j-1} X_j Z_{j+1} + J \sum_j (Z_{j-1} + Z_{j+1}) Y_j
\bigg] \\
 - \sin \theta \; \sum_j Z_j \, .
\end{multline}
Note that for $J \ll 1$ one has $J \cot 2J \simeq 1/2$, hence we retrieve the Trotter limit of the BCH formula, with $H^F \sim H_0 + \eta H_K$.

Hence, we split the first-order term $H_1$ into two orthogonal components, a domain-wall conserving and a domain-wall nonconserving part:
\be
H_1 = \DD + \VV,
\ee
with $\big[ \DD, \sum_j Z_j Z_{j+1} \big] = 0$ and $\VV$ purely off-diagonal between sectors with different number of domain walls. Explicitly we have
\begin{align}
\DD =& - \cos\theta \sum_j 
P^\uparrow_{j-1} X_{j} P^\downarrow_{j+1} + P^\downarrow_{j-1}  X_{j} P^\uparrow_{j+1} + \nonumber \\
& - \sin \theta \; \sum_j Z_j  \, ,   \\
\VV =&\\ & - \cos \theta \Bigl(  2J \cot 2J \sum_j 
P^\uparrow_{j-1} X_{j} P^\uparrow_{j+1} + P^\downarrow_{j-1}  X_{j} P^\downarrow_{j+1}
+ \nonumber
\\ & +   \; 2J \sum_j 
P_{j-1}^\uparrow Y_{j} P^\uparrow_{j+1} - P^\downarrow_{j-1}  Y_{j} P^\downarrow_{j+1}\Bigr)
 \; ,
\end{align}

Now we fix $S_1$ in such a way to exactly cancel $\VV$ in the transformed Floquet operator, i.e.,
\begin{multline}
H'_{\leq 1}  =  e^{-i \eta S_1}  H_{\leq 1} e^{i \eta S_{1}}  = J  \; \sum_j Z_j Z_{j+1}  +   \eta \DD + 
\\ +  \eta \; 
\overset{\overset{!}{=} \, 0}{\overbrace{  \bigg(  \VV -i J  \Big[  S_{1} ,  \sum_j Z_j Z_{j+1}  
\Big] \bigg)  }}   + \mathcal{O}(\eta^2).
\end{multline}
This condition for $S_1$ is solved by
\begin{multline}
 S_1 = \frac{\cos\theta}{2} \sum_j P_{j-1}^{\uparrow} [X_j - \cot(2J) Y_j] P_{j+1}^{\uparrow} +\\+  P_{j-1}^{\downarrow} [X_j + \cot(2J) Y_j] P_{j+1}^{\downarrow}
\end{multline}
which corresponds to Eq.~\eqref{eq_S1} in the main text. With such a choice, one recovers the Floquet operator $U_1' = e^{- i H_{\leq 1}'}$ in Eq.~\eqref{eq_Utilde}.

\bibliography{biblio}

\end{document}